\documentclass[11pt]{article}              

\usepackage[margin=25mm]{geometry}
\usepackage{graphicx}
\usepackage{colortbl}
\usepackage{subfigure}
 
\usepackage{fancyhdr}
\usepackage{amssymb,amsfonts,amsmath,amsthm}
\usepackage{natbib}
\usepackage{bstnotations}

\usepackage{authblk}
	
\graphicspath{{./},{./Figures/}}

\usepackage[english]{babel} 
\usepackage{graphicx}
\usepackage{float}
\usepackage{color}
\usepackage{amsthm}
\usepackage{amsmath}
\usepackage{amssymb}
\usepackage{amsfonts}
\usepackage{soul}
\usepackage{geometry}
\usepackage{natbib}
\usepackage{setspace}
\usepackage{stmaryrd}
\usepackage{placeins}
\makeatletter
\let\ftype@table\ftype@figure
\makeatother

\newcommand{\res}{{\mathcal R}}

\newcommand{\hsp}{\hspace{0.3mm}}

\newcommand{\vli}{v_l^{(i)}}
\newcommand{\vlj}{v_l^{(j)}}
\newcommand{\vri}{v_r^{(i)}}
\newcommand{\zkli}{z_{k,l}^{(i)}}

\newcommand{\Pki}{P_k^{(i)}}
\newcommand{\Pkj}{P_k^{(j)}}
\newcommand{\Imax}{I_{\rm max}}
\newcommand{\Derrmin}{\Delta\widehat{err}_{\rm min}}

\newcommand{\errCV}{\widehat{err}_{\rm CV3}}
\newcommand{\errLOO}{\widehat{err}_{\rm LOO}^*}

\newcommand{\zlio}{ z_{0,l}^{(i)}}
\newcommand{\zliop}{ z_{0,l'}^{(i)}}

\newcommand{\zljo}{ z_{0,l}^{(j)}}
\newcommand{\zljop}{ z_{0,l'}^{(j)}}

\newcommand{\zklj}{z_{k,l}^{(j)}}
\newcommand{\zkljp}{z_{k,l'}^{(j)}}

\newcommand{\zklip}{z_{k,l'}^{(i)}}

\newcommand{\vlip}{v_{l'}^{(i)}}

\newcommand{\cmu}{\cm_{\iu}(\ve X_{\iu})}

\newcommand{\cmv}{\cm_{\iv}(\ve X_{\iv})}

\newcommand{\cmtu}{\widetilde{\cm}_{\iu}(\ve X_{\iu})}
\newcommand{\cmtun}{\widetilde{\cm}_{\smallsetminus\iu}(\ve X_{\smallsetminus\iu})}

\newcommand{\Stu} {\widetilde{S}_{\iu}}
\newcommand{\StuT} {\widetilde{S}^T_{\iu}}
\newcommand{\Stun}{\widetilde{S}_{\smallsetminus\iu}}

\newcommand{\cmtuLRA}{\widetilde{\cm}_{\iu}^{\rm LRA}(\ve X_{\iu})}
\newcommand{\cmtunLRA}{\widetilde{\cm}_{\smallsetminus\iu}^{\rm LRA}(\ve X_{\smallsetminus\iu})}
\newcommand{\cmtiLRA}{\widetilde{\cm}_i^{\rm LRA}(X_i)}
\newcommand{\cmtinLRA}{\widetilde{\cm}_{\smallsetminus i}^{\rm LRA}(X_{\smallsetminus i})}

\begin{document}
\title{Global sensitivity analysis using low-rank tensor approximations} 

\author[1]{K. Konakli} \author[1]{B. Sudret} 

\affil[1]{Chair of Risk, Safety and Uncertainty Quantification,
  
  ETH Zurich, Stefano-Franscini-Platz 5, 8093 Zurich, Switzerland}

\date{}
\maketitle

\abstract{In the context of global sensitivity analysis, the Sobol' indices constitute a powerful tool for assessing the relative significance of the uncertain input parameters of a model. We herein introduce a novel approach for evaluating these indices at low computational cost, by post-processing the coefficients of polynomial meta-models belonging to the class of low-rank tensor approximations. Meta-models of this class can be particularly efficient in representing responses of high-dimensional models, because the number of unknowns in their general functional form grows only linearly with the input dimension. The proposed approach is validated in example applications, where the Sobol' indices derived from the meta-model coefficients are compared to reference indices, the latter obtained by exact analytical solutions or Monte-Carlo simulation with extremely large samples. Moreover, low-rank tensor approximations are confronted to the popular polynomial chaos expansion meta-models in case studies that involve analytical rank-one functions and finite-element models pertinent to structural mechanics and heat conduction. In the examined applications, indices based on the novel approach tend to converge faster to the reference solution with increasing size of the experimental design used to build the meta-model.\\[1em] 

}

\maketitle

\section{Introduction}

Robust predictions via computer simulation necessitate accounting for the prevailing uncertainties in the parameters of the computational model. Uncertainty quantification provides the mathematically rigorous framework for propagating the uncertainties surrounding the model input to a response quantity of interest. It comprises three fundamental steps \citep{Sudret2007, Derocquigny2012}: First, the model representing the physical system under consideration is defined; the model maps a given set of input parameters to a unique value of the response quantity of interest.  The second step involves the probabilistic description of the input parameters by incorporating available data, expert judgment or a combination of both. In the third step, the uncertainty in the input is propagated upon the response quantity of interest through repeated evaluations of the model for appropriate combinations of the input parameters. In cases when the uncertainty in the response proves excessive or when it is of interest to reduce the dimensionality of the model, sensitivity analysis may be employed to rank the input parameters with respect to their significance for the response variability. Accordingly, important parameters may be described in further detail, whereas unimportant ones may be fixed to nominal values.

Methods of sensitivity analysis can be generally classified into local and global methods. Local methods are limited to examining effects of variations of the input parameters in the vicinity of nominal values. Global methods provide more complete information by accounting for variations of the input parameters in their entire domain. Under the simplifying assumption of linear or nearly linear behavior of the model, global sensitivity measures can be computed by fitting a linear-regression model to a set of input samples and the respective responses (see \eg \citet{Iooss2014, Wei2015} for definitions of such measures). The same methods can be employed in cases with models that behave nonlinearly but monotonically, after applying a rank transformation on the available data. Variance-based methods represent a more powerful and versatile approach, also applicable to nonlinear and non-monotonic models. These methods, known as functional ANOVA (denoting ANalysis Of VAriance) techniques in statistics, rely upon the decomposition of the response variance as a sum of contributions from each input parameter or their combinations \citep{Efron:Stein:1981}. The Sobol' indices, originally proposed in \citet{Sobol1993}, constitute the most popular tool thereof. Although these indices have proven powerful in a wide range of applications, their definition is ambiguous in cases with dependent input variables, which has led to different extensions of the original framework \citep{DaVeiga2009, LiRabitz2010, Kucherenko2012, Mara2012, Zhang2015}. An alternative perspective is offered by the distribution-based indices, which are well-defined regardless of the dependence structure of the input \citep{Borgonovo2007, Liu2010, Borgonovo2014, Zhou2015, Greegar2016}. The key idea is to use an appropriate distance measure to evaluate the effect of suppressing the uncertainty in selected variables on the Probability Density Function (PDF) or the Cumulative Distribution Function (CDF) of the response. These indices are especially useful when consideration of the variance only is not deemed sufficient to characterize the response uncertainty. However, contrary to the Sobol' indices, they do not sum up to unity, which may hinder interpretation. For further information on global sensitivity analysis methods, the interested reader is referred to the review papers \citep{Saltelli2008, Iooss2014, Wei2015, Borgonovo2016}.

It should be mentioned that different classifications of sensitivity analysis techniques can be found in the literature. In cases when one needs to perform a fast exploration of the model behavior with respect to a possibly large number of uncertain input parameters, the so-called screening methods may be employed. These methods can provide a preliminary ranking of the importance of the various input parameters at low computational cost before more precise and costly methods are applied. The Cotter method \citep{Cotter1979} and the Morris method \citep{Morris1991} are widely used screening methods, with the latter covering the input space in a more exhaustive manner. The more recently proposed derivative-based global sensitivity measures can also be classified into this category, while they also provide upper bounds for the Sobol' indices \citep{Sobol2009, Lamboni2013, Sudret2015a}.

The focus of the present paper is on sensitivity analysis by means of Sobol' indices. We limit our attention to cases with independent input and address the computation of these indices for high-dimensional expensive-to-evaluate models, which are increasingly used across engineering and sciences. Various methods have been investigated for computing the Sobol' indices based on Monte Carlo simulation \citep{Archer1997, Sobol2001, Saltelli2002, Sobol2005, Saltelli2010}; because of the large number of model evaluations required, these methods are not affordable for computationally costly models. To overcome this limitation, more efficient estimators have recently been proposed \citep{Sobol2007, Janon2013, Kucherenko2015, Owen2013}. A different approach is to substitute a complex model by a \emph{meta-model}, which has similar statistical properties while maintaining a simple functional form (see \eg \citet{Oakley2004, Marrel2009, Storlie2009, Zuniga2013, Zhang2014, SudretHandbookUQ} for global sensitivity analysis with various types of meta-models). \citet{Sudret2008c} proposed to compute the Sobol' indices by post-processing the coefficients of Polynomial Chaos Expansion (PCE) meta-models. The key concept in PCE is to expand the model response onto a basis made of orthogonal multivariate polynomials in the input variables. The computational cost of the associated Sobol' indices essentially reduces to the cost of estimating the PCE coefficients, which can be curtailed by using sparse PCE \citep{Blatman2010c}. The PCE-based approach for computing the Sobol' indices is employed by a growing number of researchers in various fields including hydrogeology \citep{Fajraoui2011, Formaggia2013, Deman2015}, geotechnics \citep{Albittar2012}, ocean engineering \citep{Alexanderian2012}, biomedical engineering \citep{Huberts2014}, hybrid dynamic simulation \citep{Abbiati2015} and electromagnetism \citep{Kersaudy2014, Yuzugullu2015}. Unfortunately, the PCE approach faces the so-called \emph{curse of dimensionality}, meaning the exploding size of the candidate basis with increasing dimension.

The goal of this paper is to derive a novel approach for solving global sensitivity analysis problems in high dimensions. To this end, we make use of a recently emerged technique for building meta-models with polynomial functions based on Low-Rank Approximations (LRA) \citep{Nouy2010b, Chevreuil2013, Doostan2013, Hadigol2014partitioned, Validi2014low, Konakli2015UNCECOMP}. LRA express the model response as a sum of a small number or rank-one tensors, which are products of univariate functions in each of the input parameters. Because the number of unknown coefficients in LRA grows only linearly with the input dimension, this technique is particularly promising for dealing with cases of high dimensionality. We herein derive analytical expressions for the Sobol' sensitivity indices based on the general functional form of LRA with polynomial bases. As in the case of PCE, the computational cost of the LRA-based Sobol' indices reduces to the cost of estimating the coefficients of the meta-model. Once a LRA meta-model of the response quantity of interest is available, the Sobol' indices can be calculated with elementary operations at nearly zero additional computational cost.

The paper is organized as follows: In Section~2, we review the basic concepts of Sobol' sensitivity analysis and define the corresponding sensitivity indices. In Section~3, we describe the mathematical setup of non-intrusive meta-modeling and define error measures that characterize the meta-model accuracy. After reviewing the computation of Sobol' indices using PCE meta-models in Section~4, we introduce the LRA-based approach in Section~5. In this, we first detail a greedy algorithm for the construction of LRA in a non-intrusive manner and then, use their general functional form to derive analytical expressions for the Sobol' indices. In Section~6, we demonstrate the accuracy of the proposed method by comparing the LRA-based indices to reference ones, with the latter representing the exact solution or Monte-Carlo estimates relying on a large sample of responses of the actual model. Furthermore, we examine the comparative performance of the LRA- and PCE-based approaches in example applications that involve analytical rank-one functions and finite-element models pertinent to structural mechanics and heat conduction. The main findings are summarized in Section 7.

\section{Sobol' Sensitivity analysis}
\label{sec:Sobol}

We consider a computational model $\cm$ describing the behavior of a physical or engineered system of interest. Let $\ve X=\{X_1 \enum X_M\}$ denote the $M$-dimensional input random vector of the model with prescribed joint PDF $f_{\ve X}$. Due to the input uncertainties embodied in $\ve X$, the model response becomes random. By limiting our focus to a scalar response quantity $Y$, the computational model represents the map:
\begin{equation}
\label{eq:model}
\ve X \in \cd_{\ve X} \subset \Rr^M\longmapsto Y=\cm (\ve X) \in \Rr,
\end{equation}
where $\cd_{\ve X}$ denotes the support of $\ve X$.

Sobol' sensitivity analysis aims at apportioning the uncertainty in $Y$, described by its variance, to contributions arising from the uncertainty in individual input variables and their interactions. As explained in the Introduction, the theoretical framework described in the sequel is confined to the case when the components of $\ve X$ are independent. Under this assumption, the joint PDF $f_{\ve X}$ is the product of the marginal PDF $f_{X_i}$ of each input parameter.

\subsection{Sobol' decomposition}
Assuming that $\cm$ is a square-integrable function with respect to the probability measure associated with $f_{\ve X}$, its Sobol' decomposition in summands of increasing dimension is given by \citet{Sobol1993}:
\begin{align}
\label{eq:Sobol_decomp}
\begin{split}
\cm(\ve X)
& = \cm_0+\sum_{i=1}^M\cm_i(X_i)+\sum_{1\leq i<j\leq M}\cm_{i,j}(X_i,X_j)+\ldots+\cm_{1,2\enum M}(\ve X) \\
& = \cm_0+\sum_{\substack{\iu\subset\{1 \enum M\} \\ \iu\neq\emptyset}}\cmu,
\end{split}
\end{align}
where $\iu=\{i_1 \enum i_s\}$, $1\leq s \leq M$, denotes a subset of $\{1 \enum M\}$ and $\ve X_{\iu}=\{X_{i_1} \enum X_{i_s}\}$ is the subvector of $\ve X$ containing the variables of which the indices comprise $\iu$.

The uniqueness of the decomposition is ensured by choosing summands that satisfy the conditions:
\begin{equation}
\label{eq:M0}
\cm_0 = \Esp{\cm(\ve X)}
\end{equation}
and
\begin{equation}
\label{eq:Sobol_orth}
\Esp{\cmu~\cmv}=0 \hspace{5mm}
\forall~\iu,\iv\subset\{1 \enum M\},~\iu \neq \iv.
\end{equation}
Note that the above condition implies that all summands $\{\cm_{\iu},~\iu\neq\emptyset\}$, in Eq.~(\ref{eq:Sobol_decomp}) have zero mean values. A recursive construction of summands satisfying the above conditions is obtained as:
\begin{align}
\begin{split}
\label{eq:Sobol_recur}
\cm_i(X_i) = & \Esp{\cm(\ve X)|X_i}-\cm_0 \\
\cm_{i,j}(X_i,X_j) = & \Esp{\cm(\ve X)|X_i,X_j}-\cm_i(X_i)-\cm_j(X_j)-\cm_0
\end{split}
\end{align}
and so on. By introducing:
\begin{equation}
\label{eq:Mtild_u}
\cmtu=\Esp{\cm(\ve X)|\ve X_{\iu}},
\end{equation}
the above recursive relationship is written in the general form:
\begin{equation}
\label{eq:Sobol_recur_2}
\cmu=\cmtu-\sum_{\substack{\iv\subset\iu \\ \iv\neq\iu}} \cmv.
\end{equation}

\subsection{Sobol' sensitivity indices}
\label{sec:Sobol_ind}

The uniqueness and orthogonality properties of the Sobol' decomposition allow the following decomposition of the variance $D$ of $\cm(\ve X)$:
\begin{equation}
\label{eq:D}
D=\Var{\cm(\ve X)} = \sum_{i=1}^M D_i+\sum_{1\leq i<j\leq M}D_{i,j}+ \ldots+D_{1,2\enum M} =
\sum_{\substack{\iu\subset\{1 \enum M\}\\ \iu\neq\emptyset}} D_{\iu},
\end{equation}
where $D_{\iu}$ denotes the partial variance:
\begin{equation}
\label{eq:Du}
D_{\iu}=\Var{\cmu}=\Esp{\left(\cmu\right)^2}.
\end{equation}
The Sobol' index $S_{\iu}$, defined as:
\begin{equation}
\label{eq:Su}
S_{\iu}=\frac{D_{\iu}}{D}=\frac{\Var{\cmu}}{\Var{\cm(\ve X)}},
\end{equation}
represents the fraction of the total variance that is due to the interaction between the components of $\ve X_{\iu}$, \ie $S_{i_1 \enum i_s}$ describes the influence from the interaction between variables $\{X_{i_1} \enum X_{i_s}\}$. By definition, the Sobol' indices satisfy:
\begin{equation}
\label{eq:S_u_sum}
\sum_{i=1}^M S_i+\sum_{1\leq i<j\leq M}S_{i,j}+ \ldots+S_{1,2\enum M} = 
\sum_{\substack{\iu\subset\{1 \enum M\}\\ \iu\neq\emptyset}} S_{\iu} = 1.
\end{equation}

Accordingly, the first-order index for a single variable $X_i$ is defined as:
\begin{equation}
\label{eq:Si}
S_i=\frac{D_i}{D}=\frac{\Var{\cm_i(X_i)}}{\Var{\cm(\ve X)}}
\end{equation}
and describes the influence of $X_i$ considered separately, also called \textit{main effect}. It is noted that the first-order Sobol' indices are equivalent to the first-order indices obtained by the Fourier amplitude sensitivity test {FAST} method \citep{Cukier1978, Saltelli1999}. The total Sobol' index, denoted by $S^T_i$, represents the \textit{total effect} of $X_i$, accounting for its main effect and all interactions with other input variables. It is derived from the sum of all partial indices $S_{\iu}$ that involve the variable $X_i$:
\begin{equation}
\label{eq:SiT}
S^T_i=\sum_{\substack{\iu\subset\{1 \enum M\} \\ i \in \iu}} S_{\iu}.
\end{equation}

First-order and total Sobol' indices are also defined for groups of variables \citep{Sobol1993,Kucherenko2012,Owen2013}. We herein respectively denote by $\Stu$ and $\StuT$ the first-order and total Sobl' indices of the subvector $\ve X_{\iu}$ of $\ve X$. The first-order index $\Stu$ describes the influence of the elements of $\ve X_{\iu}$ \emph{only}, including their main effects and all interactions with other components of $\ve X_{\iu}$:
\begin{equation}
\label{eq:Stu}
\Stu=\sum_{\substack{\iv\subset\iu \\ \iv \neq \emptyset}} S_{\iv}.
\end{equation}
The total index $\StuT$ describes the total effect of the components of $\ve X_{\iu}$, including their main effects and all interactions with other variables in $\ve X$:
\begin{equation}
\label{eq:StuT}
\StuT=\sum_{\substack{\iv\subset\{1 \enum M\}\\ \iv\cap\iu\neq\emptyset}} S_{\iv}.
\end{equation}
The total index $\StuT$ can equivalently be obtained as:
\begin{equation}
\label{eq:StuT_2}
\StuT=1-\Stun,
\end{equation}
where $\smallsetminus\iu$ denotes the complementary set of $\iu$ so that $\{X_{\iu},X_{\smallsetminus\iu}\}=\ve X$. For $\iu=\{i\}$, Eq.~(\ref{eq:Stu}) and Eq.~(\ref{eq:StuT}) respectively reduce to Eq.~(\ref{eq:Si}) and Eq.~(\ref{eq:SiT}). 

To elaborate on the above definitions of the Sobol' indices, let us consider a model with three input variables \ie $\ve X=\{X_1,X_2,X_3\}$. Setting $\iu=\{1,2\}$ for instance, the first-order and total Sobol' indices of $X_{\iu}=\{X_1,X_2\}$ are respectively given by:
\begin{equation}
\label{eq:S12}
\widetilde{S}_{1,2}=S_1+S_2+S_{1,2}
\end{equation}
and
\begin{equation}
\label{eq:S12T}
\widetilde{S}^T_{1,2}=S_1+S_2+S_{1,2}+S_{1,3}+S_{2,3}+S_{1,2,3}=1-S_3.
\end{equation}
Eq.~(\ref{eq:S12}) follows from the specialization of Eq.~(\ref{eq:Stu}) to the considered case. In Eq.~(\ref{eq:S12T}), the first part is consistent with the definition in Eq.~(\ref{eq:StuT}), whereas the second part follows from Eq.~(\ref{eq:StuT_2}).

Note that the first-order index of the subvector $\ve X_{\iu}$ can alternatively be obtained as:
\begin{equation}
\label{eq:Stu_Esp}
\Stu = \frac{\Var{\Esp{\cm(\ve X)|\ve X_{\iu}}}}{\Var{\cm(\ve X)}}=\frac{\Var{\cmtu}}{\Var{\cm(\ve X)}}.
\end{equation}
The equivalence between the above equation and Eq.~(\ref{eq:Stu}) can be verified easily by considering the recursive relationship in Eq.~(\ref{eq:Sobol_recur_2}) and the orthogonality property in Eq.~(\ref{eq:Sobol_orth}).
For the total index of the subvector $\ve X_{\iu}$, Eq.~(\ref{eq:StuT_2}) in conjunction with Eq.~(\ref{eq:Stu_Esp}) leads to:
\begin{equation}
\label{eq:StuT_Esp}
\StuT =
1-\frac{\Var{\Esp{\cm(\ve X)|\ve X_{\smallsetminus\iu}}}}{\Var{\cm(\ve X)}} =
1-\frac{\Var{\cmtun}}{\Var{\cm(\ve X)}}.
\end{equation}

The Sobol' indices can be computed by estimating the variance of the summands in Eq.~(\ref{eq:Sobol_recur_2}) using sampling-based formulas \citep{Sobol1993, Saltelli2002}. However, this computation often requires a large number of model evaluations (say more than $10^4$ per index) to obtain accurate estimates; it thus becomes cumbersome or even intractable in cases when a single evaluation of the model is time consuming. In Sections \ref{sec:Sobol_PCE} and \ref{sec:Sobol_LRA}, we demonstrate that the Sobol' indices can be evaluated \emph{analytically} in terms of the coefficients of LRA or PCE meta-models. Consequently, once a PCE or LRA representation of $\cm(\ve X)$ is available, the Sobol' indices indices can be obtained at nearly zero additional computational cost. In the case of PCE meta-models, the derivation is based on Eq.~(\ref{eq:Stu}) and Eq.~(\ref{eq:StuT}), following the original idea presented in \citet{Sudret2008c}. In the case of LRA meta-models, analytical expressions for the Sobol' indices are herein derived by relying on Eq.~(\ref{eq:Stu_Esp}) and Eq.~(\ref{eq:StuT_Esp}).

\section{Non-intrusive meta-modeling and error estimation}
\label{sec:err}

A meta-model $\widehat{\cm}$ is an analytical function that mimics the behavior of $\cm$ in Eq.~(\ref{eq:model}). In non-intrusive approaches, which are of interest herein, the meta-model is developed using a set of realizations of the input vector $\ce=\{\ve \chi^{(1)} \enum \ve \chi^{(N)}\}$, called \emph{Experimental Design} (ED), and the corresponding responses of the original model $\cy=\{\cm(\ve \chi^{(1)}) \enum \cm(\ve \chi^{(N)})\}$. Thus, non-intrusive meta-modeling treats the original model as a \textit{black box}.

In the following, we describe measures of accuracy of the meta-model response $\widehat{Y}=\widehat{\cm}(\ve X)$. To this end, we introduce the discrete $L_2$ semi-norm of a function $\ve x~\in~\cd_{\ve X}~\longmapsto~a(\ve x)~\in~\Rr$, given by:
\begin{equation}
\label{eq:norm}
\norme a {\cx}=\left(\frac{1}{n}\sum_{i=1}^n a^2(\ve x_i)\right)^{1/2},
\end{equation}
where $\cx=\{\ve x_1 \enum \ve x_n\}\subset \cd_{\ve X}$ denotes a set of realizations of the input vector.

A good measure of the accuracy of the meta-model is the \emph{generalization error} $Err_G$, which represents the mean-square of the difference $(Y-\widehat Y)$. $Err_G$ can be estimated by:
\begin{equation}
\label{eq:hatErrG}
\widehat{Err}_G =\left\|\cm-\widehat {\cm}\right\|_{\cx_{\mathrm{val}}}^2,
\end{equation}
where $\cx_{\mathrm{val}}=\{{\ve x}_1 \enum {\ve x}_{n_{\mathrm{val}}}\}$ is a sufficiently large set of realizations of the input vector, called \textit{validation set}. The estimate of the relative generalization error, denoted by $\widehat{err}_G$, is obtained by normalizing $\widehat{Err}_G$ with the empirical variance of $\cy_{\mathrm{val}}=\{\cm({\ve x}_1)\enum \cm({\ve x}_{n_{\mathrm{val}}})\}$.

However, meta-models are typically used in cases when only a limited number of model evaluations is affordable. It is thus desirable to obtain an estimate of ${Err}_G$ by relying solely on the ED. One such error estimate is the \emph{empirical error} $\widehat{Err}_E$, given by:
\begin{equation}
\label{eq:ErrE}
\widehat{Err}_E =\left\|\cm-\widehat {\cm}\right\|_{\ce}^2,
\end{equation}
in which the subscript $\ce$ indicates that the semi-norm is evaluated at the points of the ED. The relative empirical error, denoted by $\widehat{err}_E$, is obtained by normalizing $\widehat{Err}_E$ with the empirical variance of $\cy=\{\cm({\ve \chi}^{(1)})\enum \cm({\ve \chi}^{(N)})\}$, the latter representing the set of model responses at the ED. Unfortunately, $\widehat{err}_E$ tends to underestimate the actual generalization error, which might be severe in cases of overfitting. Indeed, it can even decrease down to zero if the obtained meta-model interpolates the data at the ED, while it is not necessarily accurate at other points.

By using the information contained in the ED only, a fair approximation of the generalization error can be obtained with cross-validation techniques. In $k$-fold cross-validation, (i) the ED is randomly partitioned into $k$ sets of approximately equal size, (ii) for $i=1 \enum k$, a meta-model is built considering all but the $i$-th partition and the excluded set is used to evaluate the respective generalization error, (iii) the generalization error of the meta-model built with the full ED is estimated as the average of the errors of the $k$ meta-models obtained in (ii). Leave-one-out cross-validation corresponds to the case when $k=N$, \ie one point of the ED being set apart in turn \citep{Allen1971}.

\section{Sobol' sensitivity analysis using polynomial chaos expansions}
\label{sec:Sobol_PCE}

\subsection{Formulation and construction of polynomial chaos expansions}
\label{sec:PCE_form}

A meta-model of $Y=\cm(\ve X)$ in Eq.~(\ref{eq:model}) belonging to the class of PCE has the form \citep{Xiu2002wiener}:
\begin{equation}
\label{eq:PCE}
Y^{\rm PCE}=\cm^{\rm{PCE}}(\ve X)=\sum_{\ua \in \ca}{y_{\ua}} \Psi_{\ua}(\ve {X}),
\end{equation}
where $\ca$ denotes a set of multi-indices $\ua=(\alpha_1 \enum \alpha_M)$, $\{\Psi_{\ua},~\ua \in \ca\}$ is a set of multivariate polynomials that are orthonormal with respect to $f_{\ve X}$ and $\{y_{\ua},~\ua \in \ca\}$ is the set of polynomial coefficients. The orthonormality condition reads:
\begin{equation}
\label{eq:PCE_orthonorm_multi}
\Esp{\Psi_{\ua} (\ve X) \hsp \Psi_{\ve \beta} (\ve X)}=\delta_{\ua \ve \beta},
\end{equation}
where $\delta_{\ua \ve \beta}$ is the Kronecker delta, equal to one if $\ua=\ve \beta$ and zero otherwise.

The multivariate polynomials that comprise the PCE basis are obtained by tensorization of appropriate univariate polynomials:
\begin{equation}
\label{eq:multi_pol}
\Psi_{\ua}(\ve X)=\prod_{i=1}^M P^{(i)}_{\alpha_i}(X_i).
\end{equation}
In the above equation, $P^{(i)}_{\alpha_i}(X_i)$ is a polynomial of degree ${\alpha_i}$ in the $i$-th input variable belonging to a family of polynomials that are orthonormal with respect to $f_{X_i}$, thus satisfying:
\begin{equation}
\label{eq:PCE_orthonorm_uni}
\Esp{P_j^{(i)}(X_i) \hsp P_k^{(i)}(X_i)}=\delta_{jk}.
\end{equation}
For standard distributions, the associated family of orthonormal polynomials is well-known, \eg a standard normal variable is associated with the family of Hermite polynomials, whereas a uniform variable over $[-1,1]$ is associated with the family of Legendre polynomials. A general case can be treated through an isoprobabilistic transformation of the input random vector $\ve X$ to a basic random vector, \eg a vector with independent standard normal components or independent uniform components over $[-1,1]$.

The set of multi-indices $\ca$ in Eq.~(\ref{eq:PCE}) is determined by an appropriate truncation scheme. A typical scheme consists in selecting multivariate polynomials up to a total degree $p^t$, \ie $\ca=\{\ua\in\Nn^M~|~\|\ua\|_1 \leq p^t\}$, with $\|\ua\|_1=\sum_{i=1}^M{\alpha_i}$. The corresponding number of terms in the truncated series is:
\begin{equation}
\label{eq:card} 
{\rm{card}} \ca={M+p^t \choose p^t} =\frac{(M+p^t)!}{M!p^t!},
\end{equation}
which grows exponentially with $M$, thus giving rise to the so-called \emph{curse of dimensionality}. Because the terms that include interactions between input variables are usually less significant, \citet{BlatmanPEM2010} proposed to use a hyperbolic truncation scheme, where the set of retained multi-indices is defined as $\ca=\{\ua\in\Nn^M~|~\|\ua\|_q\leq p^t\}$, with the $q$-norm given by:
\begin{equation}
\label{eq:qnorm} 
\|\ua\|_q=\left(\sum_{i=1}^M{\alpha_i}^q\right)^{1/q}, \hspace{5mm} 0<q<1.
\end{equation}
According to Eq.~(\ref{eq:qnorm}), the lower the value of $q$, the smaller is the number of interaction terms in the PCE basis. The case $q=1$ corresponds to the common truncation scheme using a maximal total degree $p^t$, whereas the case $q=0$ corresponds to an additive model.

Once the basis has been specified, the set of coefficients $\ve y=\{y_{\ua},~\ua \in \ca\}$ may be computed by minimizing the mean-square error of the approximation over the ED. More efficient solution schemes can be devised by considering the respective regularized problem:
\begin{equation}
\label{eq:PCE_coef_reg}
\ve y= \mathrm{arg} \underset{\ve {\upsilon}\in\Rr^{\rm{card}\ca}}{\mathrm{min}}\left\|\cm-\sum_{\ua \in \ca}\upsilon_{\ua}\Psi_{\ua}\right\|_{\ce}^2+\lambda \cp(\ve \upsilon),
\end{equation}
in which  $\cp(\ve \upsilon)$ is an appropriate regularization functional of $\ve \upsilon=\{\upsilon_1 \enum \upsilon_{\rm{card}\ca}\}$. If $\cp(\ve \upsilon)$ is selected as the $L_1$ norm of $\ve \upsilon$, \ie $\cp(\ve \upsilon)=\sum_{i=1}^{\rm{card}\ca} \abs {\upsilon_i}$, insignificant terms may be disregarded from the set of predictors, leading to \emph{sparse} solutions. \citet{BlatmanJCP2011} proposed to use the hybrid Least Angle Regression (LAR) method for building sparse PCE. This method employs the LAR algorithm \citep{Efron2004} to select the best set of predictors and subsequently, estimates the coefficients using Ordinary Least Squares (OLS). Other techniques to derive sparse expansions can be found in \eg \citet{Doostan2011, Yang2013}.

A good measure of the PCE accuracy is the leave-one out error. As mentioned in Section~\ref{sec:err}, this corresponds to the cross-validation error for the case $k=N$. Using algebraic manipulations, the leave-one-out error can be computed based on a \emph{single} PCE that is built with the full ED. Let $h(\ve \chi^{(i)})$ denote the $i$-th diagonal term of matrix $\ve \Psi(\ve \Psi^{\rm{T}}\ve \Psi)^{-1}\ve \Psi^{\rm{T}}$, with $\ve \Psi=\{\ve \Psi_{ij}=\Psi_j(\ve \chi^{(i)}),\hspace{2mm} i=1\enum N;\hspace{1mm} j=1\enum \rm{card}\ca\}$. The leave-one-out error can then be computed as \citep{BlatmanThesis}:
\begin{equation}
\label{eq:errLOO}
\widehat{Err}_{\rm LOO}=\left\|\frac{\cm-\cm^{\rm PCE}}{1-h}\right\|_{\ce}^2.
\end{equation}
The relative leave-one-out error, denoted by $\widehat{err}_{\rm LOO}$, is obtained by normalizing $\widehat{Err}_{\rm LOO}$ with the empirical variance of $\cy=\{\cm(\ve \chi^{(1)}) \enum \cm(\ve \chi^{(N)})\}$, the latter representing the set of model responses at the ED. Because $\widehat{err}_{\rm LOO}$ may be too optimistic, the following corrected estimate may be used instead \citep{Chapelle2002}:
\begin{equation}
\label{eq:errLOO_corr}
\widehat{err}^*_{\rm LOO}=\widehat{err}_{\rm LOO}\left(1-\frac{\rm{card}\ca}{N}\right)^{-1}\left(1+\rm{tr}((\ve\Psi^{\rm{T}}\ve\Psi)^{-1})\right).
\end{equation}
This corrected leave-one-out error is a good compromise between fair error estimation and affordable computational cost.

\subsection{Computation of Sobol' sensitivity indices using polynomial chaos expansions}
\label{sec:PCE_indices}

Let us consider the PCE representation $Y^{\rm PCE}=\cm^{\rm{PCE}}(\ve X)$ in Eq.~(\ref{eq:PCE}). It is straightforward to obtain the Sobol' decomposition of $\cm^{\rm{PCE}}$ in an analytical form by observing that the summands in Eq.~(\ref{eq:Sobol_decomp}) can be written as:
\begin{equation}
\label{eq:summands_PCE}
\cm^{\rm PCE}_{\iu}(\ve X_{\iu}) = \sum_{\ua\in\ca_{\iu}}y_{\ua}\Psi_{\ua}(\ve X), \hspace{5mm}
\ca_{\iu}=\{\ua\in\ca~|~\alpha_k\neq 0 \Leftrightarrow k \in \iu\}
\end{equation}
and
\begin{equation}
\label{eq:M0_PCE}
\cm_0^{\rm PCE}=y_{\ve 0}.
\end{equation}

Because of the orthonormality of the PCE basis, the conditions in Eq.~(\ref{eq:M0}) and Eq.~(\ref{eq:Sobol_orth}) are satisfied, ensuring the uniqueness of the decomposition. Note that Eq.~(\ref{eq:M0}) in conjunction with Eq.~(\ref{eq:M0_PCE}) lead to:
\begin{equation}
\label{eq:mean_PCE}
\Esp{\cm^{\rm PCE}(\ve X)}=y_{\ve 0}.
\end{equation}
Furthermore, the orthonormality of the PCE basis allows to obtain the total variance $D^{\rm PCE}$ and any partial variance $D_{\iu}^{\rm PCE}$ of $\cm^{\rm{PCE}}(\ve X)$ by a mere combination of the squares of the PCE coefficients. These quantities are respectively given by:
\begin{equation}
\label{eq:D_PCE}
D^{\rm PCE}=\Var{\cm^{\rm PCE}(\ve X)}=\sum_{\ua\in\ca \backslash \{\ve 0\}}{y_{\ua}}^2
\end{equation}
and
\begin{equation}
\label{eq:Du_PCE}
D^{\rm PCE}_{\iu}=\Var{\cm_{\iu}^{\rm PCE}(\ve X_{\iu})}=\sum_{\ua\in\ca_{\iu}}{y_{\ua}}^2.
\end{equation}

By utilizing Eq.~(\ref{eq:D_PCE}) and Eq.~(\ref{eq:Du_PCE}), the first-order and total Sobol' indices of a subvector $\ve X_{\iu}$ of $\ve X$ are respectively obtained as:
\begin{equation}
\label{eq:Stu_PCE}
\widetilde{S}_{\iu}^{\rm PCE} = \sum_{\ua\in\widetilde{\ca}_{\iu}}{y_{\ua}}^2/D^{\rm PCE},
\hspace{5mm}
\widetilde{\ca}_{\iu}=\{\ua\in\ca\backslash\{\ve 0\}~|~\alpha_{i}=0~\forall~i\notin\iu\}
\end{equation}
and
\begin{equation}
\label{eq:StuT_PCE}
\widetilde{S}_{\iu}^{T,\hsp \rm PCE} = \sum_{\ua\in\widetilde{\ca}_{\iu}^T}{y_{\ua}}^2/D^{\rm PCE},
\hspace{5mm}
\widetilde{\ca}_{\iu}^T=\{\ua\in\ca~|~\exists~i\in\iu : \alpha_i>0 \}.
\end{equation}
To elaborate the above equations, let us consider a model with three input variables \ie $\ve X=\{X_1,X_2,X_3\}$ and focus on the case $\iu=\{1,2\}$. Then, the set $\widetilde{A}_{\iu}$ includes the multi-indices of the form $(\alpha_1, 0, 0)$, $(0,\alpha_2, 0)$ or $(\alpha_1, \alpha_2, 0)$, where $\alpha_i$, $i=1,2,3$, denotes a non-zero element. The set $\widetilde{A}_{\iu}^T$ is a superset of $\widetilde{A}_{\iu}$, additionally including the multi-indices of the form $(\alpha_1, 0, \alpha_3)$, $(0,\alpha_2, \alpha_3)$ or $(\alpha_1, \alpha_2, \alpha_3)$. Note that the set ${A}_{\iu}$ in Eq.~(\ref{eq:summands_PCE}) includes only the multi-indices of the form  $(\alpha_1, \alpha_2, 0)$.

By specializing Eq.~(\ref{eq:Stu_PCE}) and Eq.~(\ref{eq:StuT_PCE}) to the case $\iu=\{i\}$, the first-order and total Sobol' indices of a single variable $X_i$ are respectively obtained as:
\begin{equation}
\label{eq:Si_PCE}
S_i^{\rm PCE}=\sum_{\ua\in\ca_i}{y_{\ua}}^2/D^{\rm PCE},
\hspace{5mm}
\ca_i=\{\ua\in\ca~|~\alpha_i>0,~\alpha_{j\neq i}=0\}
\end{equation}
and
\begin{equation}
\label{eq:SiT_PCE}
S_i^{T, \hsp \rm PCE}=\sum_{\ua\in\ca_i^T}{y_{\ua}}^2/D^{\rm PCE},
\hspace{5mm}
\ca_i^T=\{\ua\in\ca~|~\alpha_i>0\}.
\end{equation}

\section{Sobol' sensitivity analysis using low-rank tensor approximations}
\label{sec:Sobol_LRA}

\subsection{Formulation and construction of low-rank tensor approximations}
\label{sec:LRA_form}

Let us consider again the mapping in Eq.~(\ref{eq:model}). A rank-one function of the input vector $\ve X$ has the form:
\begin{equation}
\label{eq:rank1}
w(\ve X)= \prod_{i=1}^M {v^{(i)}(X_i)},
\end{equation}
where $v^{(i)}$ denotes a univariate function of $X_i$. A representation of $\cm$ as a sum of a finite number of rank-one functions constitutes a canonical decomposition with rank equal to the number of rank-one components. Naturally, of interest are representations where the exact response is approximated with sufficient accuracy by using a relatively small number of terms, leading to the name \emph{low-rank approximations}.

A meta-model of $Y=\cm(\ve X)$ belonging to the class of LRA can be written as:
\begin{equation}
\label{eq:LRA}
Y^{\rm LRA} = \cm^{\rm LRA}(\ve X)=\sum_{l=1}^R b_l\left(\prod_{i=1}^M {\vli(X_i)}\right),
\end{equation}
where $R$ is the rank of the approximation, $\vli$ is a univariate function of $X_i$ in the $l$-th rank-one component and $\{b_l,~l=1 \enum R\}$ are scalars that can be viewed as normalizing constants. In order to obtain a representation in terms of polynomial functions, we expand $\vli$ onto a polynomial basis that is orthonormal with respect to $f_{X_i}$, \ie satisfies the condition in Eq.~(\ref{eq:PCE_orthonorm_uni}). This leads to the expression:
\begin{equation}
\label{eq:vli}
\vli(X_i)=\sum_{k=0}^{p_i} \zkli \hsp \Pki (X_i),
\end{equation}
where $\Pki$ is the $k$-th degree univariate polynomial in the $i$-th input variable, $p_i$ is the maximum degree of $\Pki$ and $\zkli$ is the coefficient of $\Pki$ in the $l$-th rank-one term. Appropriate families of univariate polynomials satisfying the orthonormality condition can be determined as discussed in Section \ref{sec:PCE_form}. By substituting Eq.~(\ref{eq:vli}) into Eq.~(\ref{eq:LRA}), we obtain:
\begin{equation}
\label{eq:LRA_pol}
Y^{\rm LRA} = \cm^{\rm LRA}(\ve X)= \sum_{l=1}^R b_l \left(\prod_{i=1}^M\left(\sum_{k=0}^{p_i} \zkli \hsp \Pki (X_i)\right)\right).
\end{equation}

Disregarding the redundant parameterization arising from the normalizing constants, the number of unknowns in the above equation is $R \cdot \sum_{i=1}^M (p_i+1)$. Note that this number grows only linearly with the input dimension $M$. A representation of $Y=\cm (\ve X)$ in the form of LRA drastically reduces the number of unknowns compared to PCE. In order to emphasize this, we consider PCE with the candidate basis determined by the truncation scheme $\ca=\{\ua \in \Nn^M~|~\alpha_i\leq p_i,~i=1 \enum M\}$, so that the expansion relies on the same polynomial functions as those used in Eq.~(\ref{eq:LRA_pol}). For the case when $p_i=p$, $i=1 \enum M$, the resulting number of unknowns is $(p+1)^M$ in PCE versus $R\cdot M\cdot (p+1)$ in LRA. Considering a typical engineering problem with $M=10$ and low-degree polynomials with $p=3$, the aforementioned formulas yield $1,048,576$ unknowns in PCE versus $40 R$ unknowns in LRA; for a low rank, say $R\leq 10$, the number of unknowns in LRA does not exceed a mere $400$. In other words, the LRA meta-model constitutes a compressed representation of a PCE meta-model, with the basis of the latter defined by constraining the maximum polynomial degree in each dimension. As will be seen in the following, the compressed formulation leads to a fundamentally different construction algorithm.

Non-intrusive algorithms proposed in the literature for building LRA in the form of Eq.~(\ref{eq:LRA_pol}) rely on Alternated Least-Squares (ALS) minimization for the computation of the polynomial coefficients \citep{Chevreuil2013, Chevreuil2013least, Doostan2013, RaiThesis}. The ALS approach consists in sequentially solving a least-squares minimization problem along each dimension ${1\enum M}$ separately, while ``freezing'' the coefficients in all remaining dimensions. We herein employ the \emph{greedy} algorithm proposed in \citet{Chevreuil2013} and further investigated in \citet{Konakli2015arxiv}, which involves progressive increase of the rank by successively adding rank-one components.

Let $Y_r^{\rm LRA} = \cm_r^{\rm LRA}(\ve X)$ denote the rank-$r$ approximation of $Y=\cm(\ve X)$:
\begin{equation}
\label{eq:Yr}
Y_r^{\rm LRA} = \cm_r^{\rm LRA}(\ve X)=\sum_{l=1}^r b_l w_l (\ve X),
\end{equation}
with:
\begin{equation}
\label{eq:wl}
w_l(\ve X)=\prod_{i=1}^M {\vli(X_i)}=\prod_{i=1}^M\left(\sum_{k=0}^{p_i} \zkli \hsp \Pki (X_i)\right).
\end{equation}
The employed algorithm comprises a sequence of pairs of a \textit{correction step} and an \textit{updating step}, so that the $r$-th correction step yields the rank-one term $w_r$ and the $r$-th updating step yields the set of coefficients $\{b_1 \enum b_r\}$. These steps are detailed in the sequel.

\textbf{Correction step}: Let $\res_r(\ve X)$ denote the residual after the completion of the  $r$-th iteration:
\begin{equation}
\label{eq:res}
\res_r(\ve X)=\cm(\ve X)-\cm^{\rm LRA}_r(\ve X).
\end{equation}
The sequence is initiated by setting $\cm^{\rm LRA}_0(\ve X)=0$ leading to $\res_0(\ve X)=\cm(\ve X)$. Based on the available experimental design, $\ce$, the rank-one tensor $w_r$ is obtained as the solution of:
\begin{equation}
\label{eq:solve_wr}
w_r=\mathrm{arg} \underset{\omega \in \cw}{\hsp\mathrm{\min}} \left\|\res_{r-1}-\omega\right\|_{\ce}^2,
\end{equation}
where $\cw$ represents the space of rank-one tensors. By employing an ALS scheme, the minimization problem in Eq.~(\ref{eq:solve_wr}) is substituted by a series of smaller minimization problems, each involving the coefficients along one dimension only:
\begin{equation}
\label{eq:solve_zr}
\ve z_r^{(j)}= \mathrm{arg}\underset{\ve \zeta \in\Rr^{p_j}}{\mathrm{min}}\left\|\res_{r-1}-\left(\prod_{i\neq j}\vri \right)\left(\sum_{k=0}^{p_j} \zeta_k \hsp \Pkj\right)\right\|_{\ce}^2,
\hspace{5mm} j=1 \enum M.
\end{equation}
The correction step is initiated by assigning arbitrary values to $\vri$, $i=1\enum M$, \eg unity values, and may involve several iterations over the set of dimensions $\{1 \enum M\}$. The stopping criterion proposed in \citet{Konakli2015arxiv} combines the number of iterations, denoted by $I_r$, with the decrease in the relative empirical error in two successive iterations, denoted by $\Delta \widehat{err}_r$, where the relative empirical error is given by:
\begin{equation}
\label{eq:err_r}
\widehat{err}_r=\frac{\left\|\res_{r-1}-w_{r}\right\|_{\ce}^2}{\Varhat{\cy}}.
\end{equation}
In the above equation, $\Varhat{\cy}$ represents the empirical variance of the set comprising the model responses at the ED. Accordingly, the algorithm exits the $r$-th correction step if either $I_r$ reaches a maximum allowable value, denoted by $\Imax$, or $\Delta \widehat{err}_r$ becomes smaller than a prescribed threshold, denoted by $\Derrmin$. \citet{Konakli2015arxiv} showed that the choice of $\Imax$ and $\Derrmin$ may have a significant effect on the accuracy of LRA; based on a number of example investigations, they proposed to use $\Imax=50$ and $\Derrmin \leq 10^{-6}$.

\textbf{Updating step}: After the completion of a correction step, the algorithm moves to an updating step, in which the set of coefficients $\ve b=\{b_1 \ldots b_r\}$ is obtained as the solution of the minimization problem:
\begin{equation}
\label{eq:solve_b}
\ve b =\mathrm{arg} \underset{\ve \beta \in\Rr^{r}}{\mathrm{min}}\left\|\cm-\sum_{l=1}^r \beta_l w_l\right\|_{\ce}^2.
\end{equation}
Note that in each updating step, the size of vector $\ve b$ is increased by one. In the $r$-th updating step, the value of the new element $b_r$ is determined for the first time, whereas the values of the existing elements $\{b_1 \enum b_{r-1}\}$ are recomputed (updated).

Construction of a rank-$R$ approximation in the form of Eq.~(\ref{eq:LRA_pol}) requires repeating pairs of a correction and an updating step for $r=1\enum R$. The algorithm is summarized below.\vspace{2mm}

\textbf{Algorithm 1}: Non-intrusive construction of a rank-$R$ approximation of $\cm$ with polynomial bases:
\begin{enumerate}
	\item Set $\cm^{\rm LRA}_0(\ve x)=0$.
	\item For $r=1\enum R$, repeat steps (a)-(d):
	\begin{enumerate}
		\item Initialize: $\vri(x_i)=1$, $i=1 \enum M$; $I_r=0$; $\Delta \widehat{err}_r=\epsilon>\Derrmin$ . 
		\item While $\Delta \widehat{err}_r>\Derrmin$ and $I_r<\Imax$, repeat steps i-iv:
		\begin{enumerate}
			\item Set $I_r \leftarrow I_r+1$.
			\item For $i=1 \enum M$, repeat steps A-B (correction step):
			\begin{enumerate}
				\item Determine  ${\ve z}_r^{(i)}=\{z_{1,r}^{(i)} \ldots z_{p_i,r}^{(i)}\}$ (Eq.~(\ref{eq:solve_zr})).
				\item Use the current values of ${\ve z}_r^{(i)}=\{z_{1,r}^{(i)} \ldots z_{p_i,r}^{(i)}\}$ to update $\vri$ (Eq.~(\ref{eq:vli})).
			\end{enumerate}
			\item Use the current functional forms of $\vri$, $i=1 \enum M$, to update $w_r$ (Eq.~(\ref{eq:wl})).
			\item Compute $\widehat{err}_r$ (Eq.~(\ref{eq:err_r})) and update $\Delta \widehat{err}_r$.
		\end{enumerate}
		\item Determine $\ve b=\{b_1 \ldots b_r\}$ (Eq.~(\ref{eq:solve_b}), updating step).
		\item Evaluate $\cm^{\rm LRA}_r$ at the points of the ED (Eq.~(\ref{eq:Yr})).
		\item Evaluate $\res_r$ at the points of the ED (Eq.~(\ref{eq:res})).
	\end{enumerate}
\end{enumerate}

Note that the above algorithm relies on the solution of several small-size minimization problems: each iteration in a correction step involves $M$ minimization problems of size $\{p_i+1, \hsp i=1 \enum M\}$ (usually $p_i<20$), whereas the $r$-th updating step involves one minimization problem of size $r$ (recall that small ranks are of interest in LRA). Because of the small number of unknowns, Eq.~(\ref{eq:solve_zr}) and Eq.~(\ref{eq:solve_b}) can be efficiently solved with OLS, as shown in \citet{Konakli2015arxiv}. An alternative approach employed in \citet{Chevreuil2013} is to substitute these equations with respective regularized problems.

In a typical application, the optimal rank $R$ is not known \textit{a priori}. As noted earlier, the progressive construction of LRA results in a set of decompositions of increasing rank. Thus, one may set $r=1 \enum r_{\rm max}$ in Step 2 of Algorithm 1, where $r_{\rm max}$ is a maximum allowable candidate rank, and at the end, select the optimal-rank approximation using error-based criteria. In the present work, we select the optimal rank $R\in\{1 \enum r_{\rm max}\}$ based on the 3-fold cross-validation error, as proposed by \citet{Chevreuil2013least} (see Section~\ref{sec:err} for details on $k$-fold cross-validation). \citet{Konakli2015arxiv} investigated the accuracy of the approach in a number of applications and showed that it leads to optimal or nearly optimal rank selection in terms of the relative generalization error estimated with a large validation set. 

\subsection{Computation of Sobol' sensitivity indices using low-rank tensor approximations}
\label{sec:LRA_indices}

In this section, we consider the LRA meta-model $Y^{\rm LRA} = \cm^{\rm LRA}(\ve X)$ in Eq.~(\ref{eq:LRA_pol}) and derive analytical expressions for the Sobol' indices in terms of the polynomial coefficients $\zkli$ and the normalizing constants $b_l$. To this end, we rely on the definitions of the first-order and total Sobol' indices for a subset $\ve X_{\iu}$ of $\ve X$ given in Eq.~(\ref{eq:Stu_Esp}) and Eq.~(\ref{eq:StuT_Esp}), respectively. In the specific case $\iu=\{i\}$, these equations yield the first-order and total Sobol' indices of a single variable $X_i$.

Employing the definition of variance and the property:
\begin{equation}
\label{eq:Mtu_mean}
\Esp{\cmtu}=\Esp{\Esp{\cm(\ve X)|\ve X_{\iu}}}=\Esp{\cm(\ve X)},
\end{equation}
Eq.~(\ref{eq:Stu_Esp}) is equivalently written as:
\begin{equation}
\label{eq:Stu_LRA}
\Stu = \frac{\Esp{\left(\cmtu\right)^2}-\left(\Esp{\cm (\ve X)}\right)^2}{\Var{\cm(\ve X)}}.
\end{equation}
We next derive analytical expressions for each of the quantities in the right-hand side of Eq.~(\ref{eq:Stu_LRA}) considering $\cm^{\rm LRA}$ in place of  $\cm$. In these derivations, we make use of the equalities:
\begin{equation}
\label{eq:mean_vli}
\Esp{\vli(X_i)} = \zlio
\end{equation}
and
\begin{equation}
\label{eq:mean_vli_vlip}
\Esp{\vli(X_i) \hsp \vlip(X_i)}=
\sum_{k=0}^{p_i} \zkli \hsp \zklip,
\end{equation}
which follow from Eq.~(\ref{eq:vli}) in conjunction with the orthonormality condition in Eq.~(\ref{eq:PCE_orthonorm_uni}).

Employing the expression for the LRA response in Eq.~(\ref{eq:LRA}), its mean value can be computed as:
\begin{align}
\begin{split}
\label{eq:mean_LRA}
\Esp{\cm^{\rm LRA}(\ve X)} = &
\sum_{l=1}^R b_l \left(\prod_{i=1}^M \Esp{{\vli(X_i)}}\right) \\ = &
\sum_{l=1}^R b_l \left(\prod_{i=1}^M \zlio \right).
\end{split}
\end{align}
In the first part of the above equation, we have utilized the property $\Esp{A\cdot B}=\Esp{A}\cdot \Esp{B}$, which holds under the condition that $A$ and $B$ are independent; note that because the components of $\ve X$ are assumed independent, the quantities $\{\vli(X_i),~i=1\enum M\}$ are also independent. The second part of Eq.~(\ref{eq:mean_LRA}) follows directly from Eq.~(\ref{eq:mean_vli}).

The mean-square of the LRA response in Eq.~(\ref{eq:LRA}) can be computed as:
\begin{align}
\begin{split}
\label{eq:ms_LRA}
\Esp{\left(\cm^{\rm LRA}(\ve X)\right)^2} = &
\sum_{l=1}^R \sum_{l'=1}^R b_l \hsp b_{l'} \left( \prod_{i=1}^M  \Esp{\vli(X_i) \hsp \vlip(X_i)} \right) \\ = &
\sum_{l=1}^R \sum_{l'=1}^R b_l \hsp b_{l'} \left( \prod_{i=1}^M \left( {\sum_{k=0}^{p_i} \zkli \hsp \zklip} \right) \right).
\end{split}
\end{align}
The first part of the above equation relies on the independence of the quantities $\{\vli(X_i) \hsp \vlip(X_i),~i=1\enum M\}$, which allows the use of the property $\Esp{A\cdot B}=\Esp{A}\cdot \Esp{B}$, whereas the second part follows directly from Eq.~(\ref{eq:mean_vli_vlip}).

Substituting Eq.~(\ref{eq:mean_LRA}) and Eq.~(\ref{eq:ms_LRA}) into the definition of the variance, we have:
\begin{align}
\begin{split}
\label{eq:var_LRA}
\Var{\cm^{\rm LRA}(\ve X)} = &
\Esp{\left(\cm^{\rm LRA}(\ve X)\right)^2}-\left(\Esp{\cm^{\rm LRA}(\ve X)}\right)^2 \\ = &
\sum_{l=1}^R \sum_{l'=1}^R b_l \hsp b_{l'} \left( \left( \prod_{i=1}^M \left( {\sum_{k=0}^{p_i} \zkli \hsp \zklip} \right) \right)- \left(\prod_{i=1}^M \zlio \hsp \zliop \right) \right).
\end{split}
\end{align}

In the following, we derive an analytical expression for $\Esp{\left(\cmtuLRA\right)^2}$, beginning with the case $\iu=\{i\}$. Noting that $\vli(X_i)$ represents a constant in the quantity $(\cm^{\rm LRA}(\ve X)|X_i)$, substitution of Eq.~(\ref{eq:LRA}) into the definition of $\cmtiLRA$ (Eq.~(\ref{eq:Mtild_u})) yields:
\begin{align}
\begin{split}
\label{eq:Mti_LRA}
\cmtiLRA = &
\sum_{l=1}^R b_l \hsp \vli(X_i) \left(\prod_{j\neq i}^M \Esp{{\vlj(X_j)}}\right) \\ = &
\sum_{l=1}^R b_l \left(\prod_{j\neq i}^M \zljo \right) \vli(X_i).
\end{split}
\end{align}
Using the above expression, we have:
\begin{align}
\begin{split}
\label{eq:ms_Mti_LRA}
\Esp{\left(\cmtiLRA\right)^2} = &
\sum_{l=1}^R \sum_{l'=1}^R b_l \hsp b_{l'} \left(\prod_{j\neq i}^M \zljo \hsp \zljop \right) \Esp{\vli(X_i) \hsp \vlip(X_i)} \\ = &
\sum_{l=1}^R \sum_{l'=1}^R b_l \hsp b_{l'} \left(\prod_{j\neq i}^M \zljo \hsp \zljop \right) \left(\sum_{k=0}^{p_i} \zkli \hsp \zklip \right).
\end{split}
\end{align}
It is straightforward to extend Eq.~(\ref{eq:Mti_LRA}) to $\cmtuLRA$ considering that the components of $\ve X_{\iu}$ represent constants in the quantity $(\cm^{\rm LRA}(\ve X)|\ve X_{\iu})$. We thereby obtain:
\begin{align}
\begin{split}
\label{eq:Mtu_LRA}
\cmtuLRA = &
\sum_{l=1}^R b_l \left(\prod_{i\in \iu} \vli(X_i) \right) \left(\prod_{j\notin \iu} \Esp{\vlj(X_j)} \right) \\ = & 
\sum_{l=1}^R b_l \left(\prod_{j\notin \iu} \zljo \right) \left(\prod_{i\in \iu} \vli(X_i) \right)
\end{split}
\end{align}
and finally:
\begin{align}
\begin{split}
\label{eq:ms_Mtu_LRA}
\Esp{\left(\cmtuLRA\right)^2} = &
\sum_{l=1}^R \sum_{l'=1}^R b_l \hsp b_{l'}
\left(\prod_{j\notin \iu} \zljo \hsp \zljop \right)
\left(\prod_{i\in \iu} \Esp{\vli(X_i) \hsp \vlip(X_i)} \right) \\ = &
\sum_{l=1}^R \sum_{l'=1}^R b_l \hsp b_{l'}
\left(\prod_{j\notin \iu} \zljo \hsp \zljop \right)
\left(\prod_{i\in \iu} \left(\sum_{k=0}^{p_i} \zkli \hsp \zklip \right) \right).
\end{split}
\end{align}

Eq.~(\ref{eq:mean_LRA}), Eq.~(\ref{eq:var_LRA}) and Eq.~(\ref{eq:ms_Mtu_LRA}) provide all the elements required to compute the first-order index for any subvector $\ve X_{\iu}$ according to Eq.~(\ref{eq:Stu_LRA}). As explained above, Eq.~(\ref{eq:ms_Mtu_LRA}) simplifies to Eq.~(\ref{eq:ms_Mti_LRA}) for the case of a single variable $X_i$.

Respective analytical expressions for the total Sobol' indices can be derived using Eq.~(\ref{eq:StuT_Esp}), which is equivalently written as:
\begin{equation}
\label{eq:StuT_LRA}
\StuT = 1-\frac{\Esp{\left(\cmtun\right)^2}-\left(\Esp{\cm (\ve X)}\right)^2}{\Var{\cm(\ve X)}}.
\end{equation}
By considering that $\smallsetminus \iu$ is the complementary set of $\iu$ with respect to $\{1 \enum M\}$, we can obtain an expression for $\Esp{\left(\cmtunLRA\right)^2}$ by simply interchanging the indices $i$ and $j$ in Eq.~(\ref{eq:ms_Mtu_LRA}):
\begin{equation}
\label{eq:ms_Mtun_LRA}
\Esp{\left(\cmtunLRA\right)^2} =
\sum_{l=1}^R \sum_{l'=1}^R b_l \hsp b_{l'}
\left(\prod_{i\in \iu} \zlio \hsp \zliop \right)
\left(\prod_{j\notin \iu} \left(\sum_{k=0}^{p_j} \zklj \hsp \zkljp \right) \right).
\end{equation}
For the special case $\iu=\{i\}$, Eq.~(\ref{eq:ms_Mtun_LRA}) reduces to:
\begin{equation}
\label{eq:ms_Mtin_LRA}
\Esp{\left(\cmtinLRA\right)^2} =
\sum_{l=1}^R \sum_{l'=1}^R b_l \hsp b_{l'}
\hsp \zlio \hsp \zliop
\left(\prod_{j\neq i} \left(\sum_{k=0}^{p_j} \zklj \hsp \zkljp \right) \right).
\end{equation}

Note that by appropriate combinations of first-order indices for single variables and groups of variables, we can obtain any higher-order index representing the effect from the interaction between a set of variables. For instance, the second-order index representing the effect from the interaction between $X_i$ and $X_j$ can be obtained as $S_{i,j}=\widetilde{S}_{i,j}-S_i-S_j$. A general expression can be derived by dividing each part of Eq.~(\ref{eq:Sobol_recur_2}) by $\Var{\cm(\ve X)}$, leading to:
\begin{equation}
\label{eq:Su_LRA}
S_{\iu}=\Stu-\sum_{\substack{\iv\subset\iu \\ \iv\neq\iu}} S_{\iv}.
\end{equation}

\section{Example applications}
\label{sec:examples}
In this section, we perform global sensitivity analysis for the responses of four models with different characteristics and dimensionality. The first two models correspond to analytical functions, namely a common benchmark function in uncertainty quantification, of dimension $M=20$, and a structural-mechanics model of dimension $M=5$; the aforementioned analytical models have a rank-one structure.  The subsequent applications involve finite-element models, namely  a structural-mechanics model of dimension $M=10$ and a heat-conduction model, with thermal conductivity described by a random field, of dimension $M=53$. The LRA-based Sobol' indices are compared to respective PCE-based indices and reference indices based on the actual model. The latter are computed either analytically or by using a Monte Carlo Simulation (MCS) approach with large samples according to \citet{Janon2013}. The computations of the PCE- and MCS-based indices  are performed with the software UQLab \citep{MarelliICVRAM2014, UQdoc_09_106}.

The meta-models are built using two types of EDs, based on Sobol pseudo-random sequences \citep{Niederreiter1992} and the so-called \emph{maximin} Latin Hypercube Sampling (LHS). Each ED of the latter type represents the best among 5 random LHS-based designs, where the selection criterion is the maximum of the minimum distance between the points. The LRA meta-models are built by implementing Algorithm 1 in Section~\ref{sec:LRA_form}. A common polynomial degree $p_1=\ldots=p_M=p$ is considered with its optimal value selected as the one leading to the minimum 3-fold cross-validation error, denoted by $\errCV$ (see \citet{Konakli2015arxiv} for an investigation of the accuracy of the approach). The involved minimization problems are solved using the OLS method. In building the PCE meta-models, a candidate basis is first determined by employing a hyperbolic truncation scheme and then, a sparse expansion is obtained by evaluating the PCE coefficients with the hybrid LAR method. The optimal combination of the maximum total polynomial degree $p^t$ and the parameter $q$ controlling the truncation, with $q\in\{0.25,0.50,0.75,1.0\}$, is selected as the one leading to the minimum corrected leave-one-out error $\errLOO$ (see Section \ref{sec:PCE_form} for details).

\subsection{Analytical functions}

\subsubsection{Sobol function}
\label{sec:Sob_LRA}

In the first example, we consider the Sobol function:
\begin{equation}
\label{eq:Sobol}
Y = \prod_{i=1}^M\frac{|4X_i-2|+c_i}{1+c_i},
\end{equation}
where $\ve X=\{X_1,\ldots, X_M\}$ are independent random variables uniformly distributed over $[0,1]$ and $\ve c=\{c_1,\ldots, c_M\}$ are non-negative constants. We examine the case with $M=20$ and the constants given by \citet{Kersaudy2015new}:
\begin{equation}
\label{eq:Sobol_c}
\ve c=\{1,2,5,10,20,50,100,500,500,500,500,500,500,500,500,500,500,500,500,500\}^\mathrm{T}.
\end{equation}
This function is a commonly used benchmark for global sensitivity analysis  with well-known analytical solutions for the Sobol' sensitivity indices. In particular, the index $S_{i_1 \enum i_s}$ is obtained as \citet{Sudret2008c}:
\begin{equation}
\label{eq:Sobol_S}
S_{i_1 \enum i_s} = \dfrac{1}{D}\prod_{i=i_1}^{i_s} D_i,
\end{equation}
with the partial and total variances respectively given by:
\begin{equation}
\label{eq:Sobol_Di}
D_i=\dfrac{1}{3(1+c_i)^2}
\end{equation}
and
\begin{equation}
\label{eq:Sobol_D}
D =\prod_{i=1}^{M} (D_i+1) -1.
\end{equation}
Then, the first-order or total indices for any single variable or groups of variables can be computed using Eq.~(\ref{eq:Si})-(\ref{eq:StuT}). Because the input variables follow a uniform distribution, we build LRA and PCE meta-models of the function using Legendre polynomials after an isoprobabilistic transformation into standard variables distributed uniformly over $[-1,1]$.

We first assess the comparative accuracy of LRA and PCE in estimating the mean and standard deviation of the response, respectively denoted by $\mu_Y$ and $\sigma_Y$. The LRA- and PCE-based estimates of $\mu_Y$ and $\sigma_Y$ are obtained in terms of the coefficients of the meta-models, as described in Sections~\ref{sec:PCE_indices} and \ref{sec:LRA_indices}. Table~\ref{tab:Sobol_moments} lists the analytical solutions for $\mu_Y$ and $\sigma_Y$ based on the actual model together with their LRA- and PCE-based estimates for two EDs of size $N=200$ and $N=500$ obtained with Sobol sequences (see the Appendix for the parameters of the respective meta-models). The relative errors $\vare$ of the estimates are given in parentheses. We observe that the mean is estimated with high accuracy in all cases, but LRA outperform PCE in the estimation of the standard deviation.

\begin{table} [!ht]
	\centering
	\caption{Sobol function: Mean and standard deviation of response for experimental designs obtained with Sobol sequences.}
	\vspace{2mm} 
	\label{tab:Sobol_moments}
	\begin{tabular}{c c c c c c}
		\hline
		{} & {} & \multicolumn{2}{c} {$N = 200$} & \multicolumn{2}{c} {$N = 500$} \\
		& Analytical & LRA ($\vare \%$) & PCE ($\vare \%$) & LRA ($\vare \%$) & PCE ($\vare \%$) \\
		\hline 
		$\mu_{Y}$ & $1.000$ & $ 1.005 $ ($0.5$) & $0.998$ ($-0.2$) & $1.000$ ($0.0$) & $0.995$ ($-0.5$)  \\
		$\sigma_{Y}$ & $0.3715$ &  $0.3820$ ($2.8$) & $0.3424$ ($-7.8$) & $0.3715$ ($0.0$) & $ 0.3536$ ($-4.8$) \\
		\hline       
	\end{tabular}
\end{table}

\begin{figure}[!ht]
	\centering
	\includegraphics[width=0.45\textwidth] {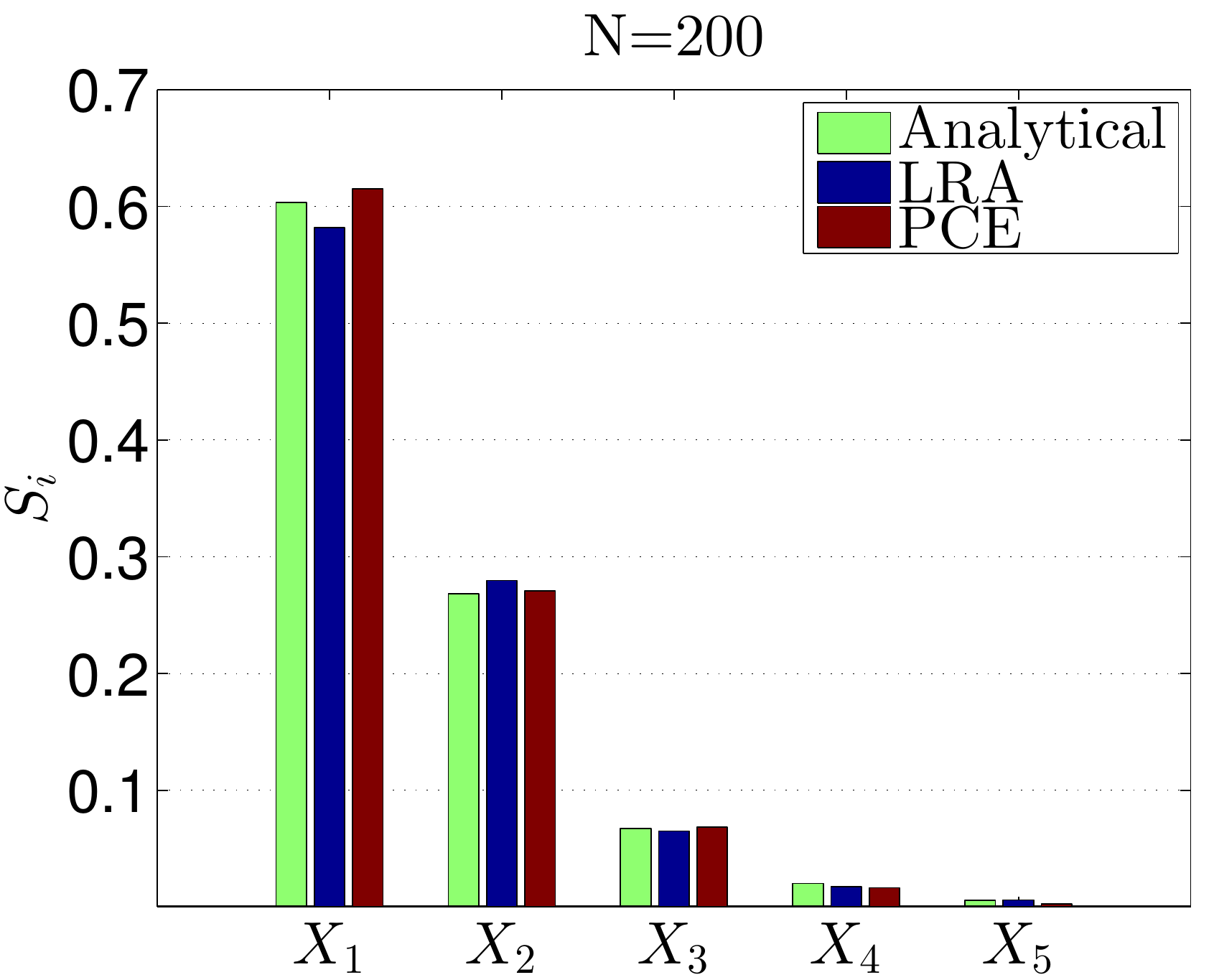}
	\includegraphics[width=0.45\textwidth] {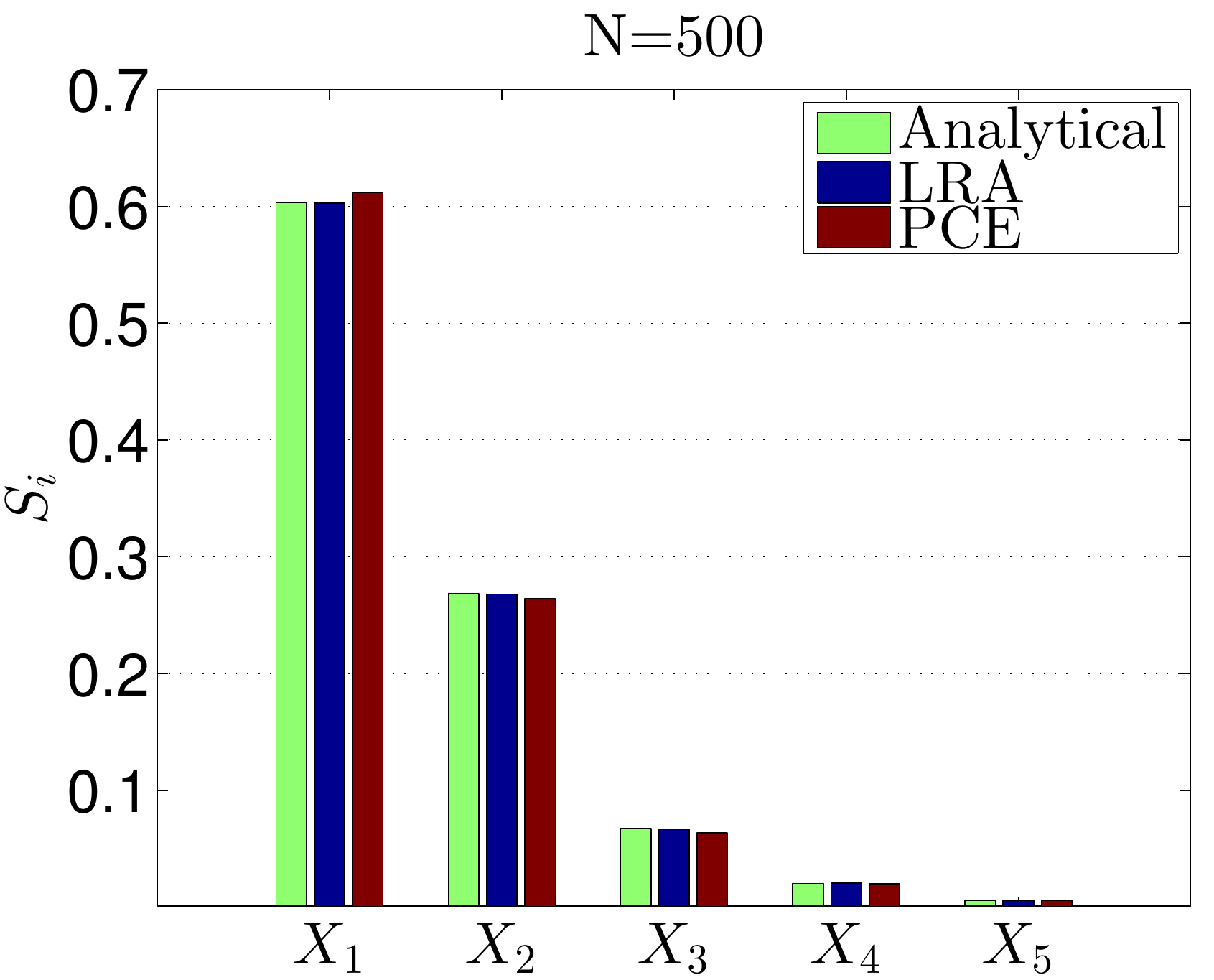}
	\caption{Sobol function:  Comparison of LRA- and PCE-based first-order Sobol' indices to their analytical values for experimental designs obtained with Sobol sequences.}
	\label{fig:Sobol_S1}
\end{figure}

\begin{figure}[!ht]
	\centering
	\includegraphics[width=0.45\textwidth] {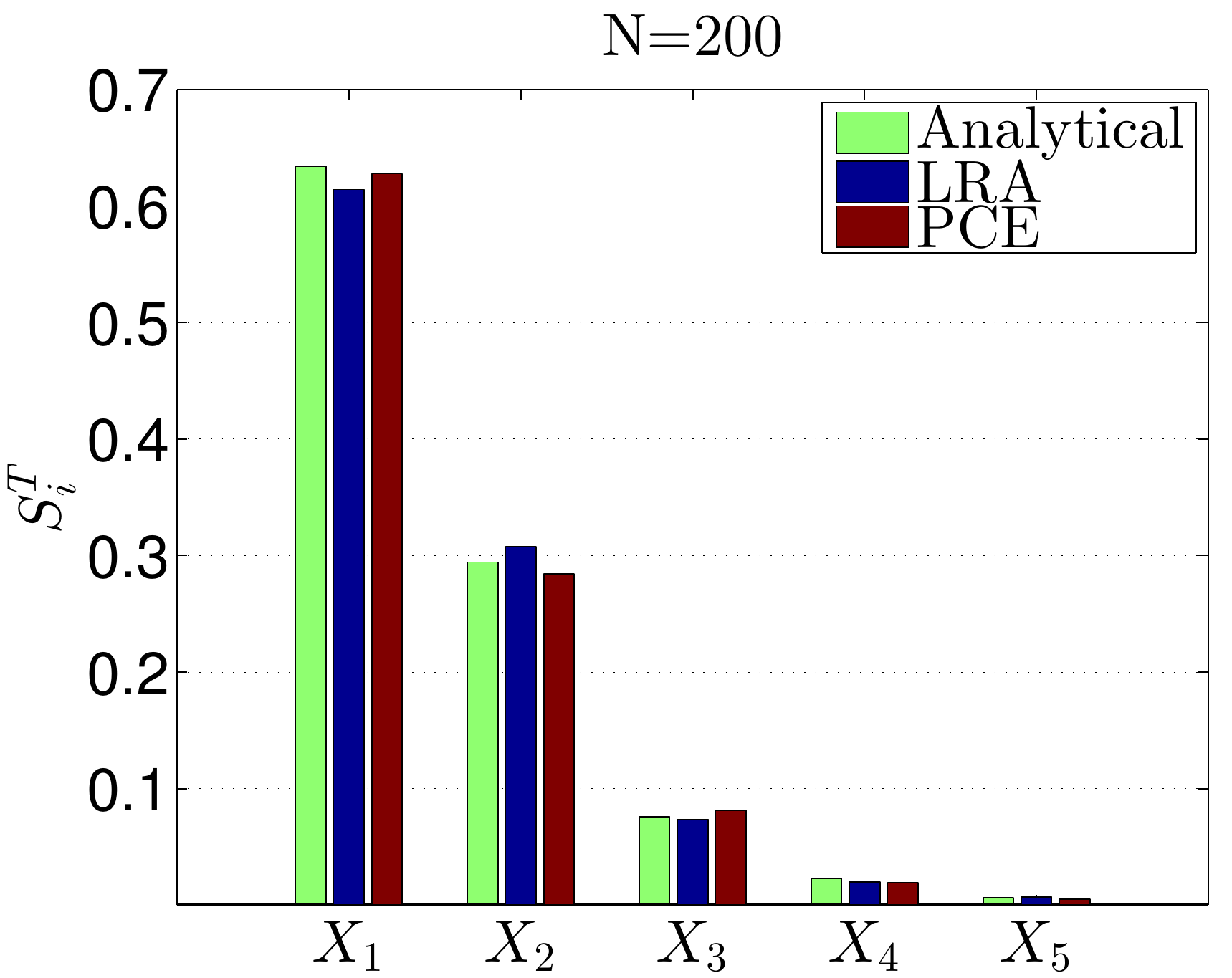}
	\includegraphics[width=0.45\textwidth] {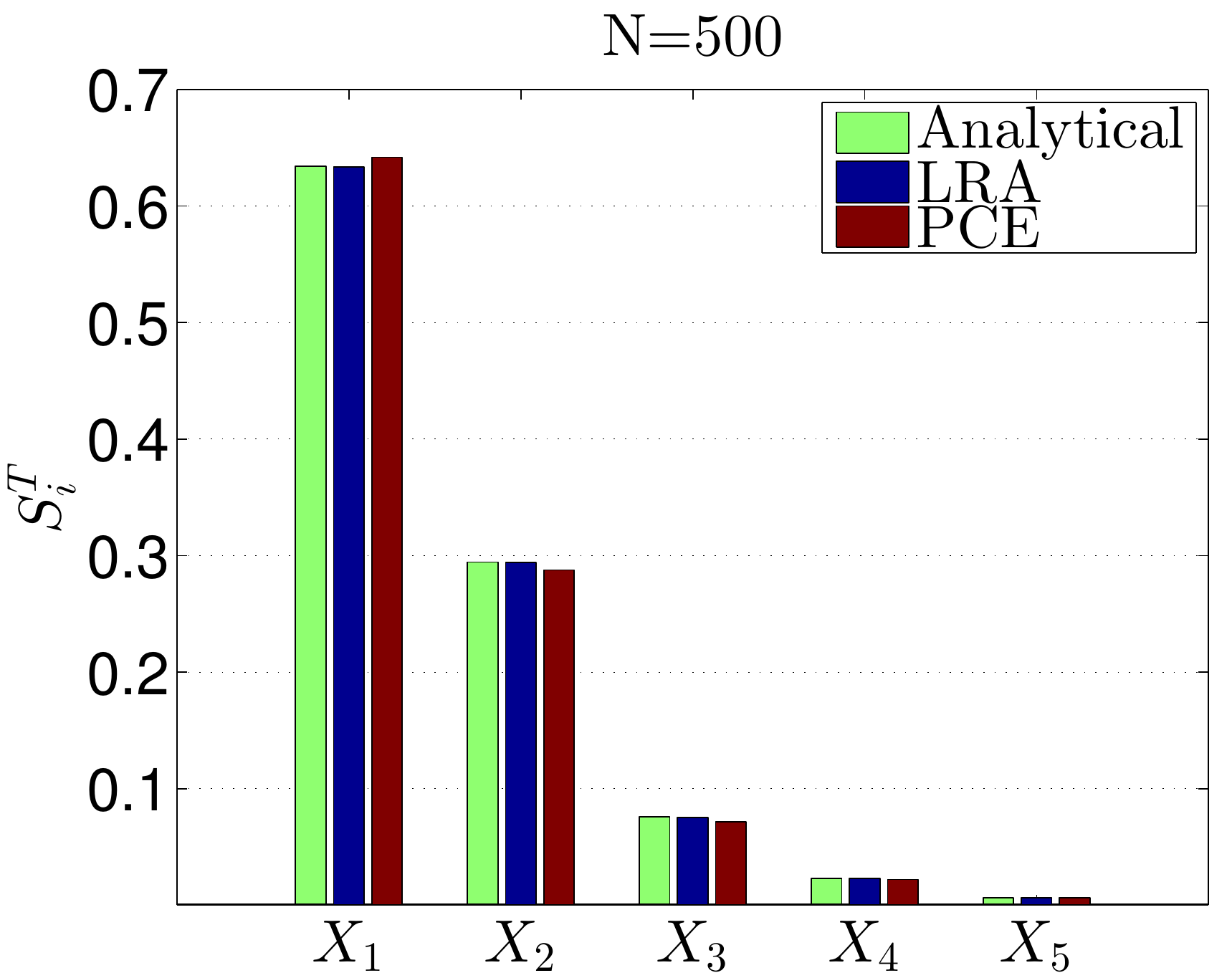}
	\caption{Sobol function: Comparison of LRA- and PCE-based total Sobol' indices to their analytical values for experimental designs obtained with Sobol sequences.}
	\label{fig:Sobol_Stot}
\end{figure}

In the sequel, we compare LRA- and PCE-based sensitivity indices, obtained by post-processing the coefficients of the meta-models according to Sections~\ref{sec:PCE_indices} and \ref{sec:LRA_indices}, with the corresponding analytical solutions for the actual model. Figures~\ref{fig:Sobol_S1} and \ref{fig:Sobol_Stot} respectively show the first-order and total Sobol' indices of the variables $X_1 \enum X_5$ considering the same EDs of size $N=200$ and $N=500$ as above. The indices of the remaining variables $X_6 \enum X_{20}$ are practically zero. The small differences between the first-order and total indices indicate only minor effects from interactions between the various variables. For $N=200$, the sensitivity indices based on the meta-models are fairly close to the reference ones, whereas for $N=500$, the agreement is nearly perfect. The values of the indices depicted in the two figures are listed in the Appendix.

\begin{figure}[!ht]
	\centering
	\includegraphics[width=0.45\textwidth] {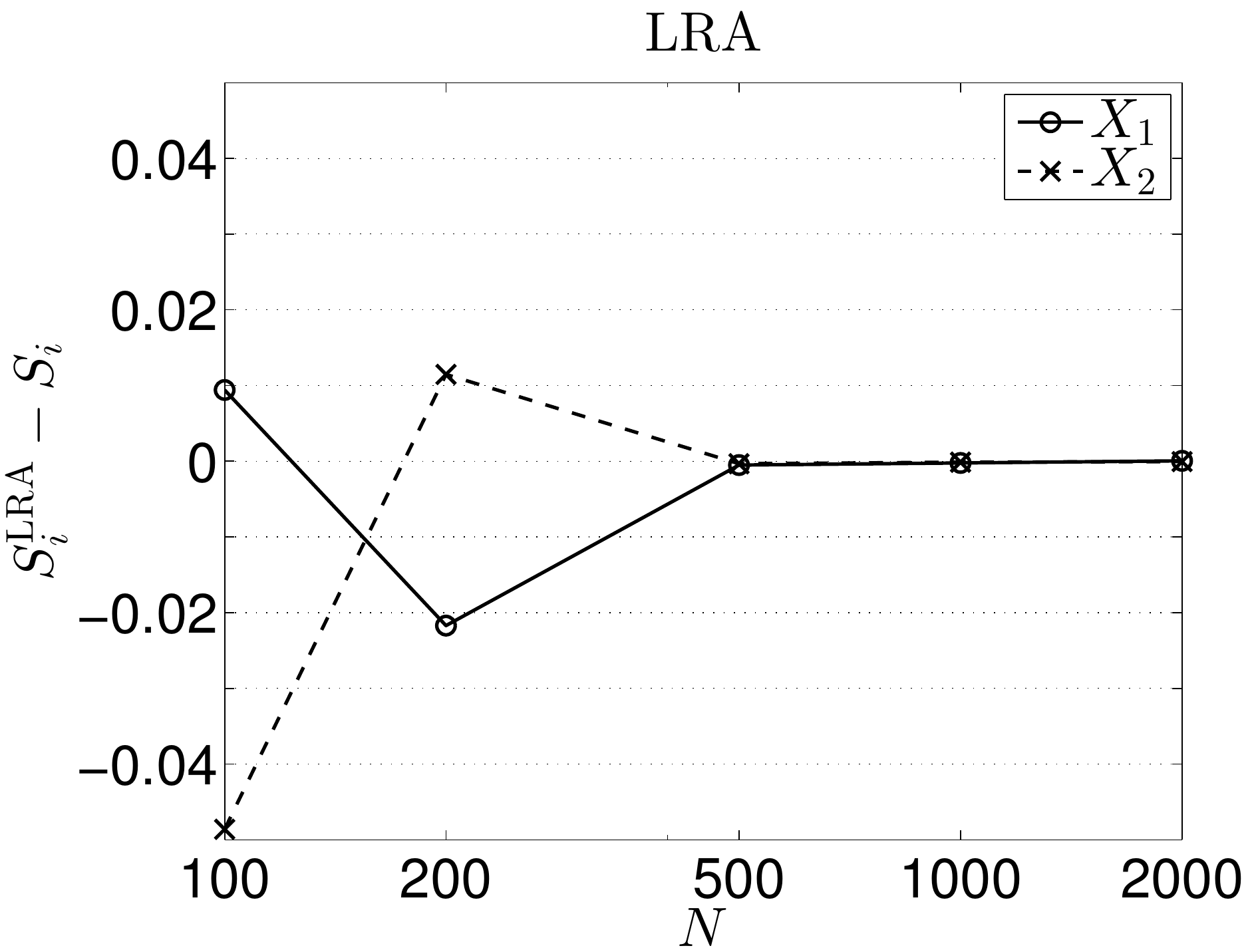}
	\includegraphics[width=0.45\textwidth] {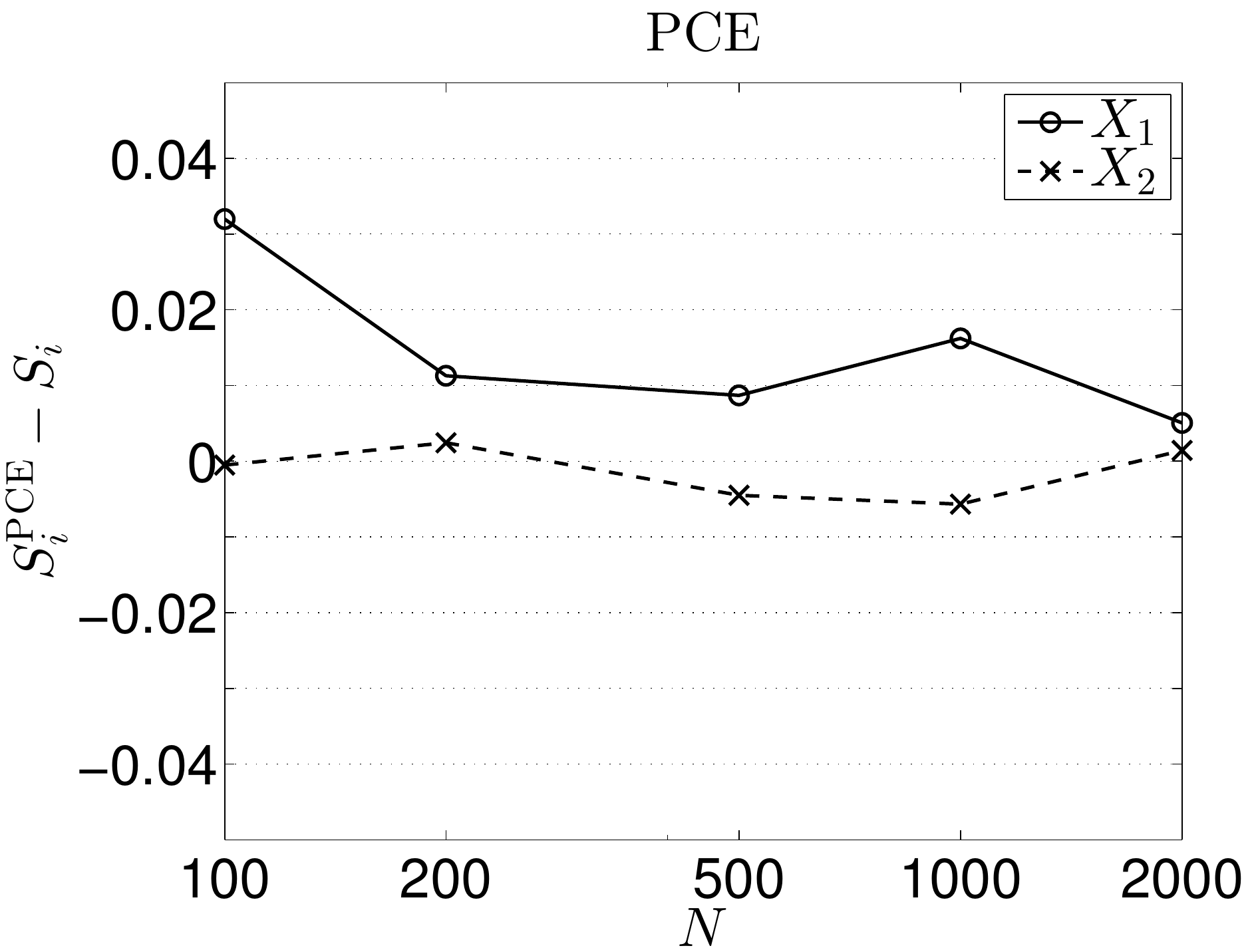}
	\caption{Sobol function: Errors of LRA- and PCE-based first-order Sobol' indices for experimental designs obtained with Sobol sequences.}
	\label{fig:Sobol_S1_N}
\end{figure}

\begin{figure}[!ht]
	\centering
	\includegraphics[width=0.45\textwidth] {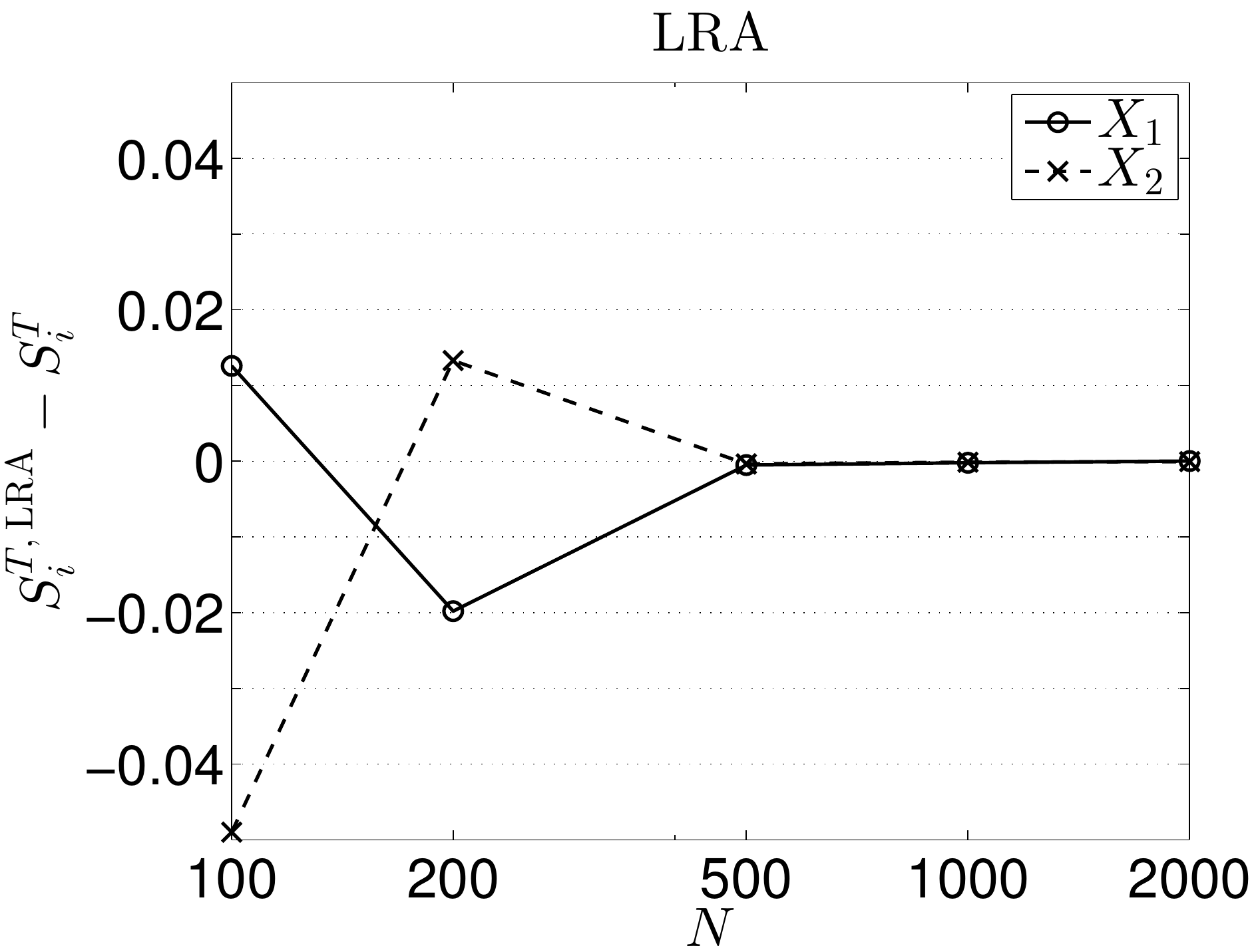}
	\includegraphics[width=0.45\textwidth] {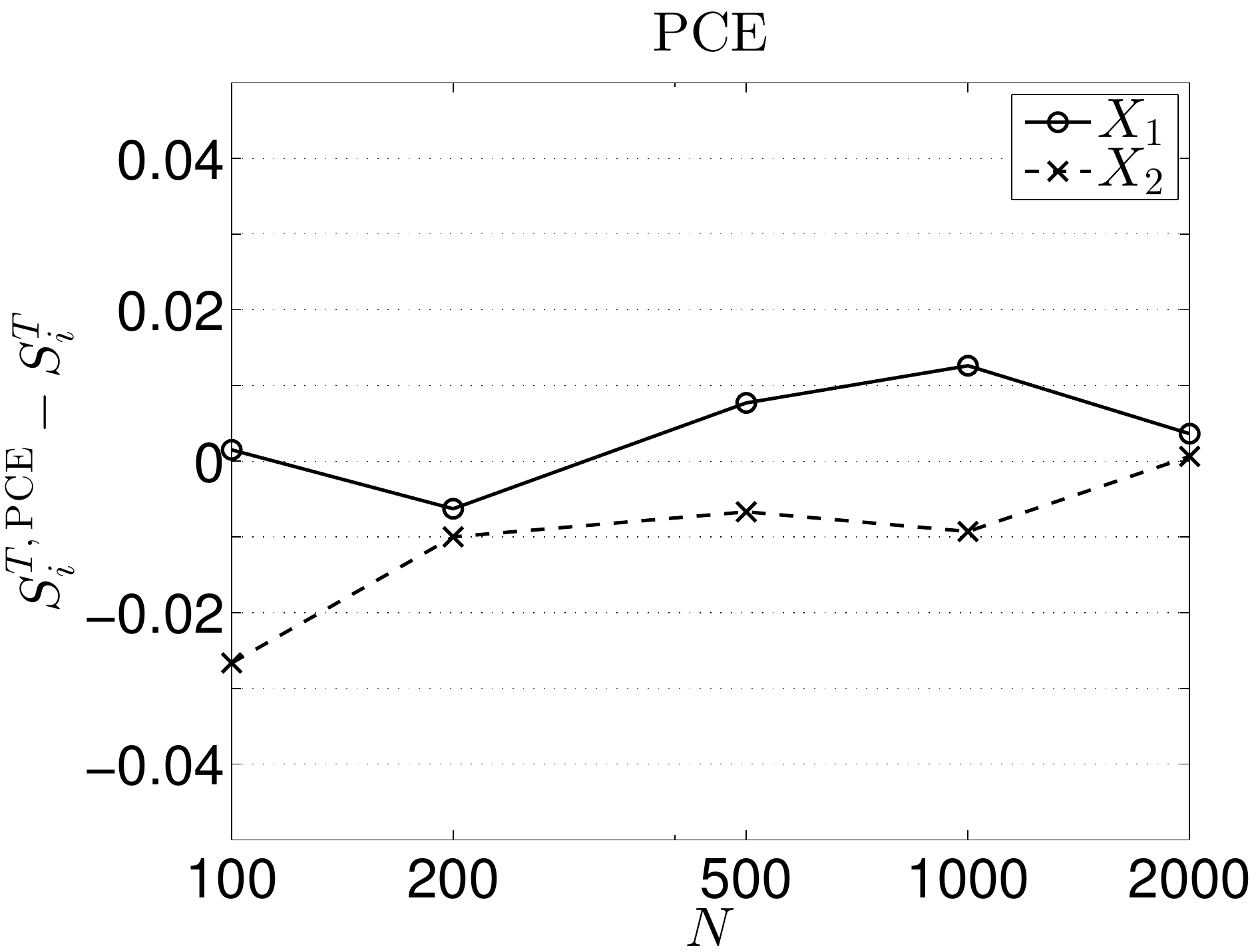}
	\caption{Sobol function: Errors of LRA- and PCE-based total Sobol' indices for experimental designs obtained with Sobol sequences.}
	\label{fig:Sobol_Stot_N}
\end{figure}

To further assess the effect of the ED size, we examine the convergence of the LRA- and PCE-based indices of the two most important variables, \ie $X_1$ and $X_2$, while $N$ varies from 100 to 2,000; as in the above investigations, the considered EDs are obtained with Sobol sequences. Figure~\ref{fig:Sobol_S1_N} shows the differences between the first-order indices obtained from the meta-model coefficients and their exact values; Figure~\ref{fig:Sobol_Stot_N} shows similar differences for the total indices. For $N\leq 200$, the PCE-based indices are overall more accurate; however, the LRA-based indices converge faster, practically reaching the exact values for $N=500$. Table~\ref{tab:Sobol_errG} lists the relative generalization errors of the considered meta-models, estimated with a MCS-based validation set comprising $10^6$ points. These errors decrease faster for LRA, which is consistent with the faster convergence of the respective sensitivity indices observed above.

\begin{table} [!ht]
	\centering
	\caption{Sobol function: Relative generalization errors of meta-models based on Sobol sequences.}
	\vspace{2mm} 
	\label{tab:Sobol_errG}
	\begin{tabular}{c c c}
		\hline
		N & $\widehat{err}_G^{\rm LRA}$ & $\widehat{err}_G^{\rm PCE}$\\
		\hline
		100    &  $8.08\cdot 10^{-2}$  &  $5.46\cdot 10^{-2}$ \\
		200    &  $2.57\cdot 10^{-2}$  &  $3.64\cdot 10^{-2}$ \\
		500    &  $2.32\cdot 10^{-3}$  &  $1.45\cdot 10^{-2}$ \\
		1,000  &  $4.68\cdot 10^{-4}$  &  $6.34\cdot 10^{-3}$ \\
		2,000  &  $2.03\cdot 10^{-4}$  &  $2.48\cdot 10^{-3}$ \\
		\hline       
	\end{tabular}
\end{table}

\begin{figure}[!ht]
	\centering
	\includegraphics[width=0.45\textwidth] {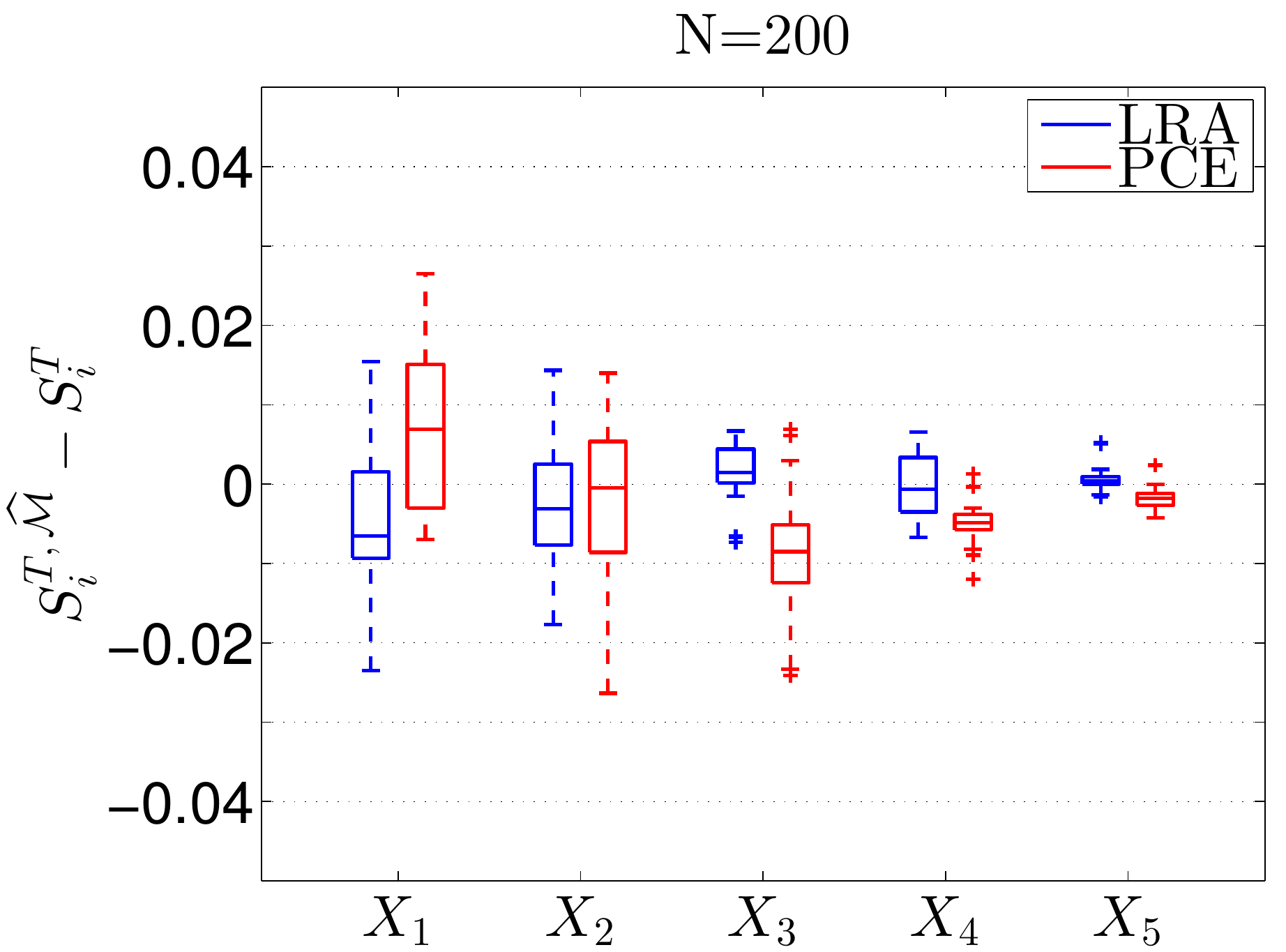}
	\includegraphics[width=0.45\textwidth] {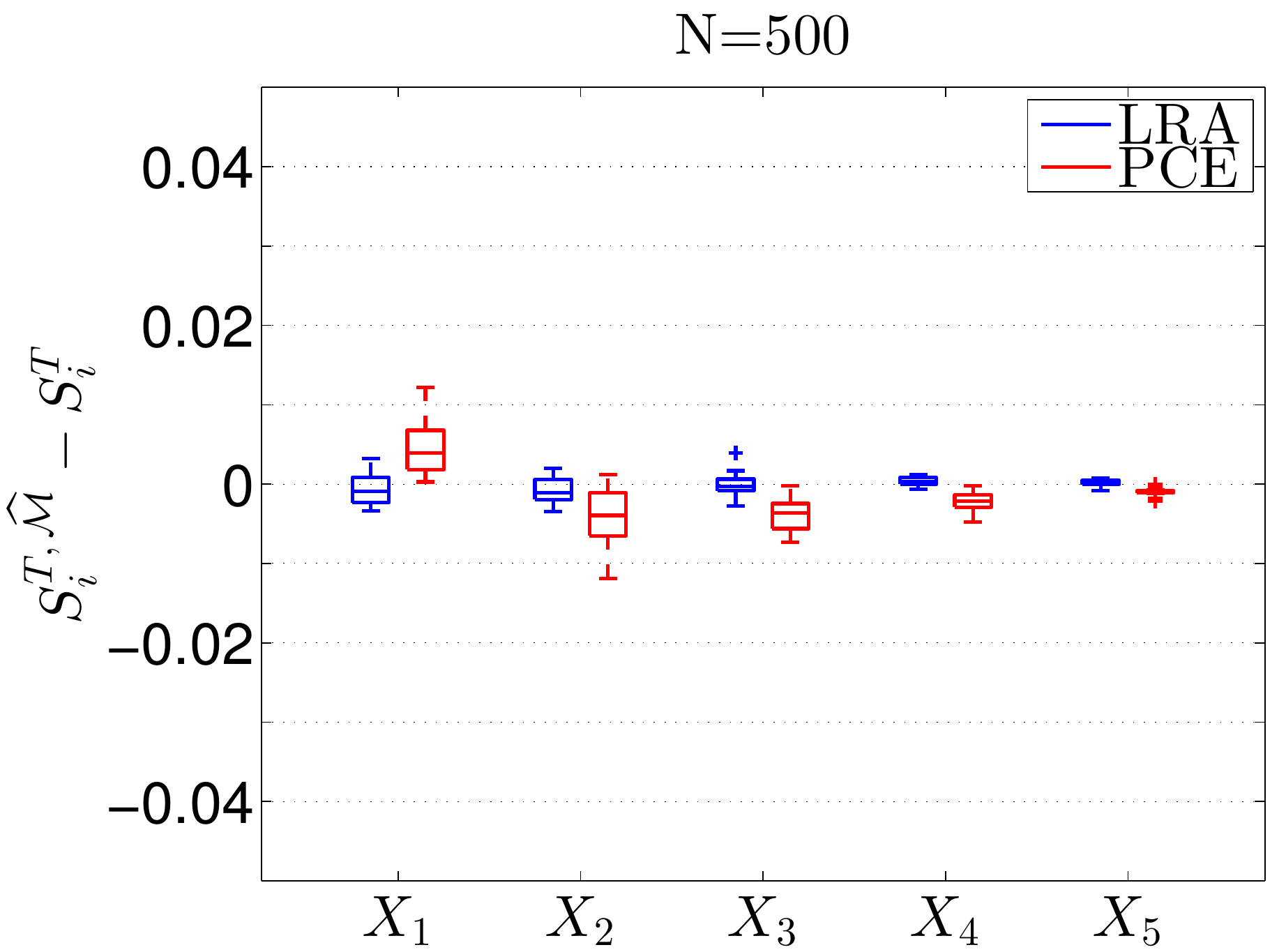}
	\caption{Sobol function: Errors of LRA- and PCE-based total Sobol' indices for experimental designs obtained with maximin LHS (20 replications).}
	\label{fig:Sobol_Stot_LHS}
\end{figure}

We conclude this example by investigating the accuracy of the LRA- and PCE-based Sobol' indices when the meta-models are built with maximin LHS designs. Figure~\ref{fig:Sobol_Stot_LHS} shows boxplots of the differences between the total Sobol' indices obtained with the meta-models and their exact values, considering 20 EDs of size $N=200$ and $N=500$, \ie for the same ED sizes examined in Figure~\ref{fig:Sobol_Stot}. Note that in all cases, the absolute median errors do not exceed $0.01$. As expected, the error spread decreases with increasing ED size.

\subsubsection{Beam deflection}
In this example, we consider a simply supported beam with a uniform rectangular cross-section subjected to a concentrated load at the mid-span. The response quantity of interest is the midspan deflection, which is obtained through basic structural mechanics as:
\begin{equation}
\label{eq:beam_u}
U=\dfrac{P L^3}{4 E b h^3},
\end{equation}
where $b$ and $h$ respectively denote the width and height of the cross section, $L$ is the length of the beam, $E$ is the Young's modulus and $P$ is the load. The aforementioned parameters are modeled as independent random variables with their distributions, mean and coefficient of variation (CoV) values listed in Table~\ref{tab:beam_input}. The dimensionality of the problem is thus $M=5$. LRA and PCE meta-models are built using Hermite polynomials after an isoprobabilistic transformation of the input variables into standard normal variables.

\begin{table} [!ht]
	\centering
	\caption{Beam deflection: Distributions of input variables.}
	\vspace{2mm} 
	\label{tab:beam_input}
	\begin{tabular}{c c c c}
		\hline Variable & Distribution & mean & CoV \\
		\hline $b$ (m) & Lognormal & 0.15 & 0.05 \\
		$h$ (m) & Lognormal & 0.3  & 0.05 \\
		$L$ (m) & Lognormal & 5    & 0.01 \\
		$E$ (MPa) & Lognormal & 30,000  & 0.15 \\
		$P$ (KN) & Lognormal & 10 & 0.20 \\
		\hline       
	\end{tabular}
\end{table}

In Table~\ref{tab:beam_moments}, we assess the accuracy of LRA and PCE meta-models in estimating the mean $\mu_U$ and standard deviation $\sigma_U$ of the response by post-processing the meta-model coefficients. To this end, we consider two EDs of size $N=30$ and $N=50$ obtained with Sobol sequences (see the Appendix for the parameters of the respective meta-models). Because $U$ in Eq.~(\ref{eq:beam_u}) is a product of lognormal random variables, it easy to obtain the mean and standard deviation of the actual model response analytically. The relative errors $\vare$ of the LRA- and PCE-based estimates with respect to the analytical values are given in parentheses. $N=30$ is sufficient to obtain highly accurate estimates of both $\mu_U$ and $\sigma_U$ with the LRA approach; for both EDs, the PCE approach appears slightly inferior in the estimation of $\sigma_U$.

\begin{table} [!ht]
	\centering
	\caption{Beam deflection: Mean and standard deviation of response for experimental designs obtained with Sobol sequences.}
	\vspace{2mm} 
	\label{tab:beam_moments}
	\begin{tabular}{c c c c c c}
		\hline
		{} & {} & \multicolumn{2}{c} {$N = 30$} & \multicolumn{2}{c} {$N = 50$} \\
		& Analytical & LRA ($\vare \%$) & PCE ($\vare \%$) & LRA ($\vare \%$) & PCE ($\vare \%$) \\
		\hline 
		$\mu_U$ (mm) & $2.677$ & $2.678$ ($0.0$) & $2.675$ ($-0.1$) & $2.677$ ($0.0$) & $2.673$ ($-0.2$) \\
		$\sigma_U$ (mm) &  $0.8088$ &  $0.8047$  ($-0.5$) & $0.7743$  ($-4.3$) &  $0.8085$ ($0.0$) & $0.7917$  ($-2.1$)\\
		\hline       
	\end{tabular}
\end{table}

\begin{figure}[!ht]
	\centering
	\includegraphics[width=0.45\textwidth] {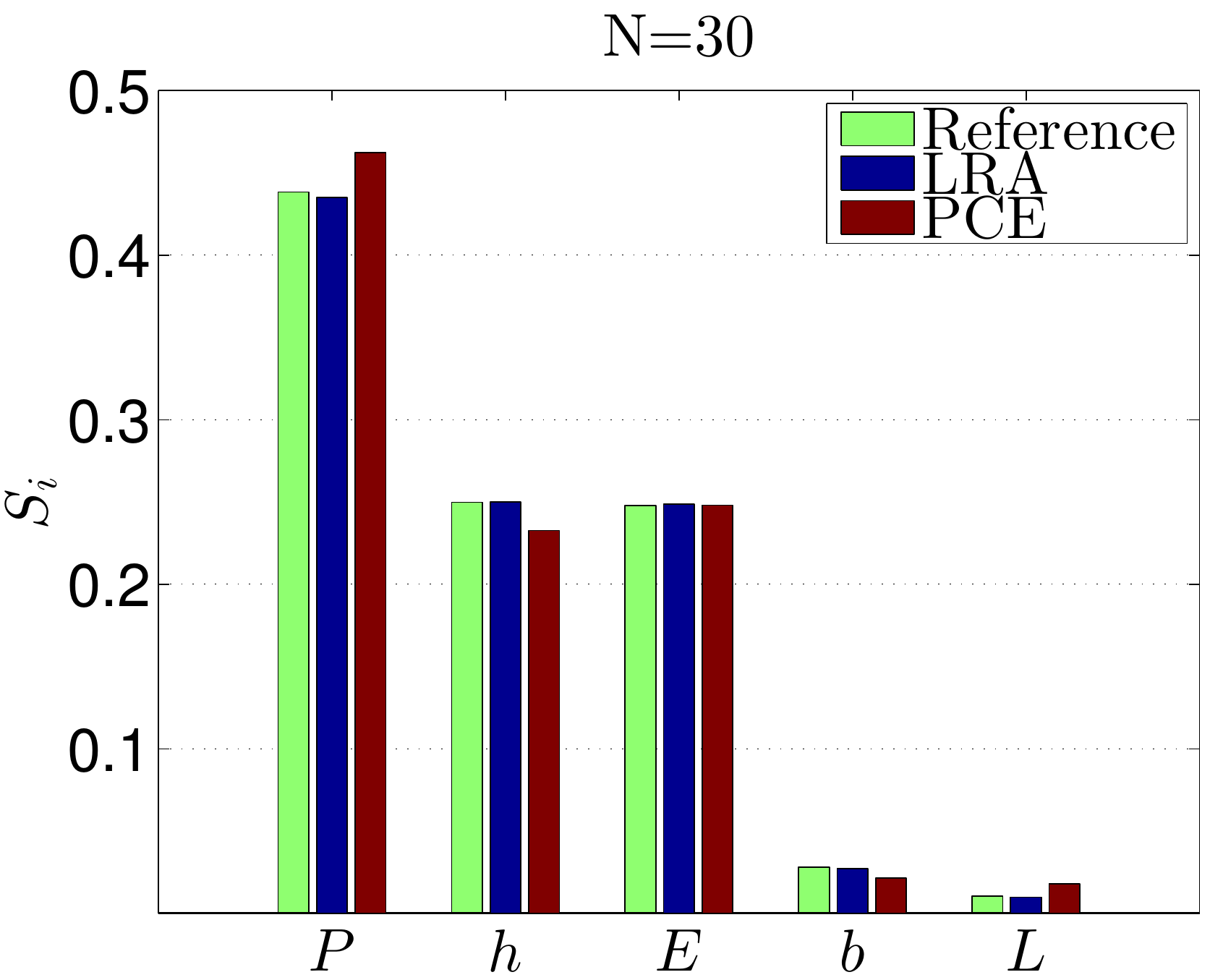}
	\includegraphics[width=0.45\textwidth] {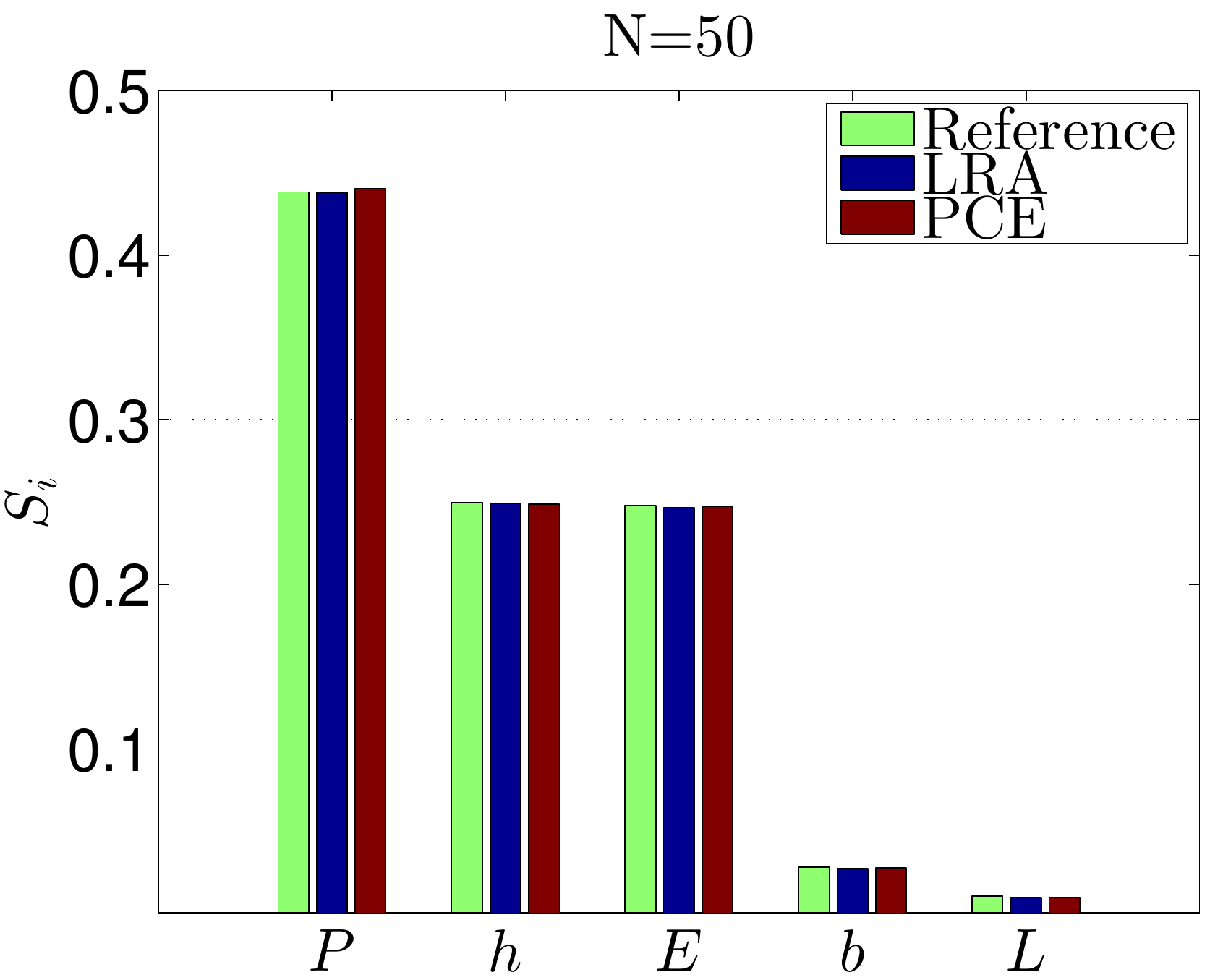}
	\caption{Beam deflection: Comparison of LRA- and PCE-based first-order Sobol' indices to their reference values for experimental designs obtained with Sobol sequences.}
	\label{fig:beam_S1}
\end{figure}

\begin{figure}[!ht]
	\centering
	\includegraphics[width=0.45\textwidth] {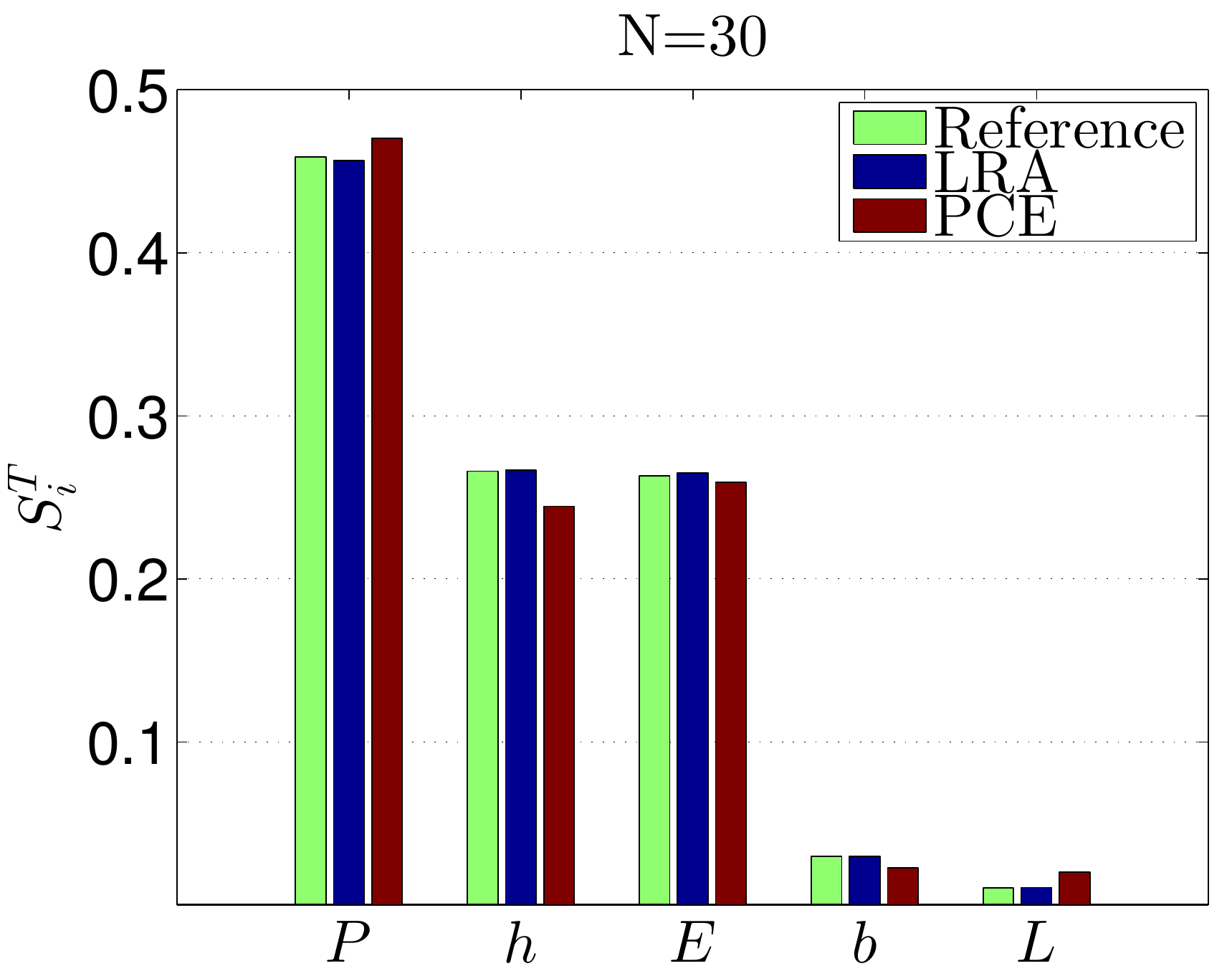}
	\includegraphics[width=0.45\textwidth] {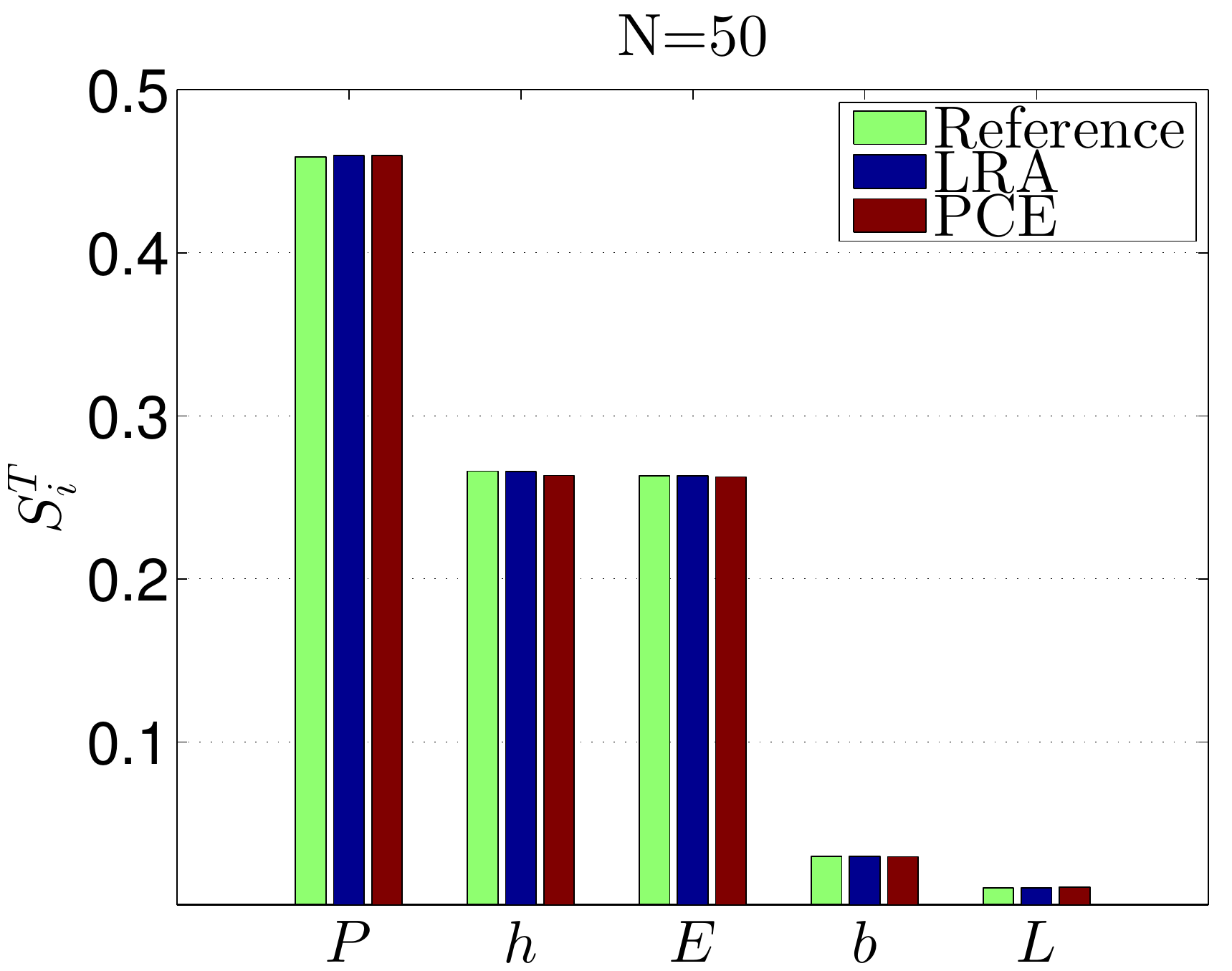}
	\caption{Beam deflection: Comparison of LRA- and PCE-based total Sobol' indices to their reference values for experimental designs obtained with Sobol sequences.}
	\label{fig:beam_Stot}
\end{figure}

Next, we examine the first-order and total Sobol' indices of the five input variables. The LRA- and PCE-based indices, obtained by post-processing the coefficients of the meta-models, are compared to respective reference values, obtained with a MCS approach using $n=10^6$ samples for each index. For the same EDs considered above, Figures~\ref{fig:beam_S1} and \ref{fig:beam_Stot} depict the first-order and total Sobol' indices, respectively, in order of total importance. It is noteworthy that the LRA-based indices are in nearly perfect agreement with the references ones even for the smaller ED, comprising only $N=30$ points. The PCE-based indices demonstrate a similar accuracy only for $N=50$. Obviously, the uncertainty in the beam deflection is driven by the uncertainty in the load, the cross-section height and the the Young's modulus, with the load having the dominant contribution. Comparisons between the first-order and the total indices indicate relatively small contributions from higher-order effects. The values of the indices depicted in the two figures are listed in the Appendix.

Table~\ref{tab:beam_errG} lists the relative generalization errors of the LRA and PCE meta-models built with the two EDs. These errors are estimated using a MCS-based validation set comprising $10^6$ points. The superior performance of LRA as compared to PCE in the estimation of the response statistics and the Sobol' indices is consistent with the higher accuracy of the former manifested in the lower generalization errors.

\begin{table} [!ht]
	\centering
	\caption{Beam deflection: Relative generalization errors of meta-models based on Sobol sequences.}
	\vspace{2mm} 
	\label{tab:beam_errG}
	\begin{tabular}{c c c}
		\hline
		N & $\widehat{err}_G^{\rm LRA}$ & $\widehat{err}_G^{\rm PCE}$\\
		\hline
		30 &  $2.32\cdot 10^{-4}$ &  $1.47\cdot 10^{-2}$ \\
		50 &  $2.63\cdot 10^{-6}$ &  $1.81\cdot 10^{-3}$ \\
		\hline       
	\end{tabular}
\end{table}

\begin{figure}[!ht]
	\centering
	\includegraphics[width=0.45\textwidth] {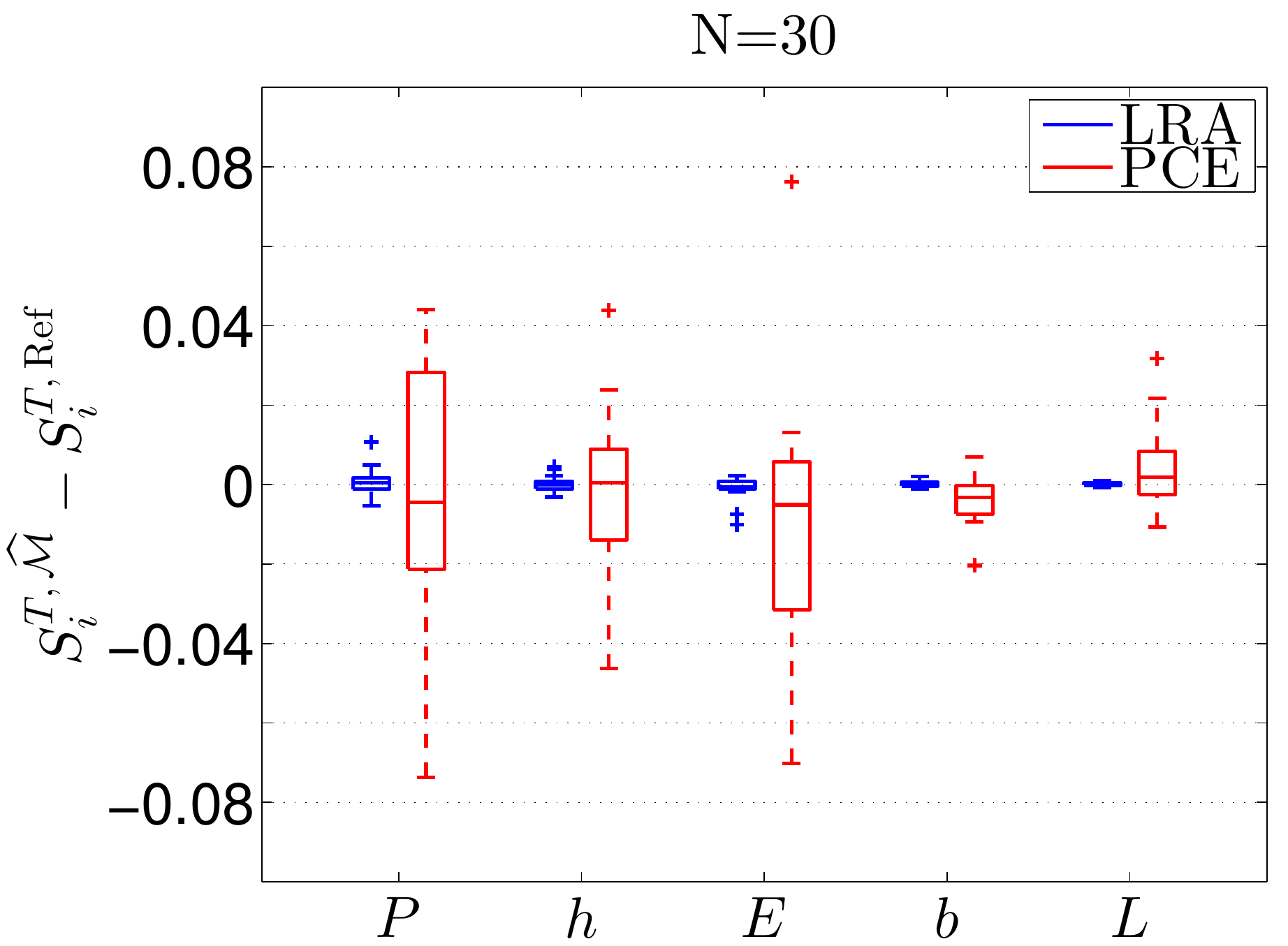}
	\includegraphics[width=0.45\textwidth] {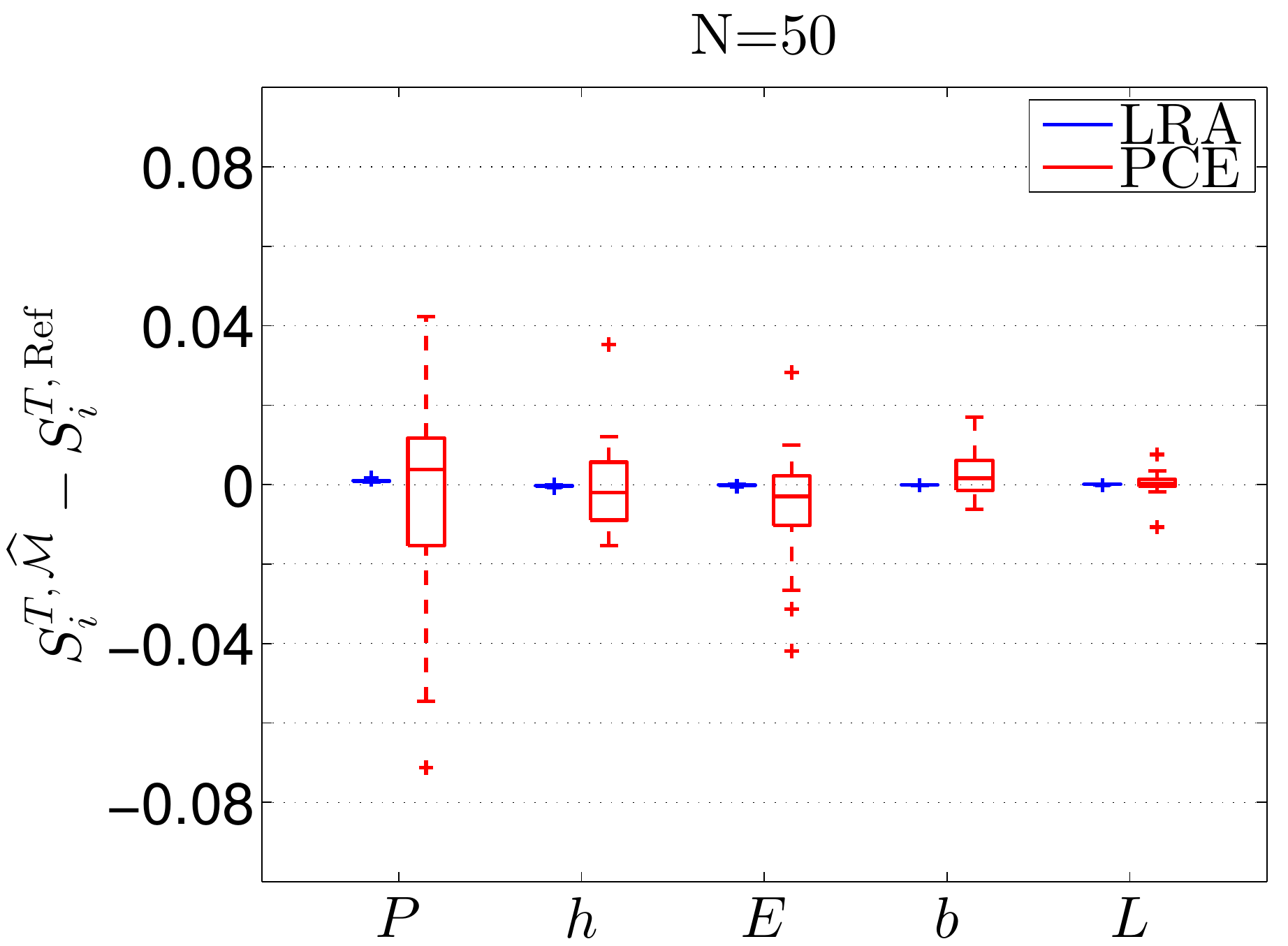}
	\caption{Beam deflection: Errors of LRA- and PCE-based total Sobol' indices for experimental designs obtained with maximin LHS (20 replications).}
	\label{fig:beam_Stot_LHS}
\end{figure}

Finally, we investigate the accuracy of the LRA- and PCE-based Sobol' indices when the meta-models are built with maximin LHS designs. Figure~\ref{fig:beam_Stot_LHS} depicts boxplots of the differences between the total indices based on the meta-models and their reference values for 20 replications with ED sizes $N=30$ and $N=50$. Although both the LRA- and the PCE-based estimates are practically unbiased, the former exhibit a significantly smaller spread, which is nearly zero for $N=50$. The notably superior performance of LRA in the present problem can be explained by the rank-one structure of the underlying model.

\subsection{Finite element models}

\subsubsection{Truss deflection}

\begin{figure}[!ht]
	\centering
	\includegraphics[trim = 15mm 70mm 15mm 70mm, width=0.6\textwidth]{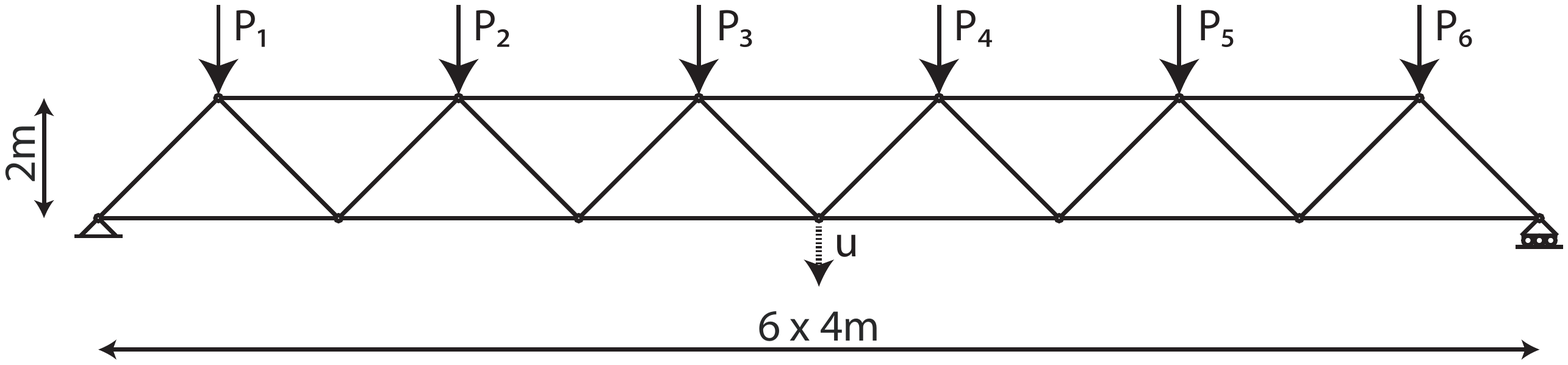}
	\caption{Truss structure.}
	\label{fig:truss}
\end{figure}

We consider the truss structure shown in Figure~\ref{fig:truss} (also studied in \citet{Sudret2007}), which is subjected to six vertical loads. The mid-span deflection, denoted by $u$, represents the response quantity of interest and is computed with an in-house finite-element analysis code developed in Matlab environment.  The random input includes $M=10$ independent random variables: the vertical loads, denoted by $P_1 \enum P_6$, the cross-sectional area and Young's modulus of the horizontal bars, respectively denoted by $A_1$ and $E_1$, and the cross-sectional area and Young's modulus of the vertical bars, respectively denoted by $A_2$ and $E_2$. The distributions of the input random variables are listed in Table~\ref{tab:truss_input}. LRA and PCE meta-models are built using Hermite polynomials after an isoprobabilistic transformation of the input variables into standard normal variables.

\begin{table} [!ht]
	\centering
	\caption{Truss deflection: Distributions of input random variables.}
	\vspace{2mm} 
	\label{tab:truss_input}
	\begin{tabular}{c c c c}
		\hline Variable & Distribution & mean & CoV \\
		\hline
		$A_1$ (m) & Lognormal & 0.002 & 0.10 \\
		$A_2$ (m) & Lognormal & 0.001  & 0.10 \\
		$E_1, E_2$ (MPa) & Lognormal & 210,000   & 0.10\\
		$P_1\enum P_6$ (KN) & Gumbel & 50  & 0.15 \\
		\hline       
	\end{tabular}
\end{table}

For two EDs of size $N=50$ and $N=200$ obtained with Sobol sequences, Table~\ref{tab:truss_moments} compares the estimates of the response mean and standard deviation obtained from the coefficients of LRA and PCE meta-models (see the Appendix for the meta-model parameters) with their respective values based on the actual model. The latter, which represent the reference values, are computed with a MCS sample comprising $n=10^6$ points. The relative errors $\vare$ of the LRA- and PCE-based estimates with respect to the reference values are given in parentheses. $N=50$ is sufficient to obtain excellent estimates of both quantities with the LRA approach; for the same ED, the PCE approach is slightly inferior in the estimation of the standard deviation. 

\begin{table} [!ht]
	\centering
	\caption{Truss deflection: Mean and standard deviation of response for experimental designs obtained with Sobol sequences.}
	\vspace{2mm} 
	\label{tab:truss_moments}
	\begin{tabular}{c c c c c c}
		\hline
		{} & {} & \multicolumn{2}{c} {$N = 50$} & \multicolumn{2}{c} {$N = 200$} \\
		& Reference & LRA ($\vare \%$) & PCE ($\vare \%$) & LRA ($\vare \%$) & PCE ($\vare \%$) \\
		\hline 
		$\mu_U$ (cm) & $7.941$   & $7.945$ ($0.0$) & $7.931$ ($-0.1$) & $7.942$ ($0.0$) & $7.939$ ($0.0$) \\
		$\sigma_U$ (cm) & $1.110$ & $1.109$ ($-0.1$) & $1.074$  ($-3.3$) & $1.114$ ($0.4$) & $1.106$  ($-0.4$) \\
		\hline       
	\end{tabular}
\end{table}

\begin{figure}[!ht]
	\centering
	\includegraphics[width=0.9\textwidth] {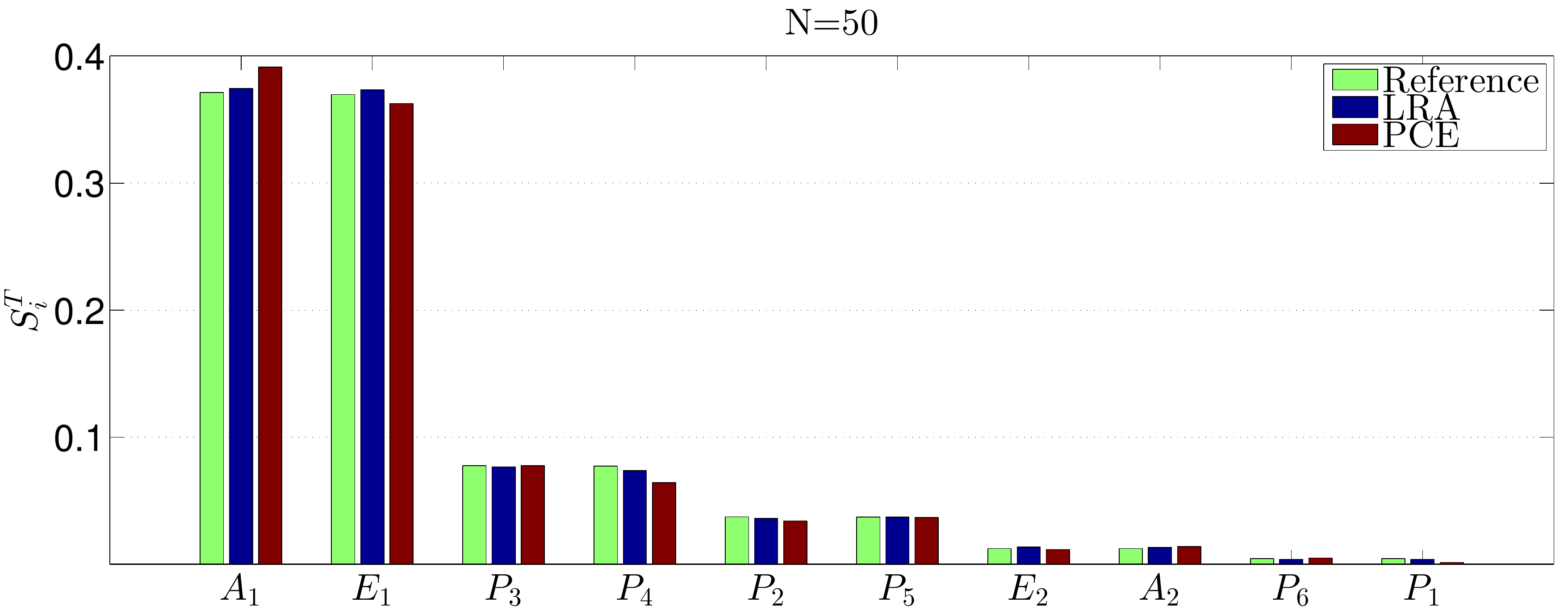}
	\includegraphics[width=0.9\textwidth] {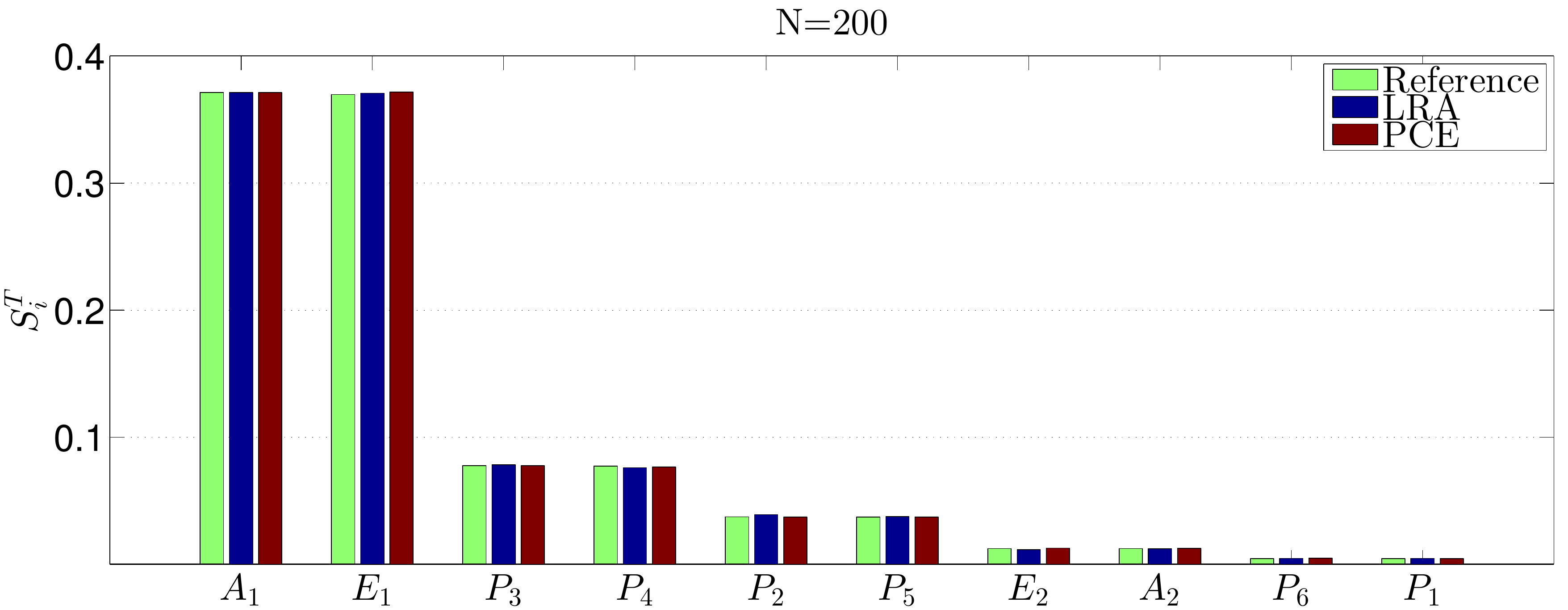}
	\caption{Truss deflection:  Comparison of LRA- and PCE-based total Sobol' indices to their reference values for experimental designs obtained with Sobol sequences.}
	\label{fig:truss_Stot}
\end{figure}

For the same EDs of size $N=50$ and $N=200$, we next compare the Sobol' indices obtained by post-processing the LRA and PCE coefficients to respective reference values. The latter are computed with a MCS approach using $n=10^6$ samples for each index. Figure~\ref{fig:truss_Stot} shows the total Sobol' indices in order of total importance. For $N=50$, the indices obtained with the meta-models are overall in fair agreement with the reference ones. In particular, the dominant contributions of $A_1$ and $E_1$ are estimated with good accuracy with the LRA approach, but with slightly lesser accuracy with the PCE approach. For $N=200$, both meta-modeling approaches yield practically perfect estimates of the reference indices. Because contributions from higher-order effects are negligible in this problem, the first-order indices are nearly equal to the total indices and are thus not shown. The values of both the total and the first-order indices for the cases examined in Figure~\ref{fig:truss_Stot} are listed in the Appendix. Note that in the reference solution, total indices of less significant variables may be marginally smaller than the corresponding first-order indices. This contradiction results from a small bias in the employed Monte-Carlo estimator, which is addressed in \citet{Owen2013}.


\begin{figure}[!ht]
	\centering
	\includegraphics[width=0.45\textwidth] {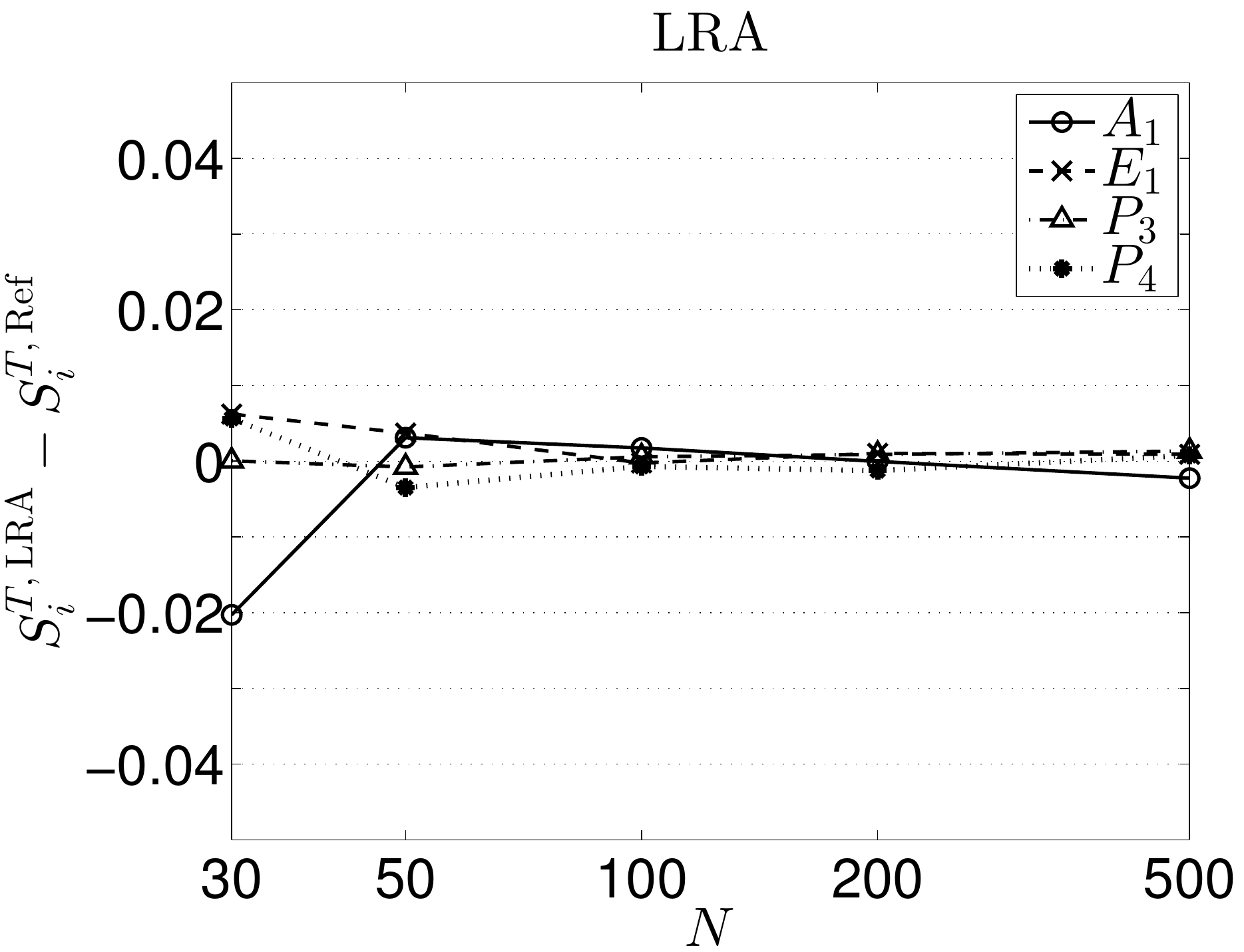}
	\includegraphics[width=0.45\textwidth] {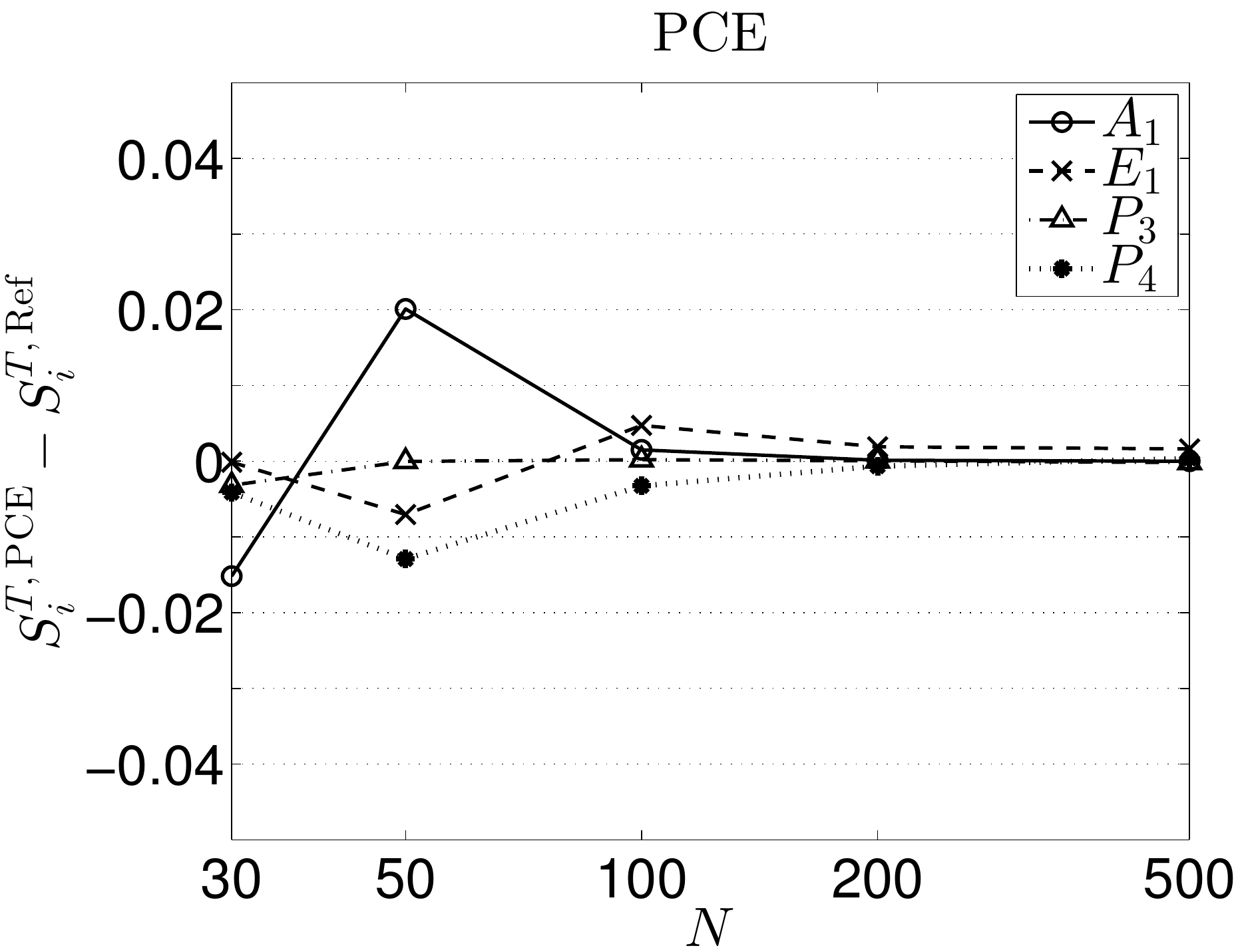}
	\caption{Truss deflection: Errors of LRA- and PCE-based total Sobol' indices for experimental designs obtained with Sobol sequences.}
	\label{fig:truss_Stot_N}
\end{figure}

We next examine the errors (differences) of the LRA- and PCE-based indices with respect to their reference values for the four most important variables, \ie $A_1$, $E_1$, $P_3$ and $P_4$, considering EDs of varying sizes obtained with Sobol sequences. Figure~\ref{fig:truss_Stot_N} shows these errors for the total indices, while $N$ varies from 30 to 500.  The PCE approach demonstrates a slightly slower convergence to the reference solution with increasing ED size. Similar results are obtained for the first-order indices (not shown herein). Table~\ref{tab:truss_errG} shows the relative generalization errors of the considered meta-models, estimated with a MCS validation set comprising $10^6$ points. For $N\geq200$, the PCE meta-models are characterized by smaller generalization errors; however, as shown above, smaller values of $N$ are sufficient to perform highly accurate sensitivity analysis with LRA.


\begin{table} [!ht]
	\centering
	\caption{Truss deflection: Relative generalization errors of meta-models based on Sobol sequences..}
	\vspace{2mm} 
	\label{tab:truss_errG}
	\begin{tabular}{c c c}
		\hline
		N & $\widehat{err}_G^{\rm LRA}$ & $\widehat{err}_G^{\rm PCE}$\\
		\hline
		30  &  $2.67\cdot 10^{-2}$ &  $3.60\cdot 10^{-2}$ \\
		50  &  $2.83\cdot 10^{-3}$ &  $1.25\cdot 10^{-2}$ \\
		100 &  $2.07\cdot 10^{-3}$ &  $2.57\cdot 10^{-3}$ \\
		200 &  $1.22\cdot 10^{-3}$ &  $1.23\cdot 10^{-4}$ \\
		500 &  $1.17\cdot 10^{-3}$ &  $9.80\cdot 10^{-6}$ \\
		\hline       
	\end{tabular}
\end{table}

\begin{figure}[!ht]
	\centering
	\includegraphics[width=0.9\textwidth] {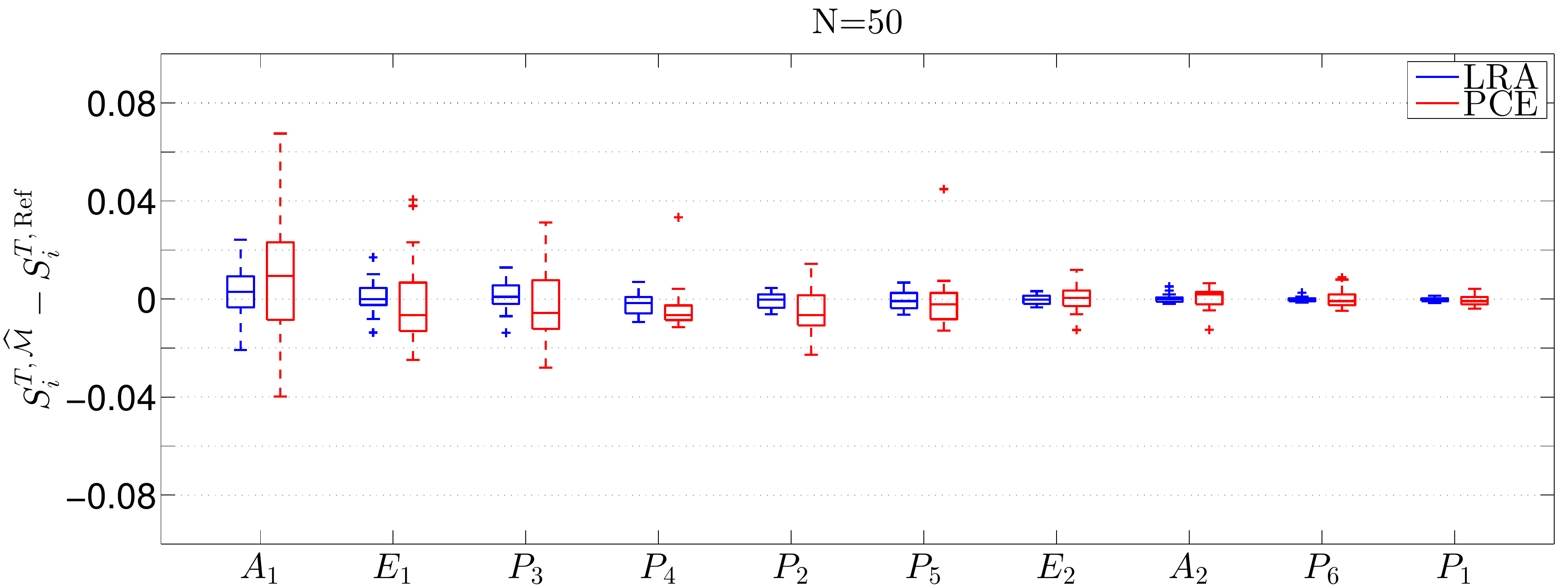}
	\caption{Truss deflection: Errors of LRA- and PCE-based total Sobol' indices for experimental designs obtained with maximin LHS (20 replications).}
	\label{fig:truss_Stot_LHS}
\end{figure}

\begin{figure}[!ht]
	\centering
	\includegraphics[width=0.45\textwidth] {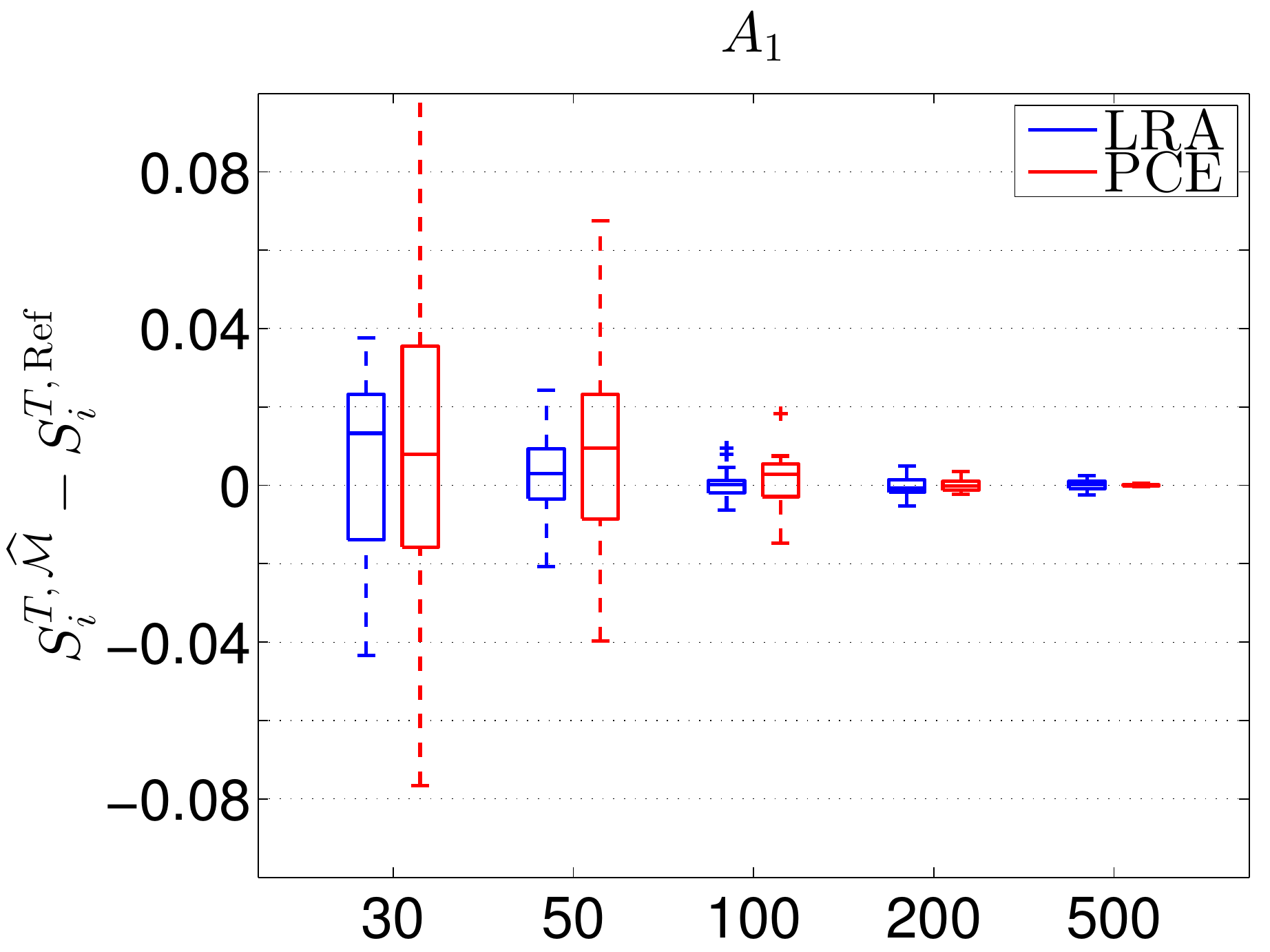}
	\includegraphics[width=0.45\textwidth] {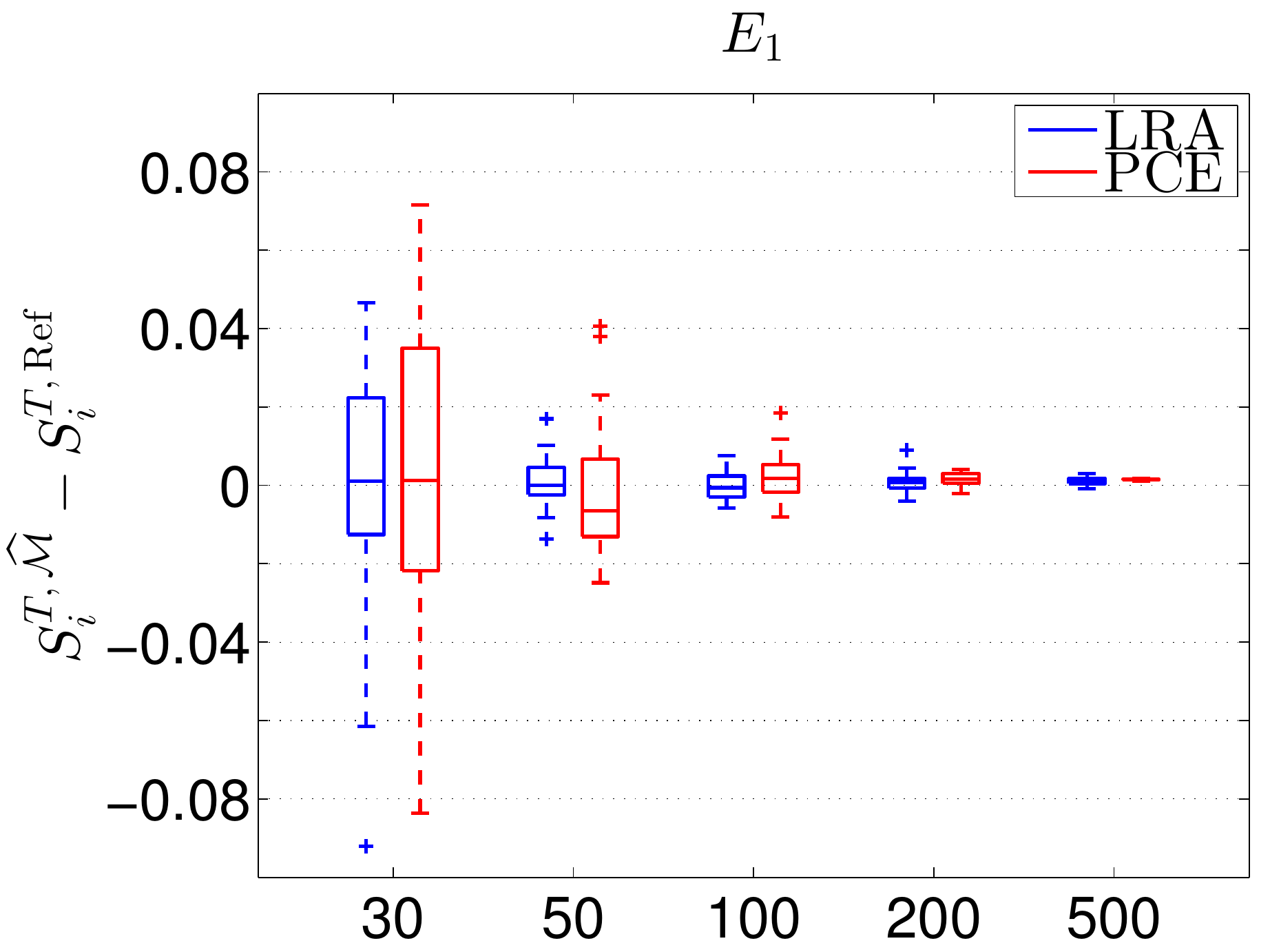}
	\caption{Truss deflection: Errors of LRA- and PCE-based total Sobol' indices of the two dominant variables for varying sizes of experimental designs obtained with maximin LHS (20 replications).}
	\label{fig:truss_Stot_LHS_N}
\end{figure}

As in the previous examples, we also investigate the accuracy of the LRA- and PCE-based Sobol' indices when the meta-models are built with maximin LHS designs. Figure~\ref{fig:truss_Stot_LHS} depicts boxplots of the differences between the total indices based on the meta-models and their reference values for 20 maximin LHS designs of size $N=50$. The LRA-based indices appear slightly superior to the PCE ones in terms of both the median value and the spread. For the two most influential variables, \ie $A_1$ and $E_1$, Figure~\ref{fig:truss_Stot_LHS_N} shows similar boxplots considering EDs of size varying from $N=30$ to $N=500$. For $N\leq 100$, the LRA-based indices exhibit a smaller spread, while for larger $N$, the spread of the estimates is nearly zero for both types of meta-models.

\subsubsection{Heat conduction with spatially varying diffusion coefficient}

\begin{figure}[!ht]
	\centering
	\includegraphics[trim = 0mm 15mm 0mm 10mm, width=0.49\textwidth] {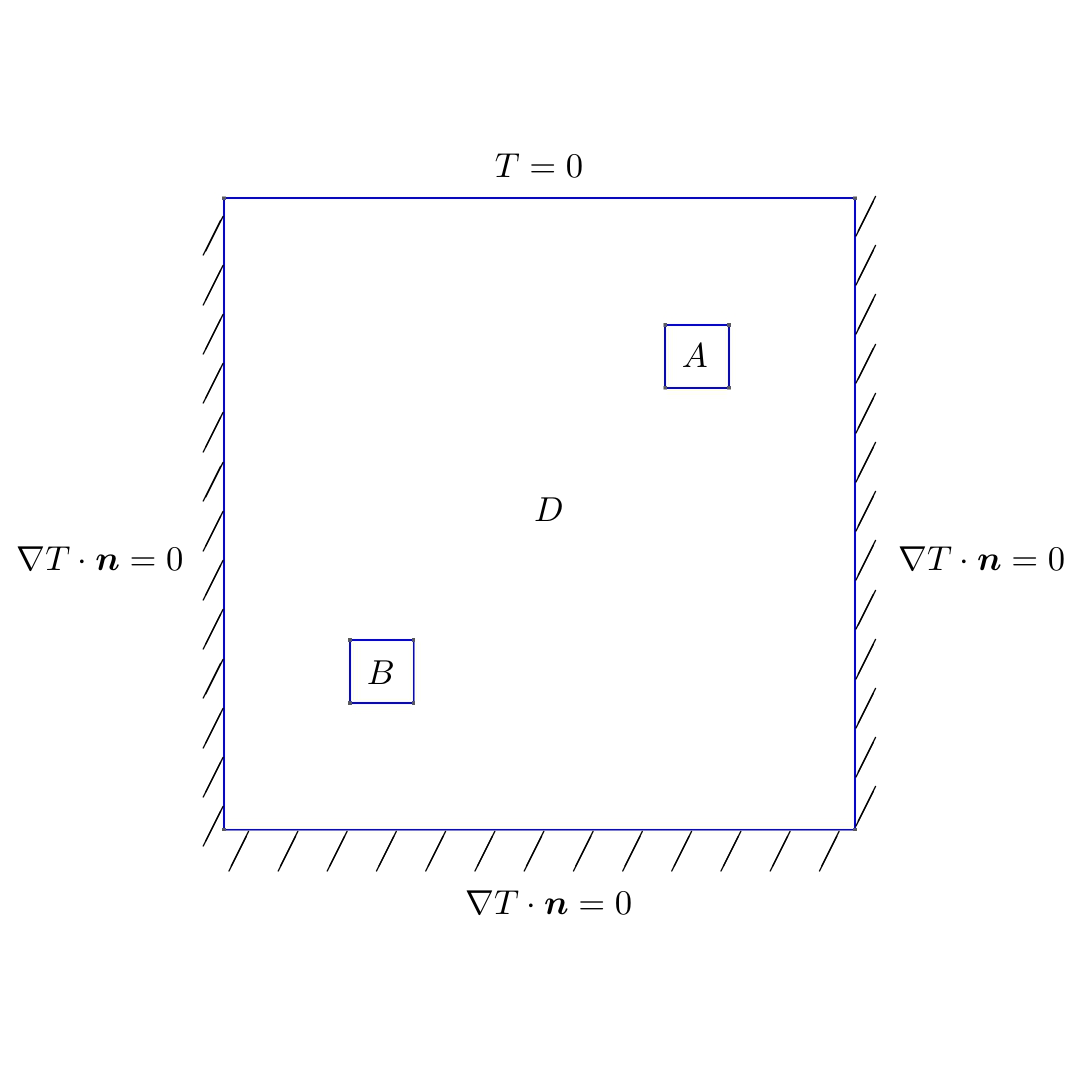}
	\includegraphics[trim = 0mm 15mm 0mm 10mm,width=0.49\textwidth] {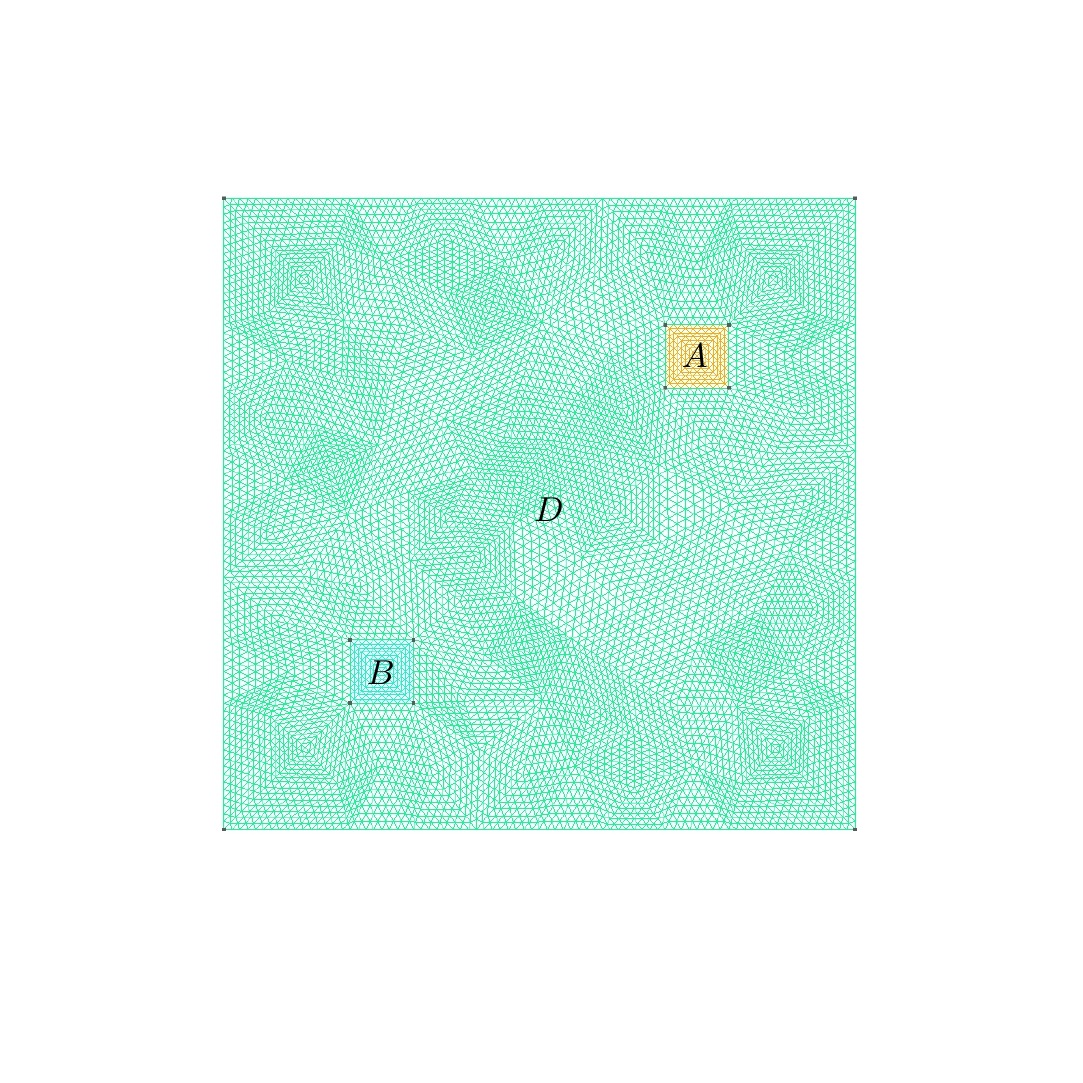}
	\caption{Heat conduction: Domain and boundary conditions (left); finite-element mesh (right).}
	\label{fig:RF_domain}
\end{figure}

We consider stationary heat conduction over the two-dimensional domain $D=(-0.5,0.5)~\rm m\times(-0.5,0.5)~\rm m$ shown in Figure~\ref{fig:RF_domain}, with the temperature field $T(\ve z)$, $\ve z \in D$, described by the partial differential equation:
\begin{equation}
\label{eq:diffusion_eq}
-\nabla\cdot(\kappa(\ve z)\hsp \nabla T(\ve z)) =Q \hsp I_A(\ve z)
\end{equation}
and boundary conditions:  $T=0$ on the top boundary; $\nabla T \cdot \ve n=0$ on the left, right and bottom boundaries, where $\ve n$ is the vector normal to the boundary. In Eq.~(\ref{eq:diffusion_eq}), $Q=500~\rm W/m^3$, $A=(0.2,0.3)~\rm m\times(0.2,0.3)~\rm m$ is a square domain within $D$ (see Figure~\ref{fig:RF_domain}) and $I_A$ is the indicator function equal to $1$ if $\ve z\in A$ and $0$ otherwise. The diffusion coefficient, $\kappa(\ve z)$, is a lognormal random field defined as:
\begin{equation}
\label{eq:diffusion_coef}
\kappa(\ve z)=\exp[a_{\kappa}+b_{\kappa} \hsp g(\ve z)],
\end{equation}
where $g(\ve z)$ denotes a standard Gaussian random field with autocorrelation function:
\begin{equation}
\label{eq:autocorr}
\rho(\ve z,\ve z')=\exp{(-\|\ve z-\ve z'\|^2/\ell^2)}.
\end{equation}
In Eq.~(\ref{eq:diffusion_coef}), the parameters $a_{\kappa}$ and $b_{\kappa}$ are such that the mean and standard deviation of $\kappa$ are $\mu_{\kappa}=1~\rm W/°C\cdot m$ and $\sigma_{\kappa}=0.3~\rm W/°C\cdot m$, respectively, while in Eq.~(\ref{eq:autocorr}), $\ell=0.2~\rm m$. The response quantity of interest is the average temperature $\overline{T}$ in the square domain $B=(-0.3,-0.2)~\rm m\times(-0.3,-0.2)~\rm m$ (see Figure \ref{fig:RF_domain}):
\begin{equation}
\label{eq:diffusion_Y}
\overline{T}=\frac{1}{|B|}\int_{\ve z \in B}T(\ve z) \hsp d\ve z.
\end{equation}

\begin{figure}[!ht]
	\centering
	\includegraphics[trim = 25mm 25mm 25mm 25mm, width=0.60\textwidth] {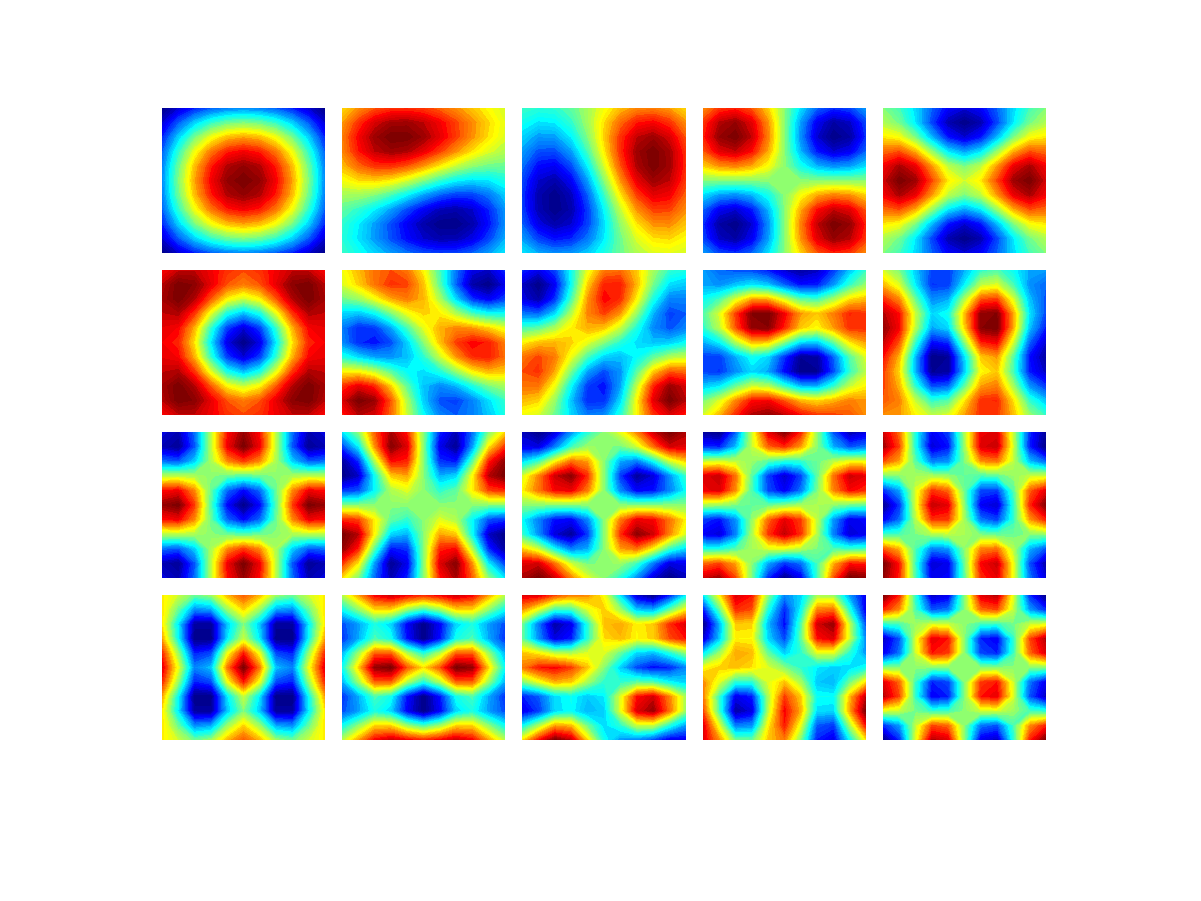}
	\caption{Heat conduction: Shapes of the first 20 basis functions in the EOLE discretization (from left-top to bottom-right row-wise).}
	\label{fig:KL_modes}
\end{figure}

Using the Expansion Optimal Linear Estimation (EOLE) method \citep{Li1993optimal}, the random field $g(\ve z)$ is approximated by:
\begin{equation}
\label{eq:EOLE}
\widehat{g}(\ve z) = \sum_{i=1}^M \frac{\xi_i}{\sqrt{l_i}}\ve {\phi}_i^{\rm{T}} \ve C_{\ve z \ve \zeta}.
\end{equation}
In the above equation, $\{\xi_1 \enum \xi_M\}$ are independent standard normal variables; $\ve C_{\ve z \ve \zeta}$ is a vector with elements $\ve C_{\ve z \ve \zeta}^{(k)}=\rho(\ve z,\ve \zeta_k)$, $k=1 \enum n$, where $\{\ve{\zeta}_1 \enum \ve{\zeta}_n\}$ are the points of an appropriately defined grid in $D$; $(l_i,\ve{\phi}_i)$ are the eigenvalues and eigenvectors of the correlation matrix $\ve C_{\ve \zeta \ve \zeta}$ with elements $\ve C_{\ve \zeta \ve \zeta}^{(k,l)}=\rho(\ve{\zeta}_k,\ve{\zeta}_l)$, $k,l=1 \enum n$. It is recommended in \citet{Sudret2000stochastic} that for a square-exponential autocorrelation function, the size of the element in the EOLE grid must be $1/2-1/3$ of $\ell$. Accordingly, in the present numerical application, we use a square grid with element size $0.01~\rm m$, thus comprising $n=121$ points. The number of terms in the EOLE series is determined according to the rule:
\begin{equation}
\label{eq:EOLE_M}
\sum_{i=1}^{M}l_i/\sum_{i=1}^{n} l_i \geq 0.99,
\end{equation}
which herein leads to $M=53$. The shapes of the first 20 basis functions $\{{\phi}_i^{\rm{T}} \ve C_{\ve z \ve \zeta}(\ve z), \hsp i=1 \enum 20\}$, are shown in Figure~\ref{fig:KL_modes}. Because the random input in this problem comprises independent standard normal variables, LRA and PCE meta-models are built using Hermite polynomials.

\begin{figure}[!ht]
	\centering
	\includegraphics[width=0.49\textwidth] {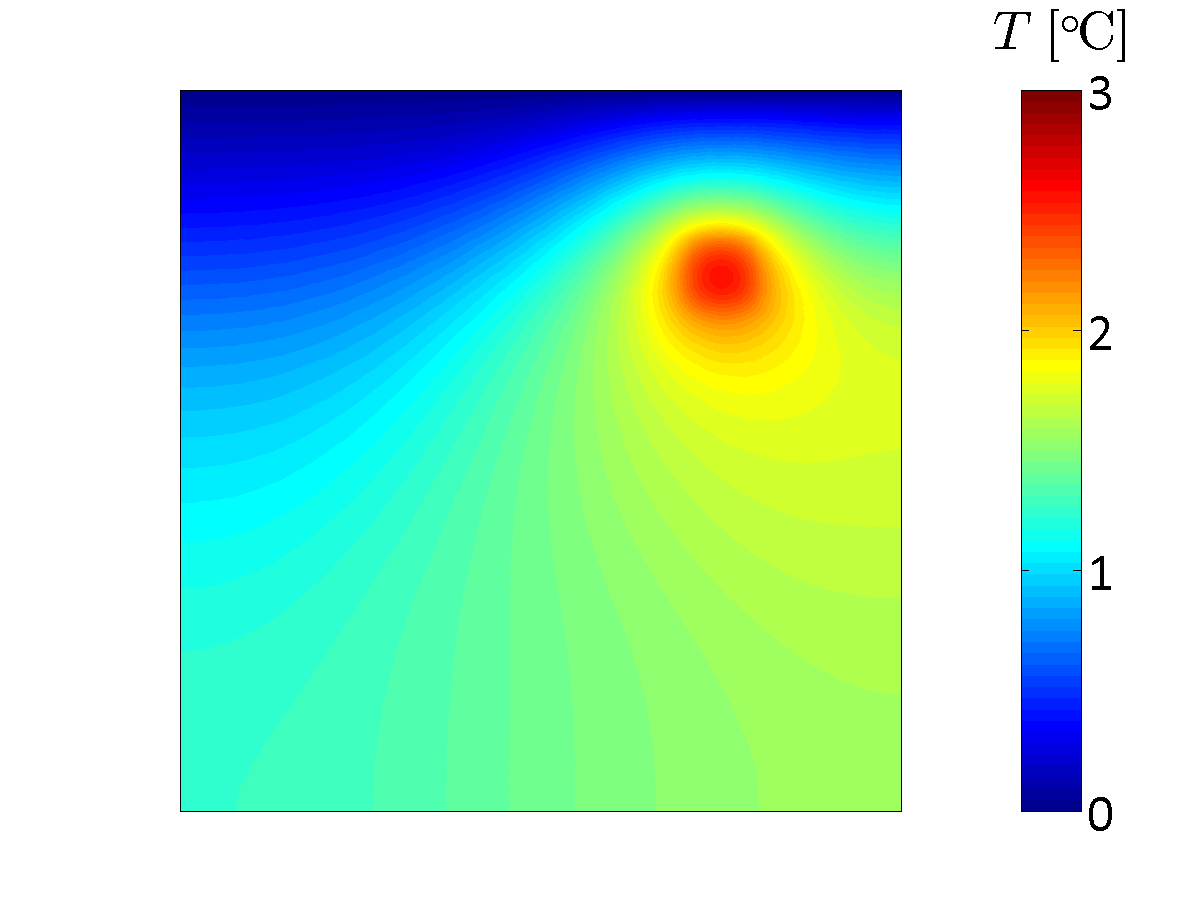}
	\includegraphics[width=0.49\textwidth] {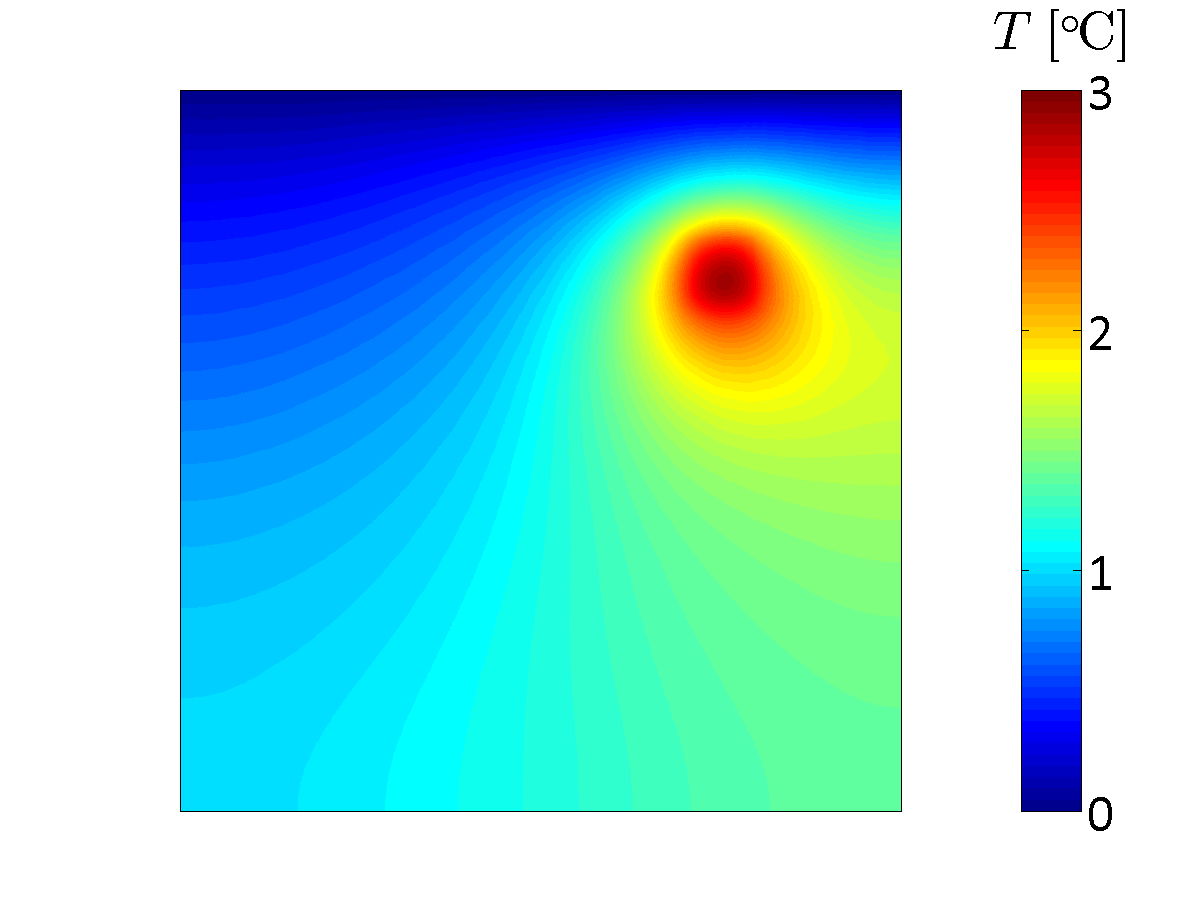}
	\caption{Heat conduction: Example realizations of the temperature field.}
	\label{fig:maps}
\end{figure}

For a given realization of $\{\xi_1 \enum \xi_M\}$, the model response considered ``exact'' in the meta-modeling application is obtained with an in-house finite-element analysis code developed in Matlab environment. The left graph of Figure~\ref{fig:RF_domain} depicts the discretization of the domain in $16,000$ triangular T3 elements, obtained with software \emph{Gmsh} \citep{Geuzaine2009gmsh}. The temperature field $T(\ve z)$ for two example realizations of the conductivity random field is shown in In Figure~\ref{fig:maps}.


\vspace{8mm}

\begin{figure}[!ht]
	\centering
	\includegraphics[width=0.9\textwidth] {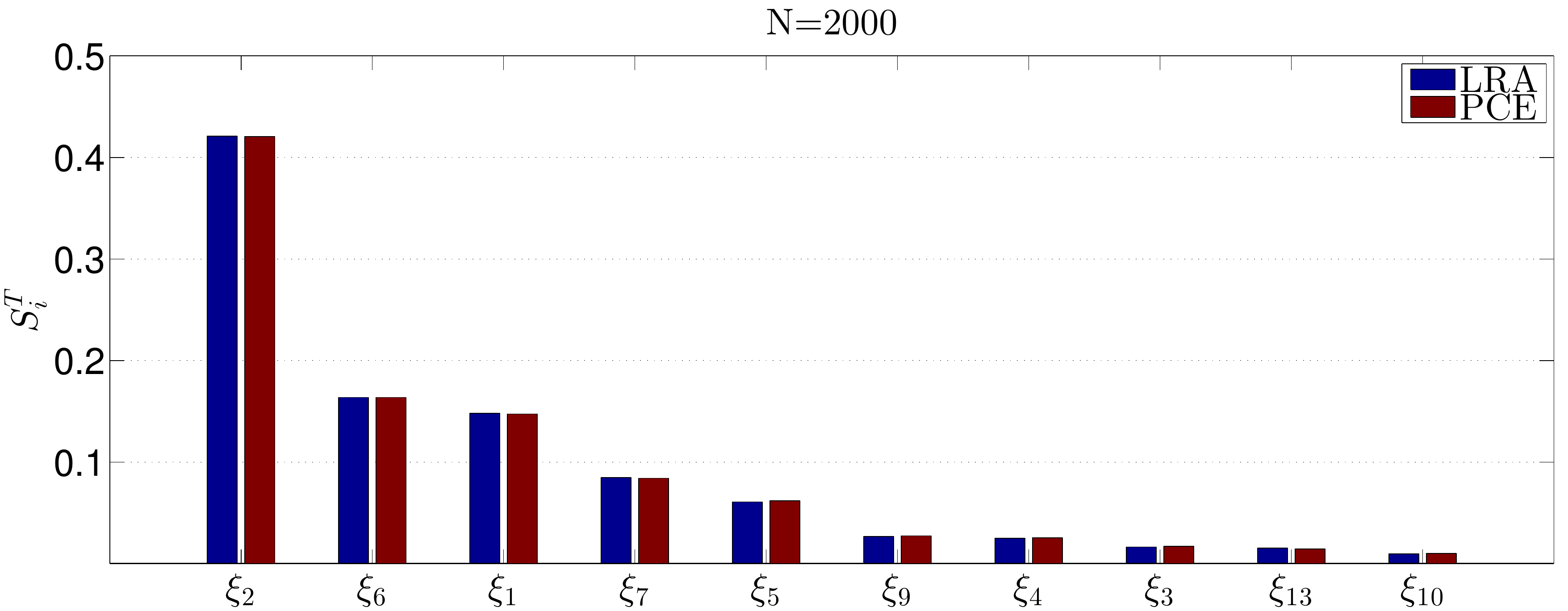}
	\caption{Heat conduction: Comparison between LRA- and PCE-based total Sobol' indices for an experimental design of size $N=2,000$ obtained with Sobol sequences.}
	\label{fig:RF_Stot}
\end{figure}

Because a single run of the considered ``exact'' model is computationally expensive (requires approximately $16~\rm sec$ with an Intel(R) Xeon(R) CPU E3-1225 v3 processor), evaluation of the Sobol' indices using MCS is impractical in this problem. Therefore, the following variance-based sensitivity analysis relies solely on the meta-modeling approach. Figure~\ref{fig:RF_Stot} shows the ten highest total Sobol' indices, obtained by post-processing the coefficients of LRA and PCE meta-models built with an ED of size $N=2,000$. The shown LRA- and PCE-based indices are practically identical and are thus, considered to represent the reference solution. (In the preceding applications, we showed that the indices based on LRA and PCE meta-models converge to the reference solution obtained with a MCS approach.) For the considered ED, a similar agreement is observed between the LRA-and PCE-based first-order indices. Because contributions from interaction effects are insignificant, the differences between the total and the first-order indices are negligible; the latter are thus not shown.

\begin{figure}[!ht]
	\centering
	\includegraphics[width=0.45\textwidth] {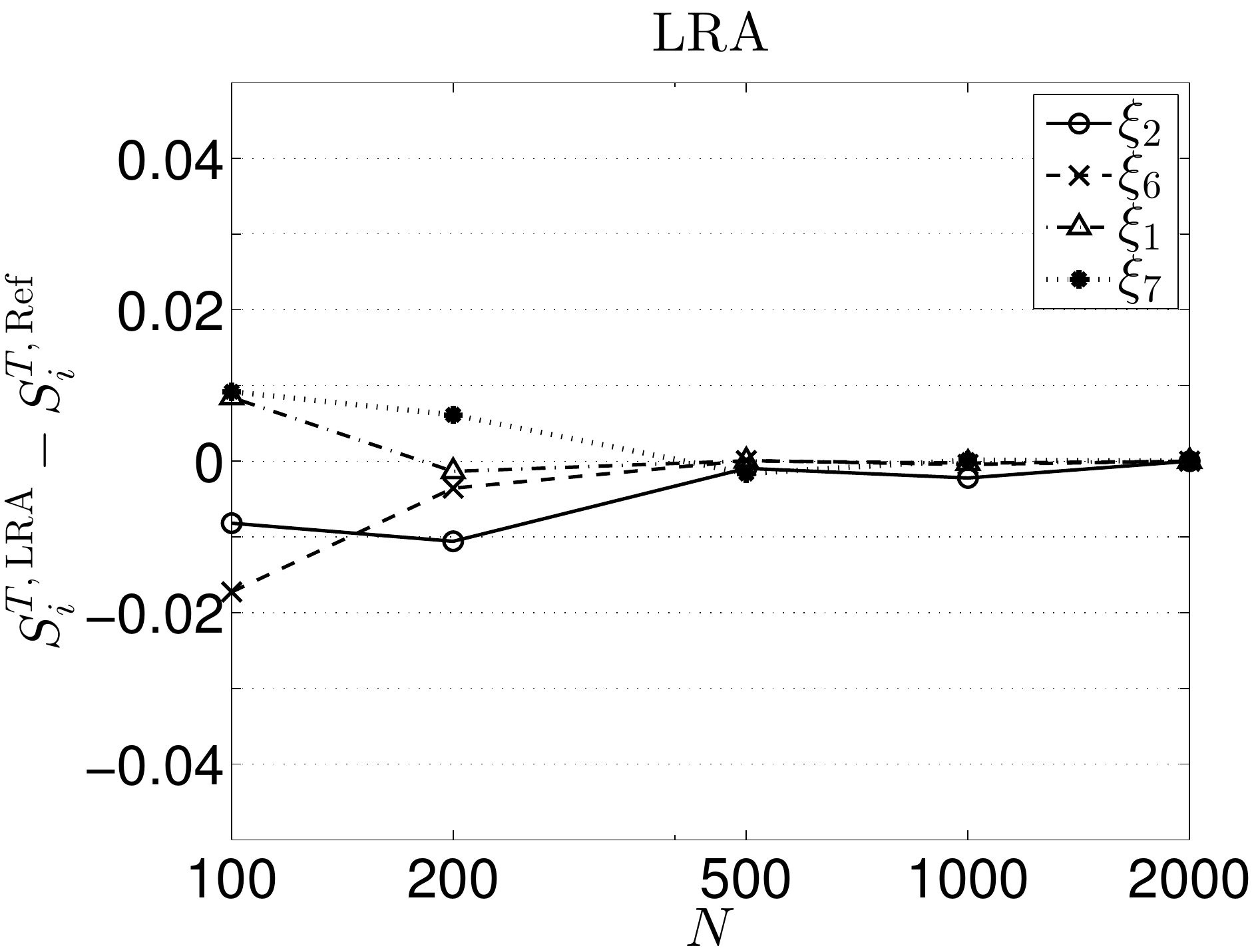}
	\includegraphics[width=0.45\textwidth] {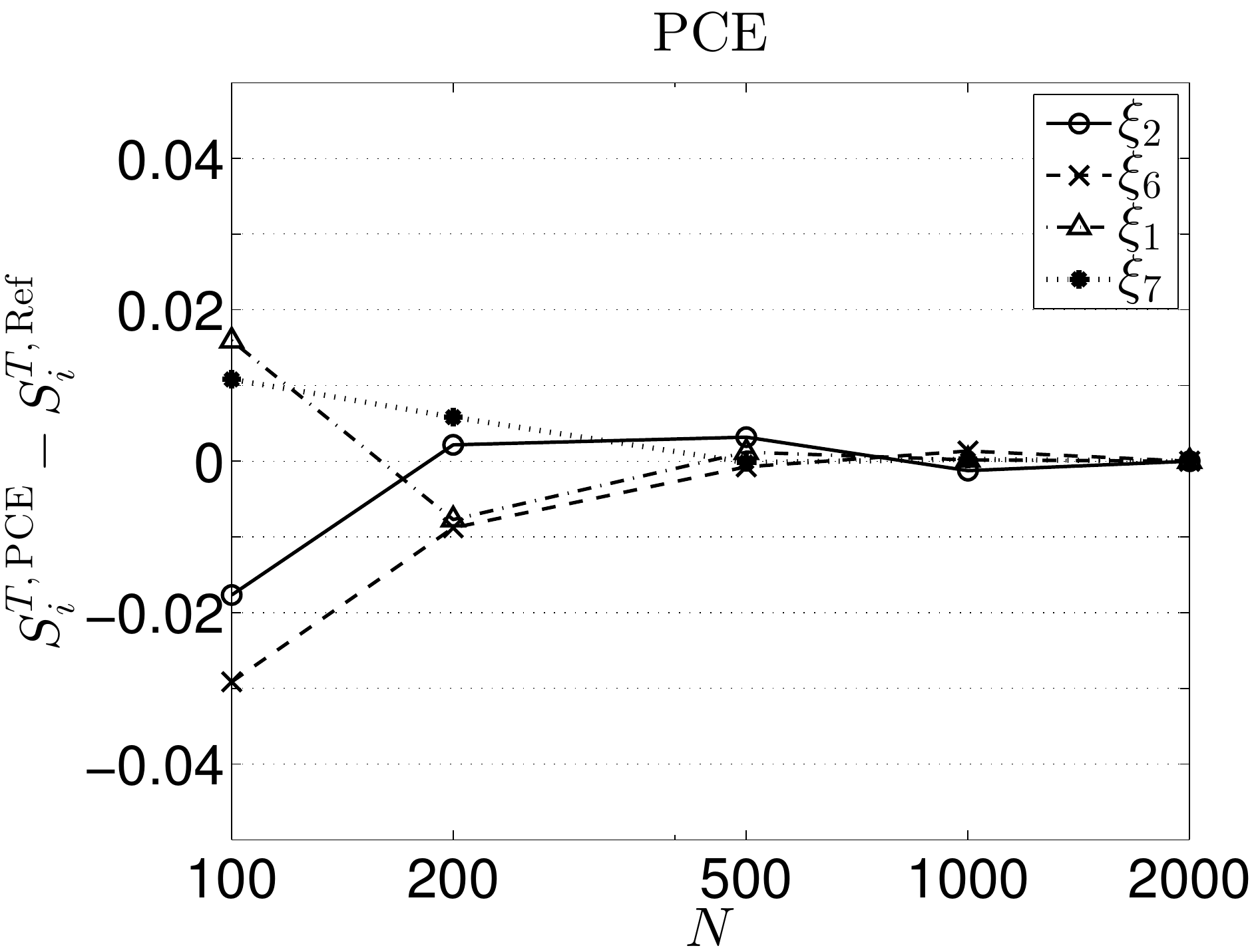}
	\caption{Heat conduction: Errors of LRA- and PCE-based total Sobol' indices for experimental designs obtained with Sobol sequences.}
	\label{fig:RF_Stot_N}
\end{figure}

In the following, we examine the convergence of LRA- and PCE-based indices to their reference values with increasing ED size. For the four most significant variables, \ie $\xi_2$, $\xi_6$, $\xi_1$ and $\xi_7$, Figure~\ref{fig:RF_Stot_N} shows the differences between the total indices and their reference values, while $N$ varies between 100 and 2,000. With the LRA approach, the indices of $\xi_6$ and $\xi_1$ practically reach their reference values at $N=200$, while the indices of $\xi_2$ and $\xi_7$ reach their reference values at $N=500$. The PCE-based indices demonstrate a slightly slower convergence. Similar results are obtained for the first-order indices (not shown herein). Table~\ref{tab:RF_errG} shows the relative generalization errors of the considered meta-models, estimated with a validation set comprising $10^4$ points. For $N\geq500$, the LRA meta-models are characterized by larger generalization errors than PCE, but as shown in Figure~\ref{fig:RF_Stot_N}, this does not essentially affect the results of the sensitivity analysis. The values of the first-order and the total indices for the example case with $N=200$ as well as for the reference case with $N=2,000$ are listed in the Appendix.


\begin{table} [!ht]
	\centering
	\caption{Heat conduction: Relative generalization errors of meta-models based on Sobol sequences.}
	\vspace{2mm} 
	\label{tab:RF_errG}
	\begin{tabular}{c c c}
		\hline
		N & $\widehat{err}_G^{\rm LRA}$ & $\widehat{err}_G^{\rm PCE}$\\
		\hline
		100  &  $2.46\cdot 10^{-2}$  &  $3.89\cdot 10^{-2}$ \\
		200  &  $1.38\cdot 10^{-2}$  &  $2.88\cdot 10^{-2}$ \\
		500  &  $9.68\cdot 10^{-3}$  &  $9.15\cdot 10^{-3}$ \\
		1,000 &  $8.19\cdot 10^{-3}$  &  $2.19\cdot 10^{-3}$ \\
		2,000 &  $7.72\cdot 10^{-3}$  &  $9.58\cdot 10^{-4}$ \\
		\hline       
	\end{tabular}
\end{table}

\begin{figure}[!ht]
	\centering
	\includegraphics[width=0.9\textwidth] {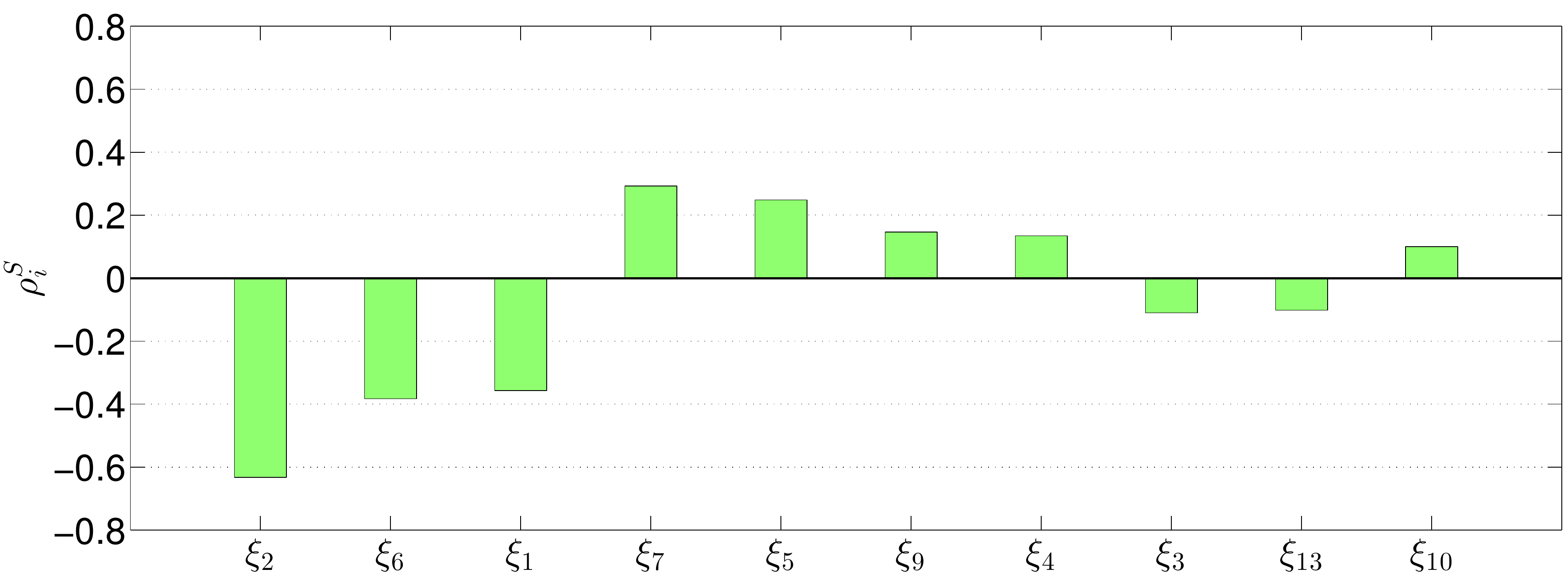}
	\caption{Heat conduction: Sensitivity measures based on the Spearman rank correlation.}
	\label{fig:RF_rhoS}
\end{figure}

Although no reference solution for the Sobol' indices is available in the present example, one may obtain a reference ranking of the input variables in terms of simpler sensitivity measures, by relying on the validation set used above to compute the meta-model errors. One such sensitivity measure is the Spearman rank correlation $\rho^S_i$ between an input variable $X_i$ and the model response of interest $Y$, describing the linear correlation between the indices (ranks) of the \emph{ordered} samples of $X_i$ and $Y$. For the considered heat-conduction model, Figure~\ref{fig:RF_rhoS} depicts the ten most significant variables according to the absolute values of $\rho^S_i$, computed with the validation set. The ranking of the input variables in terms of the Spearman correlation based on the actual model is identical to their ranking in terms of the total Sobol' indices based on the meta-models. This observation provides a validation of the sensitivity-analysis results obtained with the meta-models under the premise that the model under consideration behaves nearly monotonically.

\section{Conclusions}

In this paper, we introduce a novel approach for computing the Sobol' sensitivity indices by post-processing the coefficients of meta-models belonging to the class of low-rank approximations (LRA). The derived analytical expressions concern the particular case when LRA meta-models are built with polynomial functions due to the simplicity and versatility these offer. The proposed method can be particularly efficient in problems with high-dimensional input because: (i) the number of unknowns in LRA grows only linearly with the input dimension and (ii) the construction of LRA relies on a series of least-square minimization problems of small size, independent of the input dimension. The novel approach is presented in parallel to revisiting Sobol' sensitivity analysis by means of polynomial chaos expansions (PCE), which also relies on post-processing coefficients of meta-models with polynomial bases.

Following the detailed description of the construction of LRA in a non-intrusive manner and the derivation of the associated sensitivity indices, the accuracy and efficiency of the novel approach is investigated in example applications involving analytical models with a rank-one structure and finite-element models deriving from structural mechanics and heat conduction. The LRA-based indices are compared to PCE-based ones and to respective reference indices, representing the exact solution or Monte-Carlo estimates based on the actual model response. It is first shown that the indices obtained with the meta-models, using only a few tens or hundreds evaluations of the original model, provide excellent approximations of the exact indices or their Monte-Carlo estimates using a large number of evaluations of the original model. The heat-conduction example then demonstrates that the meta-modeling approach can be used to estimate the Sobol' indices in a case when their computation with the original model is not affordable; such situations are often encountered by practitioners in real-life problems. In the examined applications, the LRA-based indices tend to outperform the PCE-based ones by converging faster to the reference solution. 

Despite the strong promise of the proposed approach for the global sensitivity analysis of computationally costly high-dimensional models, further investigations are needed in order to establish the capacities and limitations of LRA as a general meta-modeling tool. Because evaluation of the LRA coefficients relies on a series of alternated minimizations along separate dimensions, issues of convergence may be encountered in certain cases. Investigations into these issues are currently underway.

\bibliographystyle{chicago}
\bibliography{bib}

\FloatBarrier

\section*{Appendix}

\subsection*{A.1. Sobol function}

\begin{table}[!ht]        
	\centering 
	\caption{Sobol function: Parameters and error estimates of LRA and PCE meta-models.}
	\vspace{2mm}            
	\begin{tabular}{c c c c c c c c}
		\hline
		{} & \multicolumn{3}{c} {LRA} & \multicolumn{3}{c} {PCE} \\
		{} & $R$ & $p$ & $\errCV$ & $q$ & $p^t$ & $\errLOO$\\
		\hline                                
		$N=200$ & $1$ & $6$ & $4.89\cdot10^{-2}$ & $0.5$ & $9$ &  $1.50\cdot10^{-2}$ \\
		$N=500$ & $1$ & $14$ & $7.11\cdot10^{-3}$ & $0.5$ & $13$ & $5.49\cdot10^{-3}$\\
		\hline
	\end{tabular} 
\end{table}

\begin{table}[!ht]        
	\centering 
	\caption{Sobol function: First-order Sobol' indices (see Figure \ref{fig:Sobol_S1}).}
	\vspace{2mm}             
	\begin{tabular}{c c c c c c}
		\hline
		{} & {} & \multicolumn{2}{c} {$N = 200$} & \multicolumn{2}{c} {$N = 500$} \\
		Variable & Analytical & LRA & PCE & LRA & PCE \\
		\hline                                
		$X_1$ & 0.6037 & 0.5820 & 0.6150 & 0.6032 & 0.6124 \\
		$X_2$ & 0.2683 & 0.2798 & 0.2707 & 0.2680 & 0.2638 \\
		$X_3$ & 0.0671 & 0.0650 & 0.0682 & 0.0665 & 0.0636 \\
		$X_4$ & 0.0200 & 0.0172 & 0.0165 & 0.0203 & 0.0197 \\
		$X_5$ & 0.0055 & 0.0059 & 0.0028 & 0.0055 & 0.0055 \\
		\hline
	\end{tabular} 
\end{table}

\begin{table}[!ht]        
	\centering 
	\caption{Sobol function: Total Sobol' indices (see Figure \ref{fig:Sobol_Stot}).}
	\vspace{2mm}             
	\begin{tabular}{c c c c c c}
		\hline
		{} & {} & \multicolumn{2}{c} {$N = 200$} & \multicolumn{2}{c} {$N = 500$} \\
		Variable & Analytical & LRA & PCE & LRA & PCE \\
		\hline                                
		$X_1$ & 0.6342 & 0.6144 & 0.6279 & 0.6337 & 0.6419 \\
		$X_2$ & 0.2944 & 0.3077 & 0.2844 & 0.2941 & 0.2877 \\
		$X_3$ & 0.0756 & 0.0737 & 0.0812 & 0.0750 & 0.0716 \\
		$X_4$ & 0.0226 & 0.0196 & 0.0191 & 0.0230 & 0.0220 \\
		$X_5$ & 0.0062 & 0.0068 & 0.0051 & 0.0062 & 0.0063 \\
		\hline
	\end{tabular} 
\end{table}

\subsection*{A.2. Beam deflection}

\begin{table}[!ht]        
	\centering 
	\caption{Beam deflection: Parameters and error estimates of LRA and PCE meta-models.}
	\vspace{2mm}             
	\begin{tabular}{c c c c c c c c}
		\hline
		{} & \multicolumn{3}{c} {LRA} & \multicolumn{3}{c} {PCE} \\
		{} & $R$ & $p$ & $\errCV$ & $q$ & $p^t$ & $\errLOO$\\
		\hline                                
		$N=30$ & $1$ & $2$ & $1.21\cdot10^{-4}$ & $1$ & $2$ &  $3.13\cdot10^{-2}$ \\
		$N=50$ & $1$ & $3$ & $ 3.14\cdot10^{-7}$& $1$ & $2$ & $1.56\cdot10^{-3}$\\
		\hline
	\end{tabular} 
\end{table}

\begin{table}[!ht]        
	\centering 
	\caption{Beam deflection: First-order Sobol' indices (see Figure \ref{fig:beam_S1}).}
	\vspace{2mm}             
	\begin{tabular}{c c c c c c}
		\hline
		{} & {} & \multicolumn{2}{c} {$N = 30$} & \multicolumn{2}{c} {$N = 50$} \\
		Variable & Reference & LRA & PCE & LRA & PCE \\
		\hline                                
		$P$ & 0.4383 & 0.4351 & 0.4623 & 0.4381 & 0.4403 \\
		$h$ & 0.2499 & 0.2501 & 0.2328 & 0.2490 & 0.2485 \\
		$E$ & 0.2479 & 0.2486 & 0.2480 & 0.2467 & 0.2476 \\
		$b$ & 0.0282 & 0.0274 & 0.0217 & 0.0274 & 0.0276 \\
		$L$ & 0.0105 & 0.0100 & 0.0181 & 0.0099 & 0.0099 \\
		\hline
	\end{tabular} 
\end{table}

\begin{table}[!ht]        
	\centering 
	\caption{Beam deflection: Total Sobol' indices (see Figure \ref{fig:beam_Stot}).}
	\vspace{2mm}            
	\begin{tabular}{c c c c c c}
		\hline
		{} & {} & \multicolumn{2}{c} {$N = 30$} & \multicolumn{2}{c} {$N = 50$} \\
		Variable & Reference & LRA & PCE & LRA & PCE \\
		\hline                                
		$P$ & 0.4589 & 0.4565 & 0.4703 & 0.4597 & 0.4595 \\
		$h$ & 0.2661 & 0.2666 & 0.2445 & 0.2657 & 0.2635 \\
		$E$ & 0.2633 & 0.2651 & 0.2593 & 0.2632 & 0.2625 \\
		$b$ & 0.0299 & 0.0298 & 0.0228 & 0.0298 & 0.0296 \\
		$L$ & 0.0107 & 0.0109 & 0.0203 & 0.0108 & 0.0110 \\
		\hline
	\end{tabular} 
\end{table}

\subsection*{A.3 Truss deflection}

\begin{table}[!ht]        
	\centering 
	\caption{Truss deflection: Parameters and error estimates of LRA and PCE meta-models.}
	\vspace{2mm}             
	\begin{tabular}{c c c c c c c c}
		\hline
		{} & \multicolumn{3}{c} {LRA} & \multicolumn{3}{c} {PCE} \\
		{} & $R$ & $p$ & $\errCV$ & $q$ & $p^t$ & $\errLOO$\\
		\hline                                
		$N=50$ & $1$ & $2$ & $6.26\cdot10^{-3}$ & $0.25$ & $2$ &  $4.25\cdot10^{-2}$ \\
		$N=200$ & $1$ & $3$ & $ 1.52\cdot10^{-3}$& $0.75$ & $4$ & $ 3.99\cdot10^{-4}$\\
		\hline
	\end{tabular} 
\end{table}

\begin{table}[!ht]        
	\centering 
	\caption{Truss deflection: First-order Sobol' indices.}
	\vspace{2mm}             
	\begin{tabular}{c c c c c c}
		\hline
		{} & {} & \multicolumn{2}{c} {$N = 50$} & \multicolumn{2}{c} {$N = 200$} \\
		Variable & Reference & LRA & PCE & LRA & PCE \\
		\hline                                
		$A_1$ & 0.3664 & 0.3698 & 0.3913 & 0.3666 & 0.3662 \\
		$E_1$ & 0.3662 & 0.3688 & 0.3626 & 0.3661 & 0.3664 \\
		$P_3$ & 0.0777 & 0.0755 & 0.0776 & 0.0771 & 0.0766 \\
		$P_4$ & 0.0770 & 0.0725 & 0.0644 & 0.0747 & 0.0756 \\
		$P_2$ & 0.0383 & 0.0357 & 0.0344 & 0.0385 & 0.0368 \\
		$P_5$ & 0.0380 & 0.0367 & 0.0370 & 0.0370 & 0.0368 \\
		$E_2$ & 0.0137 & 0.0136 & 0.0118 & 0.0115 & 0.0126 \\
		$A_2$ & 0.0138 & 0.0131 & 0.0141 & 0.0120 & 0.0124 \\
		$P_6$ & 0.0059 & 0.0037 & 0.0051 & 0.0048 & 0.0048 \\
		$P_1$ & 0.0060 & 0.0037 & 0.0017 & 0.0047 & 0.0046 \\
		\hline
	\end{tabular} 
\end{table}

\begin{table}[!ht]        
	\centering 
	\caption{Truss deflection: Total Sobol' indices (see Figure \ref{fig:truss_Stot}).}
	\vspace{2mm}             
	\begin{tabular}{c c c c c c}
		\hline
		{} & {} & \multicolumn{2}{c} {$N = 50$} & \multicolumn{2}{c} {$N = 200$} \\
		Variable & Reference & LRA & PCE & LRA & PCE \\
		\hline                                
		$A_1$ & 0.3712 & 0.3743 & 0.3913 & 0.3712 & 0.3713 \\
		$E_1$ & 0.3696 & 0.3733 & 0.3626 & 0.3707 & 0.3715 \\
		$P_3$ & 0.0776 & 0.0769 & 0.0776 & 0.0785 & 0.0777 \\
		$P_4$ & 0.0773 & 0.0739 & 0.0644 & 0.0761 & 0.0767 \\
		$P_2$ & 0.0374 & 0.0364 & 0.0344 & 0.0392 & 0.0374 \\
		$P_5$ & 0.0374 & 0.0374 & 0.0370 & 0.0377 & 0.0373 \\
		$E_2$ & 0.0126 & 0.0138 & 0.0118 & 0.0117 & 0.0128 \\
		$A_2$ & 0.0126 & 0.0134 & 0.0141 & 0.0122 & 0.0126 \\
		$P_6$ & 0.0048 & 0.0038 & 0.0051 & 0.0049 & 0.0049 \\
		$P_1$ & 0.0047 & 0.0038 & 0.0017 & 0.0048 & 0.0047 \\
		\hline
	\end{tabular}
\end{table}

\subsection*{A.4 Heat conduction with spatially varying diffusion coefficient}

\begin{table}[!ht]        
	\centering 
	\caption{Heat conduction: Parameters and error estimates of LRA and PCE meta-models.}
	\vspace{2mm}            
	\begin{tabular}{c c c c c c c c}
		\hline
		{} & \multicolumn{3}{c} {LRA} & \multicolumn{3}{c} {PCE} \\
		{} & $R$ & $p$ & $\errCV$ & $q$ & $p^t$ & $\errLOO$\\
		\hline                                
		$N=200$ & $1$ & $1$ &  $1.84\cdot 10^{-2}$ & $0.25$ & $3$ &  $4.08\cdot 10^{-2}$ \\
		$N=2,000$ & $1$ & $2$ & $ 7.76\cdot 10^{-3}$ & $0.75$ & $3$ & $1.22\cdot 10^{-3}$\\
		\hline
	\end{tabular} 
\end{table}

\begin{table}[!ht]        
	\centering 
	\caption{Heat conduction: First-order Sobol' indices.}
	\vspace{2mm} 
	\vspace{2mm}            
	\begin{tabular}{c c c c c c}
		\hline
		{} & \multicolumn{2}{c} {$N = 200$} & \multicolumn{2}{c} {$N = 2,000$} \\
		Variable & LRA & PCE & LRA & PCE \\
		\hline                                
		$\xi_2$ & 0.4034 & 0.4227 & 0.4138 & 0.4132 \\   
		$\xi_6$ & 0.1563 & 0.1548 & 0.1597 & 0.1593 \\   
		$\xi_1$ & 0.1432 & 0.1398 & 0.1445 & 0.1433 \\   
		$\xi_7$ & 0.0885 & 0.0900 & 0.0825 & 0.0813 \\   
		$\xi_5$ & 0.0643 & 0.0667 & 0.0594 & 0.0593 \\   
		$\xi_9$ & 0.0277 & 0.0278 & 0.0260 & 0.0265 \\   
		$\xi_4$ & 0.0214 & 0.0188 & 0.0243 & 0.0243 \\   
		$\xi_3$ & 0.0154 & 0.0123 & 0.0162 & 0.0160 \\   
		$\xi_{13}$ & 0.0165 & 0.0179 & 0.0149 & 0.0139 \\
		$\xi_{10}$ & 0.0091 & 0.0072 & 0.0095 & 0.0094 \\
		\hline
	\end{tabular} 
\end{table}

\begin{table}[!ht]        
	\centering 
	\caption{Heat conduction: Total Sobol' indices (see Figure \ref{fig:RF_Stot}).}
	\vspace{2mm}            
	\begin{tabular}{c c c c c c}
		\hline
		{} & \multicolumn{2}{c} {$N = 200$} & \multicolumn{2}{c} {$N = 2,000$} \\
		Variable & LRA & PCE & LRA & PCE \\
		\hline                                
		$\xi_2$ & 0.4103 & 0.4227 & 0.4209 & 0.4206 \\   
		$\xi_6$ & 0.1601 & 0.1548 & 0.1637 & 0.1636 \\   
		$\xi_1$ & 0.1468 & 0.1398 & 0.1481 & 0.1476 \\   
		$\xi_7$ & 0.0908 & 0.0900 & 0.0847 & 0.0842 \\   
		$\xi_5$ & 0.0660 & 0.0667 & 0.0610 & 0.0619 \\   
		$\xi_9$ & 0.0285 & 0.0278 & 0.0267 & 0.0275 \\   
		$\xi_4$ & 0.0220 & 0.0188 & 0.0250 & 0.0256 \\   
		$\xi_3$ & 0.0158 & 0.0123 & 0.0166 & 0.0173 \\   
		$\xi_{13}$ & 0.0170 & 0.0179 & 0.0153 & 0.0147 \\
		$\xi_{10}$ & 0.0093 & 0.0072 & 0.0098 & 0.0101 \\
		\hline
	\end{tabular} 
\end{table}

\end{document}